\documentclass[12pt,fleqn]{book} 

\usepackage[top=3cm,bottom=3cm,left=3.2cm,right=3.2cm,headsep=10pt,letterpaper]{geometry} 

\usepackage{xcolor} 
\definecolor{ocre}{RGB}{100,70,201} 

\usepackage{avant} 
\usepackage{mathptmx} 

\usepackage{siunitx}

\usepackage{microtype} 
\usepackage[utf8]{inputenc} 
\usepackage[T1]{fontenc} 
\usepackage{rotating}

\usepackage{hyperref}
\usepackage{footnote}
\usepackage{biblatex}
\usepackage{hyperref}
\hypersetup{
    colorlinks=true,
    linkcolor=blue,
    filecolor=magenta,      
    urlcolor=cyan,
    pdftitle={Overleaf Example},
    pdfpagemode=FullScreen,
    }

\usepackage{caption}
\usepackage{subcaption}

\usepackage[nonumberlist,sort=standard]{glossaries}

\usepackage{multirow}
\makeglossaries

\definecolor{bittersweet}{rgb}{1.0, 0.44, 0.37}

\usepackage{titlesec} 

\usepackage{graphicx} 
\graphicspath{{Pictures/}} 

\usepackage{lipsum} 

\usepackage{tikz} 

\usepackage[english]{babel} 

\usepackage{enumitem} 
\setlist{nolistsep} 

\usepackage{booktabs} 

\usepackage{eso-pic} 

\usepackage{titletoc} 

\contentsmargin{0cm} 
\titlecontents{chapter}[1.25cm] 
{\addvspace{15pt}\large\sffamily\bfseries} 
{\color{ocre!60}\contentslabel[\Large\thecontentslabel]{1.25cm}\color{ocre}} 
{}  
{\color{ocre!60}\normalsize\sffamily\bfseries\;\titlerule*[.5pc]{.}\;\thecontentspage} 
\titlecontents{section}[1.25cm] 
{\addvspace{5pt}\sffamily\bfseries} 
{\contentslabel[\thecontentslabel]{1.25cm}} 
{}
{\sffamily\hfill\color{black}\thecontentspage} 
[]
\titlecontents{subsection}[1.25cm] 
{\addvspace{1pt}\sffamily\small} 
{\contentslabel[\thecontentslabel]{1.25cm}} 
{}
{\sffamily\;\titlerule*[.5pc]{.}\;\thecontentspage} 
[] 

\titlecontents{lsection}[0em] 
{\footnotesize\sffamily} 
{}
{}
{}

\titlecontents{lsubsection}[.5em] 
{\normalfont\footnotesize\sffamily} 
{}
{}
{}
 
\usepackage{fancyhdr} 

\fancypagestyle{plain}{\fancyhead{}} 

\makeatletter
\renewcommand{\cleardoublepage}{
\clearpage\ifodd\c@page\else
\hbox{}
\vspace*{\fill}
\thispagestyle{empty}
\newpage
\fi}

\usepackage{amsmath,amsfonts,amssymb,amsthm} 

\newtheoremstyle{ocrenumbox}
{0pt}
{0pt}
{\normalfont}
{}
{\small\bf\sffamily\color{ocre}}
{\;}
{0.25em}
{\small\sffamily\color{ocre}\thmname{#1}\nobreakspace\thmnumber{\@ifnotempty{#1}{}\@upn{#2}}
\thmnote{\nobreakspace\the\thm@notefont\sffamily\bfseries\color{black}---\nobreakspace#3.}} 

\newtheoremstyle{blacknumex}
{5pt}
{5pt}
{\normalfont}
{} 
{\small\bf\sffamily}
{\;}
{0.25em}
{\small\sffamily{\tiny\ensuremath{\blacksquare}}\nobreakspace\thmname{#1}\nobreakspace\thmnumber{\@ifnotempty{#1}{}\@upn{#2}}
\thmnote{\nobreakspace\the\thm@notefont\sffamily\bfseries---\nobreakspace#3.}}

\newtheoremstyle{blacknumbox} 
{0pt}
{0pt}
{\normalfont}
{}
{\small\bf\sffamily}
{\;}
{0.25em}
{\small\sffamily\thmname{#1}\nobreakspace\thmnumber{\@ifnotempty{#1}{}\@upn{#2}}
\thmnote{\nobreakspace\the\thm@notefont\sffamily\bfseries---\nobreakspace#3.}}

\newtheoremstyle{ocrenum}
{5pt}
{5pt}
{\normalfont}
{}
{\small\bf\sffamily\color{ocre}}
{\;}
{0.25em}
{\small\sffamily\color{ocre}\thmname{#1}\nobreakspace\thmnumber{\@ifnotempty{#1}{}\@upn{#2}}
\thmnote{\nobreakspace\the\thm@notefont\sffamily\bfseries\color{black}---\nobreakspace#3.}} 
\makeatother

\newcounter{dummy} 
\numberwithin{dummy}{section}
\theoremstyle{ocrenumbox}
\newtheorem{theoremeT}[dummy]{Theorem}

\newtheorem{exerciseT}{Exercise}[chapter]
\theoremstyle{blacknumex}
\newtheorem{exampleT}{Example}[chapter]
\theoremstyle{blacknumbox}

\newtheorem{definitionT}{Definition}[section]
\newtheorem{corollaryT}[dummy]{Corollary}
\theoremstyle{ocrenum}

\RequirePackage[framemethod=default]{mdframed} 

\newmdenv[skipabove=7pt,
skipbelow=7pt,
backgroundcolor=black!5,
linecolor=ocre,
innerleftmargin=5pt,
innerrightmargin=5pt,
innertopmargin=5pt,
leftmargin=0cm,
rightmargin=0cm,
innerbottommargin=5pt]{tBox}

\newmdenv[skipabove=7pt,
skipbelow=7pt,
rightline=false,
leftline=true,
topline=false,
bottomline=false,
backgroundcolor=ocre!10,
linecolor=ocre,
innerleftmargin=5pt,
innerrightmargin=5pt,
innertopmargin=5pt,
innerbottommargin=5pt,
leftmargin=0cm,
rightmargin=0cm,
linewidth=4pt]{eBox}	

\newmdenv[skipabove=7pt,
skipbelow=7pt,
rightline=false,
leftline=true,
topline=false,
bottomline=false,
linecolor=ocre,
innerleftmargin=5pt,
innerrightmargin=5pt,
innertopmargin=0pt,
leftmargin=0cm,
rightmargin=0cm,
linewidth=4pt,
innerbottommargin=0pt]{dBox}	

\newmdenv[skipabove=7pt,
skipbelow=7pt,
rightline=false,
leftline=true,
topline=false,
bottomline=false,
linecolor=gray,
backgroundcolor=black!5,
innerleftmargin=5pt,
innerrightmargin=5pt,
innertopmargin=5pt,
leftmargin=0cm,
rightmargin=0cm,
linewidth=4pt,
innerbottommargin=5pt]{cBox}

\makeatletter
\renewcommand{\@seccntformat}[1]{\llap{\textcolor{ocre}{\csname the#1\endcsname}\hspace{1em}}}                    
\renewcommand{\section}{\@startsection{section}{1}{\z@}
{-4ex \@plus -1ex \@minus -.4ex}
{1ex \@plus.2ex }
{\normalfont\large\sffamily\bfseries}}
\renewcommand{\subsection}{\@startsection {subsection}{2}{\z@}
{-3ex \@plus -0.1ex \@minus -.4ex}
{0.5ex \@plus.2ex }
{\normalfont\sffamily\bfseries}}
\renewcommand{\subsubsection}{\@startsection {subsubsection}{3}{\z@}
{-2ex \@plus -0.1ex \@minus -.2ex}
{.2ex \@plus.2ex }
{\normalfont\small\sffamily\bfseries}}                        
\renewcommand\paragraph{\@startsection{paragraph}{4}{\z@}
{-2ex \@plus-.2ex \@minus .2ex}
{.1ex}
{\normalfont\small\sffamily\bfseries}}

\usepackage{hyperref}
\hypersetup{hidelinks,backref=true,pagebackref=true,hyperindex=true,colorlinks=false,breaklinks=true,urlcolor= ocre,bookmarks=true,bookmarksopen=false,pdftitle={Title},pdfauthor={Author}}

\newcommand{\thechapterimage}{}
\newcommand{\chapterimage}[1]{\renewcommand{\thechapterimage}{#1}}

\def\thechapter{\arabic{chapter}}
\def\@makechapterhead#1{
\thispagestyle{empty}
{\centering \normalfont\sffamily
\ifnum \c@secnumdepth >\m@ne
\if@mainmatter
\startcontents
\begin{tikzpicture}[remember picture,overlay]
\node at (current page.north west)
{\begin{tikzpicture}[remember picture,overlay]
\node[anchor=north west,inner sep=0pt] at (0,0) {\includegraphics[width=\paperwidth]{\thechapterimage}};
\draw[anchor=west] (5cm,-9cm) node [rounded corners=20pt,fill=ocre!10!white,text opacity=1,draw=ocre,draw opacity=1,line width=1.5pt,fill opacity=.6,inner sep=12pt]{\huge\sffamily\bfseries\textcolor{black}{\thechapter. #1\strut\makebox[22cm]{}}};
\end{tikzpicture}};
\end{tikzpicture}}
\par\vspace*{230\p@}
\fi
\fi}

\def\@makeschapterhead#1{
\thispagestyle{empty}
{\centering \normalfont\sffamily
\ifnum \c@secnumdepth >\m@ne
\if@mainmatter
\begin{tikzpicture}[remember picture,overlay]
\node at (current page.north west)
{\begin{tikzpicture}[remember picture,overlay]
\node[anchor=north west,inner sep=0pt] at (0,0) {\includegraphics[width=\paperwidth]{\thechapterimage}};
\draw[anchor=west] (5cm,-9cm) node [rounded corners=20pt,fill=ocre!10!white,fill opacity=.6,inner sep=12pt,text opacity=1,draw=ocre,draw opacity=1,line width=1.5pt]{\huge\sffamily\bfseries\textcolor{black}{#1\strut\makebox[22cm]{}}};
\end{tikzpicture}};
\end{tikzpicture}}
\par\vspace*{230\p@}
\fi
\fi
}
\makeatother 
\begin{document}

\pagestyle{fancy}
\begingroup
\thispagestyle{empty}
\AddToShipoutPicture*{\put(0,0){\includegraphics[scale=1]
{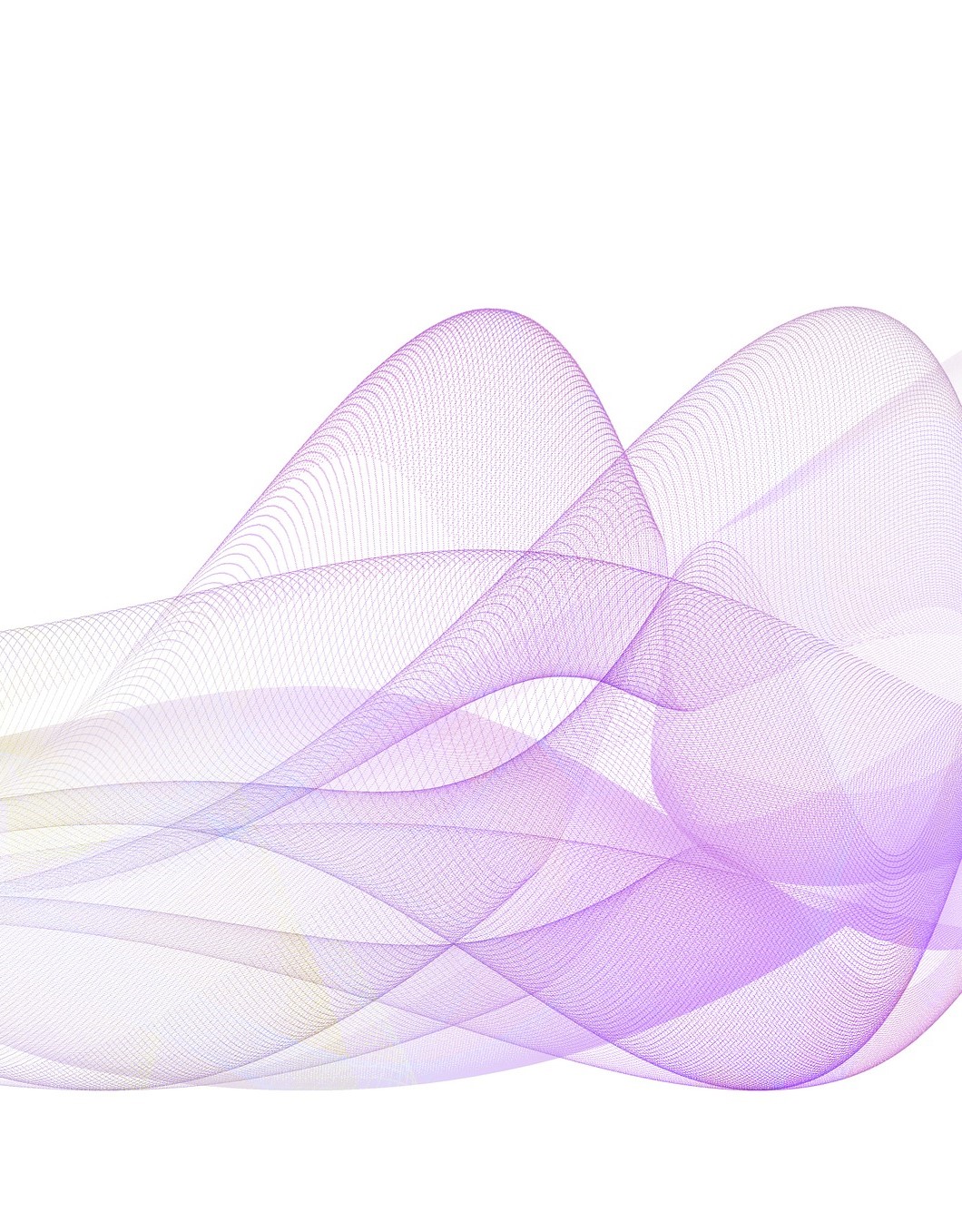}}} 
\centering
\vspace*{5cm}
\par\normalfont\fontsize{35}{35}\sffamily\selectfont
\textbf{Report on the Advanced Linear Collider Study Group (ALEGRO) Workshop 2026}\\
\vspace*{1cm}
{\huge ALEGRO 2026}\par 
\endgroup


\newpage
~\vfill
\thispagestyle{empty}

\noindent Copyright \copyright\ 2026 \\ 

\noindent \textsc{Report on the Advanced Linear Collider Study Group Workshop}\\

\noindent \href{https://agenda.infn.it/event/47329/overview}{\textsc{URL https://agenda.infn.it/event/47329/overview}
}\\ 


\noindent This report is the outcome of the ALEGRO2026 workshop held in March 2026 at INFN-LNF and supported by LNF.\\ 

\noindent \textit{First release date: July 12, 2026} 


\newpage

\begingroup
\thispagestyle{empty}
\centering
\vspace*{5cm}
\par\normalfont\fontsize{35}{35}\sffamily\selectfont
\textbf{Report on the Advanced Linear Collider Study Group (ALEGRO) Workshop 2026}\\
\vspace*{0.5cm}
{\Huge Livio Verra}\par 
\vspace{0.3cm}
\large{Chair of the ALEGRO 2026 workshop}\\
\vspace{0.5cm}
{\Huge Brigitte Cros, Patric Muggli}\par 
\vspace{0.3cm}
\large{for the ICFA Advanced and Novel Accelerators Panel}\\
\endgroup
\newpage

\thispagestyle{empty}
\vspace*{5cm}




{\large 'The best way to predict the future is to create it.'
\vspace{0.5cm}

Peter Drucker}












\newpage
\thispagestyle{empty}
 \begin{figure}[h]
	\begin{center}	
\includegraphics[width=7cm]{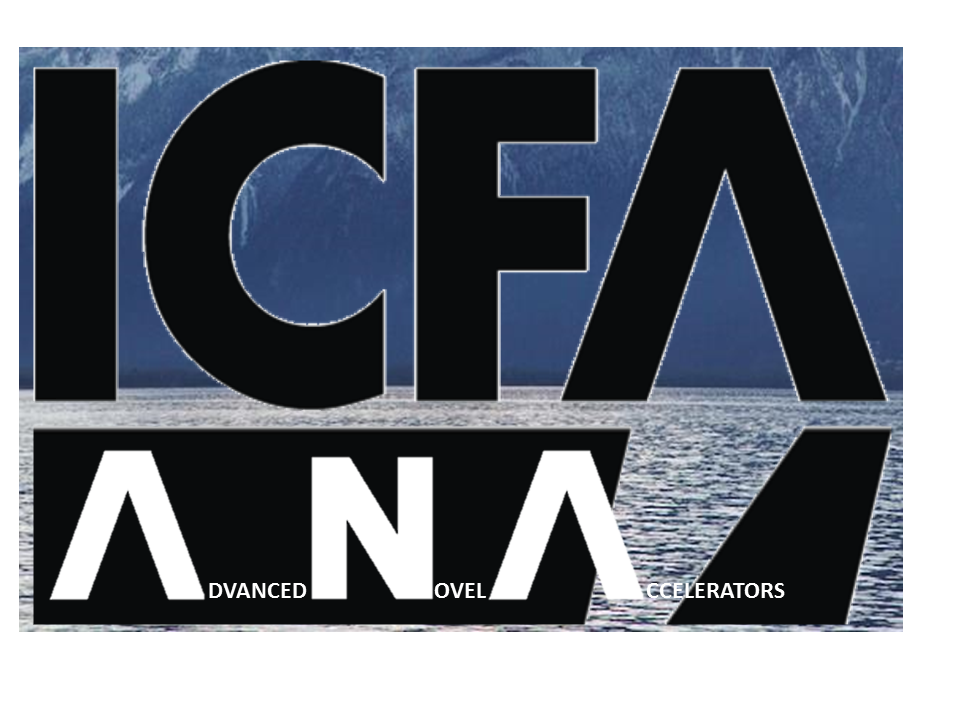}				\end{center}
\end{figure}
\begingroup
\par\normalfont\fontsize{20}{20}\sffamily\selectfont
\hspace*{2cm} \textbf{Executive Summary}\\
\endgroup


\newglossaryentry{ESPPU}{name={ESPPU},description={European Strategy for Particle Physics Update}}

As for previous editions, the workshop focused on the application of Advanced Novel Accelerators \newglossaryentry{ANA}{name={ANA},description={Advanced Novel Accelerators}} (ANA) to particle physics
\newglossaryentry{PP}{name={PP},description={Particle Physics}} (PP) and %
high-energy physics
\newglossaryentry{HEP}{name={HEP},description={High-Energy Physics}} (HEP), %
keeping in mind the ultimate goal of a collider at the energy frontier (10\,TeV, e$^+$e$^-$, e$^-$e$^-$, or $\gamma\gamma$). %
\newglossaryentry{e-}{name={e$^-$},description={electron}}
\newglossaryentry{e+}{name={e$^+$},description={positron}}
\newglossaryentry{p+}{name={p$^+$},description={proton}}
\newglossaryentry{IFEL}{name={IFEL},description={Inverse Free Electron lasers}}

The development of ANA is conducted at universities and national laboratories worldwide. %
The community is thematically broad and diverse, in particular since lasers suitable for ANA research (multi-hundred-terawatt peak power, a few tens of femtosecond-long pulses) and acceleration of electrons to hundreds of megaelectronvolt to multi-gigaelectronvolt became commercially available. %
The community spans  several continents (Europe, America, Asia), including more than 62 laboratories in more than 20 countries. %

It is one of the missions of the ICFA-ANA panel to feature the amazing progress made with ANA, to provide international coordination and to foster international collaborations towards a future HEP collider. %

The ALEGRO workshop was hosted March 3-5, 2026, by INFN-Frascati National Laboratory and chaired by Livio Verra. %

This workshop, the 7$^{th}$ of its kind,  focused on 
ANA relying on the excitation of wakefields by laser pulses or particle bunches in plasma or in corrugated cavities. %
In the following we thus refer to these concepts as advanced wakefield accelerators, or AWA. %
\newglossaryentry{AWA}{name={AWA},description={Advanced Wakefield Accelerators}}

The program was organized around the mid- and long-term possible contributions of the AWA community to PP and HEP. %

ALEGRO was formed around the concept of a linear collider at the energy frontier (10\,TeV and beyond). %
Such a collider is considered as the high-energy evolution of a (linear) Higgs factory. %
Following the latest recommendation of the P5 in the US (December 2023), a 10\,TeV collider design study was undertaken in 2025 and was one of the main features of the 2025 and 2026 workshops. %
A whole day was devoted to this project, and presentations covered all the aspects of such a collider, from the acceleration schemes to the physics case. %

However, its relevance strongly depends on the outcome of the European Strategy for Particle Physics Update (ESPPU). %
At this time, the lead next flagship project is the circular FCCee. %
\newglossaryentry{FCCee}{name={FCCee},description={Future Circular Collider e$^-$e$^-$}}%

Plasma-based Higgs factories have been proposed: HALHF \newglossaryentry{HALHF}{name={HALHF},description={Hybrid Asymmetric Linear Higgs Factory}} driven by an e$^-$ beam for the e$^-$ accelerator, and ALiVE \newglossaryentry{ALiVE}{name={ALiVE},description={Advanced Linear accelerator for Very high Energies}}, driven by a p$^+$ bunch. 

Independently of the outcome of the ESPPU, AWA also have shorter term applications. %
These include fixed target experiments (AWAKE\newglossaryentry{AWAKE}{name={AWAKE},description={Advanced Wakefield Experiment}}), and injectors for synchrotrons (FCCee, CEPC), but also for the EIC. %
\newglossaryentry{EIC}{name={EIC},description={Electron-Ion Collider}}%
In addition, proton-driven plasma wakefield acceleration could be used to boost the energy of the electron beam of LEP3, a possible alternative to FCCee, to  increase its energy reach to the $t\bar{t}$ center-of-mass energy. %
\newglossaryentry{LEP3}{name={LEP3},description={Large Electron Positron (Collider) 3}}%

Applications of ANA to photon sources, the other users of high-energy electron beams, have always been considered as first demonstrators of the ability of ANA to operate as reliable accelerators. %
These applications require challenging parameters, but much less so than collider ones. %
The workshop therefore also featured other applications, to free electron lasers \newglossaryentry{FEL}{name={FEL},description={Free-Electron Laser}}(FEL) such as EuPRAXIA@SPARC\_LAB, to injector for synchrotron light sources such as PETRA IV, and to strong-field quantum electro-dynamics (SF QED) studies. %
It is worth noting that lasing with electron beams produced by laser wakefield or plasma wakefield accelerators was recently observed in a number of FELs, demonstrating progress in electron beam quality. %

Challenges related to operation of user facilities, needing a laser (to drive a laser wakefield accelerator) or an electron beam source (to drive an FEL) were also presented as an indication of what it entails to operate an accelerator on a 24/7 basis, as required by most PP and HEP applications. %

Opportunities that Artificial Intelligence \newglossaryentry{AI}{name={AI},description={Artificial Intelligence}}(AI) and Machine Learning \newglossaryentry{ML}{name={ML},description={Machine Learning}}(ML) can bring to the field, and to the optimization of advanced accelerators were discussed in a general context. %

This document presents a summary of the workshop, as seen by the co-chairs, as well as short "two-pagers" written by the presenters at the workshop. %

The summary expresses the opinions of the chair and co-chairs of the workshop, and the "two-pagers" the opinion of their author(s). %
This report is an overview and offers the reader entry points for the different topics covered. %

\vspace{1cm}
Brigitte Cros, Patric Muggli, Livio Verra

On behalf of the ICFA-ANA panel

\newpage
\vspace{0.5cm}
\noindent \textbf{Members of the ICFA ANA panel, September 2025} 
\\

\noindent Laura Corner, \textit{ University of Liverpool, UK}

\noindent Brigitte Cros, \textit{Centre National de la Recherche Scientifique (CNRS) -- Université Paris Saclay, France}

\noindent Massimo Ferrario, \textit{Istituto Nazionale di Fisica Nucleare (INFN), Italy}

\noindent Spencer Gessner, \textit{SLAC National Accelerator Laboratory , USA}

\noindent Carl Lindstrøm, \textit{University of Oslo, Norway}

\noindent Masaki Masaki Kando, \textit{National Institutes for Quantum and Radiological Science and Technology, Japan}

\noindent Patric Muggli (Chair), \textit{Max Planck Institute for Physics (MPP), Germany, and CERN, Switzerland}

\noindent Pietro Musumeci, \textit{University of California, Los Angeles (UCLA), USA}

\noindent Inhyuk Nam, \textit{Ulsan National Institute of Science and Technology (UNIST), Korea}

\noindent Jens Osterhoff, \textit{Lawrence Berkeley National Laboratory (LBNL), USA}

\noindent John Power, \textit{Argonne National Laboratory (ANL), USA}

\noindent Chuanxiang Tang (previous chair), \textit{Tsinghua University, China}

\vspace{2.0cm}
\noindent \textbf{Acknowledgements}
\\
\\
\noindent This report is the result of the work of many people, including the ICFA-ANA panel members, the workshop organizing committee, the workshop participants, and the authors and co-authors of the written contributions. %
In addition, the contributions of the INFN - LNF administrative staff is acknowledged and greatly appreciated. %
\textbf{Thank you!} %
\newpage
\newglossaryentry{CERN}{name={CERN},description={European Organization for Nuclear Research}}
\newglossaryentry{SLAC}{name={SLAC},description={Stanford Linear Accelerator Center}}
\newglossaryentry{LBNL}{name={LBNL},description={Lawrence Berkeley National Laboratory}}
\newglossaryentry{DESY}{name={DESY},description={Deutsches Elektronen-SYnchrotron}}
\newglossaryentry{INFN}{name={INFN},description={Istituto Nazionale di Fisica Nucleare}}
\newglossaryentry{RD}{name={R\&D},description={Research and Development}}
\newglossaryentry{CLIC}{name={CLIC},description={Compact LInear Collider}}

\newpage


\chapterimage{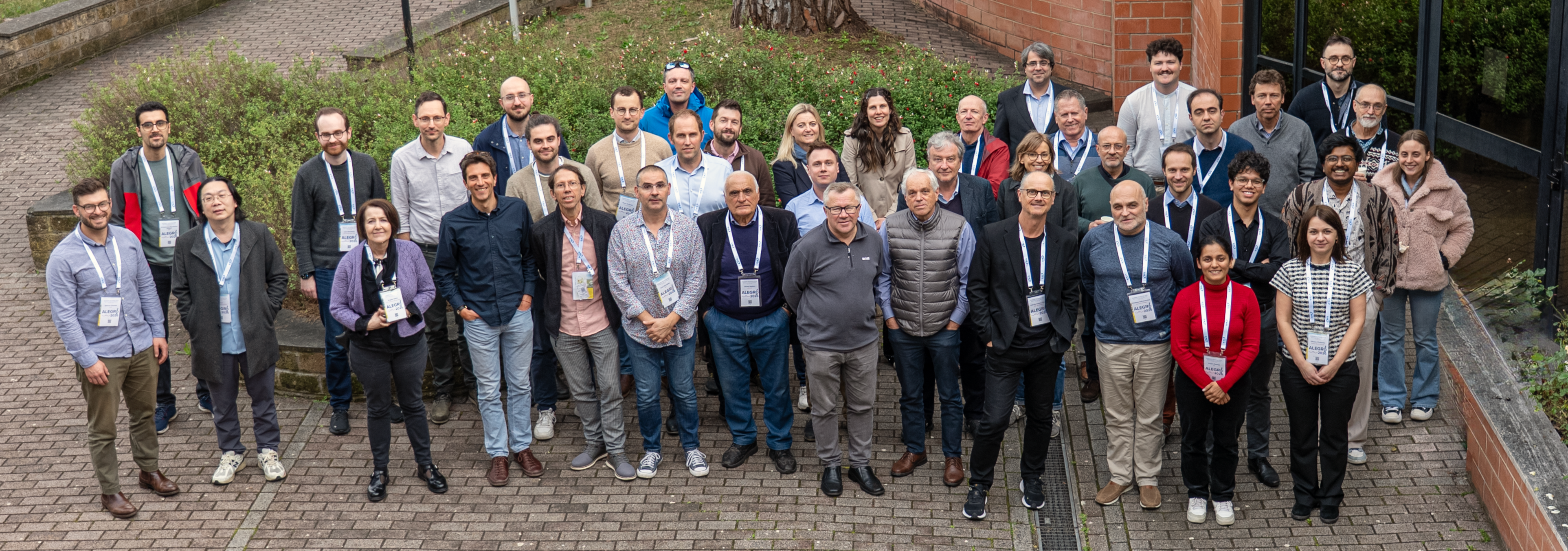} 

\pagestyle{empty} 

\cleardoublepage      
\phantomsection
\label{tocpage}
\tableofcontents 

\cleardoublepage 


\pagestyle{fancy} 
\fancyhead[RE,LO]{\hyperref[tocpage]{ALEGRO Workshop 2026}}


\chapterimage{figures/images_head_plantp}
\chapter{Overview of the workshop}


We provide here our \textit{brief overview} of the workshop. %
We refer to the name of the presenters, though the views expressed are ours. %

Each speaker was invited to submit a short summary about their presentation. %
These are available in the next chapter, in the order of presentation at the workshop. %
"Clickable" links to the respective slides are also provided in each section. %
In the main text, reference to a section is "clickable". 
Clicking on "ALEGRO Workshop 2026" in the header of each page brings to the Table of Contents.

\section{Purpose of the Workshop}
This \href{https://agenda.infn.it/event/47329/overview}{ALEGRO 2026 workshop} was the 7$^{th}$ workshop (\href{https://indico.cern.ch/event/569406/}{ANAR2017 at CERN}, \href{https://indico.cern.ch/event/677640/}{Oxford John Adams Institute 2018}, \href{https://indico.cern.ch/event/732810/}{CERN 2019}, \href{https://indico.cern.ch/event/1193719/}{Hamburg DESY 2023}, \href{https://indico.cern.ch/event/1364999/overview}{Lisbon IST 2024}, \href{https://indico.slac.stanford.edu/event/9402/}{SLAC 2025}) organized by the \href{https://www.lpgp.u-psud.fr/icfaana}{ICFA-ANA panel} and a local organizing committee. %

\newglossaryentry{ALEGRO}{name={ALEGRO},description={Advanced LinEar collider study GROup}}ALEGRO is the Advanced LinEar collider study GROup: an international study group to promote Advanced and Novel Accelerators (ANA) for High-Energy Physics (HEP). 
The purpose of the workshop was to gather the world-wide members of the ANA community interested in developing ANA for particle and high-energy physics applications. %
It was timely because the ESPP process~\cite{PPGBriefingBook} was in its final stretch. %
The process and its current status were presented. %
The report of the US P5~\cite{P5:2023report} was released in December 2023. %

The workshop program (\url{https://agenda.infn.it/event/47329/overview}) was organized around mid- and long-term applications of ANA, with presentations of all the relevant projects. %

A major endeavor that started shortly before the 2025 ALEGRO meeting at SLAC is the \href{https://10tev-wakefield-collider.github.io/}{10 TeV Wakefield Collider Design Study}~\cite{10TeV}. %
A whole day of the workshop was dedicated to this project and to summaries of the work performed since the last workshop. %
Emphasis was put on the physics cases for a 10\,TeV collider (e$^+$e$^-$, e$^-$e$^-$, $\gamma\gamma$), and also for AWAKE. %
Higgs factories (HALHF, ALiVE) have established physics cases. %

The workshop also highlighted mid-term applications of ANA to particle physics (PP) (AWAKE~\cite{AWAKE}, CEPC~\cite{CEPC} and EIC~\cite{pwfa-alive-EIC} injectors, LEP3 booster~\cite{HEP-LEP3-EPPSU}) and HEP (LCVision~\cite{LCV}, HALHF~\cite{halhf}, ALiVE~\cite{alive}, ...) and to new topics that have emerged in the community, such as the "plasma physics" at the interaction point. %

Applications to light sources, injector for synchrotrons such as PETRA IV~\cite{PETRAIVl}, and free-electron-lasers, such as EuPRAXIA@SPARC\_LAB~\cite{tdr2026}, and SF-QED were also featured. %

ANA are in the exploratory and R\&D phase of their evolution, and among requirements for applications is a high operational level: reproducibility, high repetition rate, and availability  24/7. %
Facilities already operate with AWA for FEL and for laser-plasma interaction studies, and with parameters approaching those  for PP and HEP applications. %
Operational challenges were presented.

Impact of AI, not only on operation, but also on optimization of the machine, and on all aspects all the way to the writing up of the results was presented.


Ample time was scheduled for discussions since the workshop audience gathered representatives of many of the major players in the ANA field. %

We note here that remarkable progress has been made with plasma-based ANA. %
They have produced beams with sufficient quality for FELs to lase~\cite{pompili2022_pwfa_fel,GALLETTI:2022,Wang_2022,LABAT:2023,GALLETTI:2024,BARBER:2025}. %

Here after we offer a brief summary of the presentations at the workshop. %
We mention the name of the presenters and refer the reader to the respective summary included in this document (Contributions Chapter). %

\section{Editorial Note}
We note here that the accelerator concepts ALEGRO supports are referred to as ANA. %
However, application to high-energy physics is focusing on driving wakefields in metallic and dielectric structures and in plasmas. %
In this case, a more specific acronym may be used to refer to these accelerators is AWA: Advanced Wakefield Accelerators. %
We use the two acronyms ANA and AWA interchangeably in this document. %

\section{ESPPU, Applications of AWA}
The ESPPU process has reached its end with a very strong recommendation for FCCee as the next flagship project at CERN, and a final decision in 2028 (M.~Turner, \hyperref[sec:Turner]{Section~\ref{sec:Turner}}). %

AWA have application for linear accelerators, and are thus a natural long-term evolution for a linear Higgs factory. %
However, they have a number of mid-term applications, for example as possible plasma-based injector, as proposed for CEPC (W.~Lu) and that could be adapted for the FCCee, and other synchrotrons, and for fixed target experiments, as for example with the AWAKE acceleration scheme (M.~Wing, \hyperref[sec:Wing]{Section~\ref{sec:Wing}}). %

Applications to light sources have always been considered as stepping stones towards those for HEP and PP. %
This is because, in general, light sources are less demanding in terms of beam parameters. %
They also operate at lower energies ($<$20\,GeV). %
However, they require reproducibility and reliability similar to those of accelerators for HEP (L.~Giannessi, \hyperref[sec:Giannessi]{Section~\ref{sec:Giannessi}}). %

There is now a project for a LWFA-based injector for the PETRA IV synchrotron light source at DESY~\cite{PETRAIV} (A.~Maier). %

In addition, EuPRAXIA@SPARC\_LAB (M.~Ferrario, \hyperref[sec:Ferrario]{Section~\ref{sec:Ferrario}}) is an upcoming free electron laser (FEL) in the water window driven by a 1\,GeV electron beam emerging from a combination of X-band and plasma accelerators. %
The goal is to demonstrate that an FEL user facility can be based on a plasma-based accelerator, and that the resulting facility is more compact, more affordable, and consumes less energy than a more conventional facility. %
This is an important project in the more general context of EuPRAXIA, and that is on the ESFRI roadmap. %
\newglossaryentry{ESFRI}{name={ESFRI},description={European Strategy Forum on Research Infrastructures}}

\section{10\,TeV Collider Design}
The 10\,TeV collider design study is considered as a green-field design because of the format of the beam, envisaged as periodic bunches
with tens to hundreds of kHz repetition rate,  that is quite different from that of warm RF accelerators, operating in burst mode.
However, it is a natural evolution of a (linear) Higgs factory and TeV collider to reach the 10\,TeV energy range. %
This is in principle possible by the large average (or geographical) accelerating gradient ($>$500\,MeV/m) that makes a AWA 10\,TeV collider fit within the tunnel length of a TeV RF collider. %
This evolution is included in the LCVision project (E.~Adli). %

The 10\,TeV collider design study (J.~Osterhoff, \hyperref[sec:Osterhoff]{Section~\ref{sec:Osterhoff}}) has strengthened the link between the AWA/ANA community and the collider (beam delivery systems, detector) and the particle physics community (interaction point physics, physics case, etc.). %

\subsection{Physics case}
The physics case for an e$^-$e$^-$ linear collider up to the TeV energy scale has been developed in the context of the LCVision project~\cite{LCVision}. %
It needs to be extended to the 10\,TeV energy scale (E.~Bagnaschi, T.~Opferkuch, \hyperref[sec:Opferkuch]{Section~\ref{sec:Opferkuch}}) and has to take into account the challenges associated with quality acceleration of positron bunches in plasma, and thus also include cases for e$^-$e$^-$ and $\gamma\gamma$ collisions (T.~Barklow, \hyperref[sec:Barklow]{Section~\ref{sec:Barklow}}). %
Studies indicate that an e$^+$e$^-$ collider yields larger event numbers than and e$^-$e$^-$ collider, but also that a $\gamma\gamma$ collider is preferable to an e$^-$e$^-$ collider. %
The $\gamma\gamma$ collider requires relatively low energy per laser pulse and accommodates the round beams produced by plasma-based AWA. %

It was noted that high-energy lepton colliders provide complementary physics reach not only to low-energy lepton colliders, but also to high-energy hadron colliders. %

\subsection{Beam delivery system and detector}
The beam delivery system (K.~Downham, \hyperref[sec:Downham]{Section~\ref{sec:Downham}}) and detector (A.~Rastogi, \hyperref[sec:Rastogi]{Section~\ref{sec:Rastogi}}) must accommodate the high beam energy, the possible background emerging from the plasma, and the large disruption expected at the interaction point caused by the ultra-short and intense bunches. %

Another important feature is the flat beams for optimum interaction between colliding beam that plasma-based accelerators do not naturally accommodate (M.~Thevenet, \hyperref[sec:Thevenet]{Section~\ref{sec:Thevenet}}). %
This has important consequences on the entire collider, from the injector to the physics case for the plasma-based acceleration schemes. %

\subsection{Drivers}
AWA can be driven by electron beams in structures (J.~Power, SWFA, \hyperref[sec:Power]{Section~\ref{sec:Power}}) or in plasma (A.~Knetsch and D.~Storey, %
PWFA, \hyperref[sec:Storey]{Section~\ref{sec:Storey}}), and by laser pulses in plasma (C.~Benedetti and F.~Massimo, LWFA, \hyperref[sec:Benedetti]{Section~\ref{sec:Benedetti}}). %
These therefore require drivers such as high-power linacs (O.~Chubenko, \hyperref[sec:Chubenko]{Section~\ref{sec:Chubenko})}, and high-power, short pulse lasers (A.~Galvanauskas). %
\newglossaryentry{PWFA}{name={PWFA},description={Beam-Driven Plasma Wakefield Acceleration}}
\newglossaryentry{LWFA}{name={LWFA},description={Laser Wakefield Acceleration}}

PWFA require drive and witness bunches with quite different parameters, possibly with shaped temporal profiles and a polarized witness bunch (L.~Reichwein, \hyperref[sec:Reichwein]{Section~\ref{sec:Reichwein}}). %

Electron linac driver systems have been developed for example for CLIC and can be used for SWFA. %
However, PWFA requires periodically spaced particle bunches,
and sources must be adapted accordingly. %

The main challenges for laser drivers are the high repetition rate (0.1 to 1 kHz) and associated large average power, as well as the energy efficiency. %
The quality of the laser beam (phase front) is also paramount for the quality of the accelerated bunch. %
Coherent beam combining (CBC) fiber lasers may be the most promising candidate for a collider. %

Significant progress in stability and reproducibility has been made by including feedback loops in the laser driving  LWFA (A.~Maier), loops usually absent in short pulse laser developed for research purposes. %

\subsection{Particle sources}
Plasmas can produced polarized electrons by photo-dissociation of pre-polarized molecules (L.~Reichwein, \hyperref[sec:Reichwein]{Section~\ref{sec:Reichwein}}). %

Positrons can be extracted from pair production at the focus of very intense laser pulses (M.~Vranic, \hyperref[sec:Vranic]{Section~\ref{sec:Vranic}}). %

These new sources of particles could provide beams to test preservation of polarization of electron bunches, and of positron bunches quality  upon acceleration in plasma. %

\subsection{Accelerating structure}
Significant progress has been made in LWFA using hydrodynamic, optically-field-ionized (HOFI) channels. %
These have the advantage of not needing a material structure to contain the plasma and are thus not affected by heat deposition and possible laser damage. %
They are formed afresh for each event. %
Presented results suggest that these offer the best option for a collider (C.~Benedetti, F.~Massimo, \hyperref[sec:Benedetti]{Section~\ref{sec:Benedetti}}). %
Staging remains a major challenge (R.~D'Arcy, \hyperref[sec:Darcy]{Section~\ref{sec:Darcy}}), both on the ability of plasma lenses and magnetic optics~\cite{verra2026effect} to preserve the emittance of the witness bunch. %

RF structures powered by very short RF pulses ($\sim$ns) extracted from drive bunches through wakefields can sustain large surface and accelerating fields (J.~Power, \hyperref[sec:Power]{Section~\ref{sec:Power}}). %
While the peak accelerating gradients are lower than in plasma, the average or geographical gradient they can reach can in principle also reach 500\,MeV/m. %
These AWA can accelerate electron as well as positron bunches, an advantage over plasma-based AWA. %

\section{Mid-Term Applications}
A possible 10\,TeV collider is clearly a long-term, post Higgs factory, application for AWA. %
This is probably the right time scale for the AWA field to mature to be ready for the design, building, and operation of such a collider with very demanding parameters, mostly well-beyond those achieved so far. %

However, there are mid-term possible applications, with less demanding parameters.
These include fixed target experiments (M.~Wing, \hyperref[sec:Wing]{Section~\ref{sec:Wing}}), Higgs factories (HALHF, R.~D'Arcy, \hyperref[sec:Darcy]{Section~\ref{sec:Darcy}}; E.~Adli; ALiVE, A.~Caldwell, \hyperref[sec:Caldwell]{Section~\ref{sec:Caldwell}}), injectors for synchrotrons (CEPC, W.~Lu; FCCee). %

\section{Plasma Physics at the Interaction Point}
A new topic that emerged is the extension of use of electromagnetic (instead of electrostatic), particle in cell (PIC) %
\newglossaryentry{PIC}{name={PIC},description={Particle in cell}} %
codes, usually used to simulate the excitation of wakefields and the acceleration of particles in plasma, to describe the beam-beam interaction or the "plasma physics" at the interaction point (IP) %
\newglossaryentry{IP}{name={IP},description={Interaction Point}}(T.~Grismayer, \hyperref[sec:Grismayer]{Section~\ref{sec:Grismayer}}, S.~Gessner, \hyperref[sec:Gessner]{Section~\ref{sec:Gessner}}). %
These codes (VLPL~\cite{PUKHOV_1999}, OSIRIS~\cite{OSIRIS}, Warp-X~\cite{Vay_2018}) now include quantum electro-dynamic effects (QED) and pair production, so they can be used to simulate possible strong-field QED experiments that are planned with the multi-petawatt laser systems that become available all around the world (S.~Meuren). %
But they can also simulate the interaction between colliding beams at the IP, including effect not described by currently used codes (GUINEA-PIG~\cite{Schulte}, CAIN~\cite{CAIN}). %
In addition, these fully electro-magnetic codes include longitudinal fields and the longitudinal motion of beam particles during the collision that play a significant role when the bunches are short, and the fields and disruption parameter are large, as is typically the case with bunches emerging from plasma-based accelerators. %
These codes are also relevant for the projected colliders (e.g., FCCee) that also reach extreme regimes at the IP, highlighting another mid-term contribution from this community. %

\section{Operation of Facilities}
Most AWAs operate Today as research facilities, delivering laser or particle beams a few hours per day. %
However, applications to PP and HEP require operation 24/7 for long periods of time. %
This has strong implications in particular on the equipment and the need for reliability. %

Though not for PP or HEP applications, there are user facilities driven by laser (P.~Oliveira, \hyperref[sec:Oliveira]{Section~\ref{sec:Oliveira}}, A.~Maier) and electron beams (L.~Giannessi, \hyperref[sec:Giannessi]{Section~\ref{sec:Giannessi}}) operating with high reliability and reproducibility. %
These point the way at the steps the community must take for the transition from acceleration to accelerator. %

It is clear that AI already plays an important role in the operation of accelerators. %
It is also clear that AI should be included in all the processes related to new accelerators, from their design, to their operation, optimization, and even to the publication of the scientific results (V.~Kain, \hyperref[sec:Kain]{Section~\ref{sec:Kain}}). %

\section{Emerging Projects}
The AWA community is very dynamic and explores new applications for AWA providing more affordable and more sustainable alternatives to RF acceleration systems. %
A general theme is that of the injector for a synchrotron, be it a light source such as PETRA IV (A.~Maier), or a collider such as CEPC (W.~Lu) or the EIC (A.~Caldwell, \hyperref[sec:Caldwell]{Section~\ref{sec:Caldwell}}, J.~Farmer, \hyperref[sec:Farmer]{Section~\ref{sec:Farmer}}). %
The latest proposal is for an energy booster for LEP3 (A.~Caldwell, \hyperref[sec:Caldwell]{Section~\ref{sec:Caldwell}}). %

Considering the wide success and broad impact of FELs on may fields of science and technology, it is likely that more FEL driven by beams produced by LWFA will be proposed. %

\section{Global Questions}
The workshop highlighted progress in the AWA world towards applications to PP and HEP. %
Much progress has been made towards mid-term applications and applications to light sources. %
However, it is also clear that major global questions need to be answered towards a 10\,TeV collider, seen as the ultimate challenge. %
\begin{itemize}
    \item Can plasma-based schemes accelerate positron bunches with sufficient quality to envisage a plasma-based e$^+$e$^-$ collider? %
    \item Are round or flat accelerated beams globally optimum for a plasma-based collider? %
    \item Can staging of plasmas preserve the very low emittance required for the accelerated bunch? %
    \item Should the driver of wakefields in plasma be a particle bunch (PWFA) or a laser pulse (LWFA)? %
    \item What scheme SWFA, LWFA, or PWFA, or combination of schemes can reach the highest average/geographical accelerating gradient, and minimize the size and cost, while maximizing the sustainability of a collider?
\end{itemize}
These global questions, as well as all the detailed ones will continue to be discussed, and hopefully answered, at the next ALEGRO workshops. %

\section{Role of ALEGRO}
ALEGRO sees its role in fostering progress towards all applications of AWA-based accelerators for applications to particle physics and high-energy physics. %
ALEGRO also supports the development of ANA towards application to light sources, either as injector (for synchrotrons), or as the beam source itself (FELs). %
Meeting the challenges these applications offer is a necessary step towards applications with much more demanding beam parameters. %
However, currently ALEGRO has no source of funding to support any work. %
 At this time, the allocation of funding dedicated  to ANA accelerator design would be suitable to cover important numerical simulation studies for the design of demonstrator facilities. %

An important experimental demonstration towards a linear collider would be that of staging of two, coupled multi-GeV accelerating structures that would preserve the beam quality, only adding energy. %
This demonstrator needs to be designed, simulated start-to-end (plasma, staging optics), and operated in a dedicated facility, so that rapid progress can be made and tolerances to parameters evaluated. %
Considering the size and cost of such a demonstrator, it can only be envisaged in the context of a long-term, large-scale project such as the LCVision or a Higgs factory (HALHF). %

In parallel, developments of AWA as components of colliders, whether circular or linear (injector, energy booster, etc.) must be strongly supported. %

All these developments require close collaborations between members of the ANA, accelerator, and high-energy physics, collaborations that ALEGRO workshops hope to advance. %

\newglossaryentry{CDR}{name={CDR},description={Conceptual Design Report}}
\newglossaryentry{LPA}{name={LPA},description={Laser Plasma Accelerator}}
\newglossaryentry{SWFA}{name={SWFA},description={Structure WakeField Acceleration}}

\newpage

\chapterimage{figures/images_head_sandwhitep}
\chapter{Contributions}

The following summaries are intended to provide a concise overview of the workshop contributions. %
They are based on the submitted material, whereas the details can be found in the corresponding references. %
We provide the link to the slides for each contribution. %

The findings and opinions presented in these summaries are those of the respective authors and should not be interpreted as necessarily representing the views of the workshop organizers and/or affiliated or funding institutions. %
\newpage
\section{Report on the ESPP Update - Contributions of the Advanced Accelerator Concepts to HEP}
\label{sec:Turner}
\href{https://agenda.infn.it/event/47329/contributions/280875/}{Slides: \url{https://agenda.infn.it/event/47329/contributions/280875/}}
\subsection*{\textit{Marlene Turner\textsuperscript}}

\noindent \textit{CERN, 1211 Geneva 23, Switzerland}

\subsection{Overview}
The \textit{European Strategy for Particle Physics Update} (ESPPU) is a process through which the European particle physics community defines its long-term scientific priorities. Its main purpose is to build a shared vision for the future of particle physics in Europe. Given the international nature and high cost of major particle physics projects, coordinated decision-making is essential to avoid duplication, align priorities, and make the best use of available resources.
The current (3$^{rd}$) ESSPU was launched by the CERN Council on March 21, 2024, and is expected to conclude in June 2026. Its central objective is to reach consensus on CERN’s next major accelerator project, recommending the preferred option and possible alternatives.

\subsection{ESSPU Input}
The 3rd ESPPU gathered around 260 community submissions, with 47 of those on accelerators and/or accelerator technology. Notably, 12 of these 47 submissions mentioned \textit{Advanced Accelerator Concepts} (AAC), highlighting their importance in the field. 
\begin{itemize} 
    \item Among these, three wakefield-based collider design studies:
        \begin{itemize}
            \item \textbf{\href{https://indico.cern.ch/event/1439855/abstracts/190833/}{HALHF}~\cite{halhf}:} A hybrid asymmetric linear Higgs factory concept.
            \item \textbf{\href{https://indico.cern.ch/event/1439855/contributions/6461625/}{ALiVE}~\cite{alive}:} A linear accelerator design for very high energies.
            \item \textbf{\href{https://indico.cern.ch/event/1439855/contributions/6461496/}{10 TeV Design Study}~\cite{10TeV}:} A design effort towards a 10 TeV center-of-mass wakefield collider.
        \end{itemize}
    \item Further submissions from the AAC community include:
        \begin{itemize}
            \item \textbf{\href{https://indico.cern.ch/event/1439855/contributions/6461556/}{AWAKE}~\cite{AWAKE}:} A proton-driven plasma wakefield acceleration R\&D program.
            \item \textbf{\href{https://indico.cern.ch/event/1439855/contributions/6461571/}{ALEGRO}~\cite{ALEGRO}:} Strategic coordination of collider efforts through the \textit{Advanced LinEar collider study GROup}.
        \end{itemize}
    \item Additional mentions of AACs are in the following submissions: \href{indico.cern.ch/event/1439855/contributions/6461469/contribution.pdf}{Future Opportunities with Lepton-Hadron Collisions}~\cite{leptonhadron}, \href{https://indico.cern.ch/event/1439855/abstracts/190997/}{A Linear Collider Vision for the Future of Particle Physics}~\cite{LCV}, \href{https://indico.cern.ch/event/1439855/contributions/6461564/}{European Accelerator R\&D Platform Review for the Laboratory Directors Group}~\cite{LDG}, \href{https://indico.cern.ch/event/1439855/contributions/6461600/}{Input to the European Strategy for Particle Physics: Strong-Field Quantum Electrodynamics}~\cite{SFQED}, \href{https://indico.cern.ch/event/1439855/contributions/6461627/}{Advanced Accelerator and HEP Developments through Networking between the Large Particle Physics Laboratories and CERN}~\cite{coord}, \href{https://indico.cern.ch/event/1439855/contributions/6461651/attachments/3046030/5417862/Report_of_Physics_Beyond_Colliders_at_CERN.pdf_v02.pdf}{Summary Report of the Physics Beyond Colliders Study at CERN}~\cite{PBC} and \href{https://indico.cern.ch/event/1439855/contributions/6461672/}{A Flexible Strategy for the Future of Particle Physics at CERN}~\cite{flexible}.
\end{itemize}
\vspace{0.5cm}

Additionally, large-scale accelerator projects at CERN were asked to provide parameter tables for a comparative evalutation. Invited were: FCCee, FCChh, LCF, LEP3, CLIC, LHeC and the Muon Collider. 
\vspace{1cm}

CERN Member and Associate Member States were asked to provide input on:
\begin{itemize}
    \item What is the preferred large-scale post LHC accelerator at CERN?
    \item What is the preferred alternative if the preferred option were not feasible?
    \item What is the preferred alternative if the preferred option were not competitive?
\end{itemize}

\subsection{Evaluation Process}
The input submitted by the community was reviewed by the \textit{Physics Preparatory Group} (PPG), which organized an \href{https://agenda.infn.it/event/44943/timetable/}{Open Symposium} in Venice~\cite{OpenSymposium} in June 2025 and prepared the \href{https://cds.cern.ch/record/2944678/files/Publication.pdf}{PPG Briefing Book}~\cite{PPGBriefingBook}. In parallel, European Strategy Group (ESG) Working Group 2a conducted a \href{https://cds.cern.ch/record/2947728/files/ESG_WG2a_Final.pdf}{comparative assessment}~\cite{ComparativeSummary} of large-scale accelerator projects at CERN. A summary of the answers to the questions from the CERN Member and Associate Member States was also prepared and can be found in Ref.~\cite{nationalinput}. 

\subsection{Outcomes}
The \href{https://cds.cern.ch/record/2950671/files/CERN-ESU-2025-002.pdf}{draft
strategy}~\cite{DraftStrategy} prepared by the ESG expresses a clear preference for the FCCee as CERN’s next flagship project, with a scaled-down FCCee option identified as an alternative. Other alternatives are not ranked. Criticism of linear collider options focuses on their cost and the absence of a credible upgrade path to the \SI{10}{TeV} energy scale. This last point is directly relevant to the AAC community, which could help provide such a path once the technology reaches greater maturity.

Looking ahead, the main question is whether the CERN Council will adopt the draft strategy. If adopted, the Council could move toward approval of the FCCee project around 2028, although this remains uncertain in light of outstanding financial and political considerations. If necessary, a lighter strategy update could be performed in 2028.

\subsubsection{Implications for AACs}
The \href{https://cds.cern.ch/record/2950671/files/CERN-ESU-2025-002.pdf}{draft
 strategy}~\cite{DraftStrategy} calls for 'support at the appropriate level' for AAC technologies in the context of Accelerator R\&D within accelerator technology R\&D.


\newpage
\section{Emittance Mixing of Flat Beams in Plasma Accelerators}
\label{sec:Thevenet}

\href{https://agenda.infn.it/event/47329/contributions/286824/}{Slides: \url{https://agenda.infn.it/event/47329/contributions/286824/}}

\subsection*{\textit{
S.~Diederichs\textsuperscript{1,2},
C.~Benedetti\textsuperscript{3},
A.~Ferran Pousa\textsuperscript{1},
A.~Sinn\textsuperscript{1},
J.~Osterhoff\textsuperscript{1,3},
C.~B.~Schroeder\textsuperscript{3,4},
M.~Thévenet\textsuperscript{1}
}}

\noindent \textit{\textsuperscript{1}Deutsches Elektronen-Synchrotron DESY, Notkestr. 85, 22607 Hamburg, Germany}

\noindent \textit{\textsuperscript{2}CERN, Espl. des Particules 1, 1211 Geneva, Switzerland}

\noindent \textit{\textsuperscript{3}Lawrence Berkeley National Laboratory, 1 Cyclotron Rd, Berkeley, California 94720, USA}

\noindent \textit{\textsuperscript{4}Department of Nuclear Engineering, University of California, Berkeley, California 94720, USA}

\subsection{Context}

Future linear colliders require high luminosity while minimizing beamstrahlung effects. Since the luminosity scales as $1/(\sigma_x\sigma_y)$~\cite{Schulte:2016}, whereas beamstrahlung scales as $1/(\sigma_x+\sigma_y)$~\cite{Schroeder:2022}, both requirements can be simultaneously satisfied by operating with flat beams ($\sigma_x\gg\sigma_y$). This motivates the generation of beams with a large transverse emittance ratio, $\epsilon_x/\epsilon_y\gg1$, and the preservation of this flatness throughout the acceleration process.

In beam-driven plasma wakefield accelerators operated in the ideal blowout regime, the focusing force generated by the stationary ion column is linear, suggesting that the flatness and transverse quality of the accelerated beam can, in principle, be preserved. However, the acceleration of collider-relevant flat beams under realistic plasma conditions, where nonlinear wakefield effects become significant, had not been investigated in detail.



\subsection{Objectives}

This project aims to understand the dynamics of flat beams in plasma wakefield accelerators operating beyond the ideal blowout regime. In particular, it investigates how nonlinear wakefields, arising from effects such as ion motion, influence the preservation of beam flatness, emittance, and ultimately the luminosity of a plasma-based linear collider. The results are applied to the HALHF beam-driven collider concept, where an alternative configuration employing vertically larger flat driver beams is explored. While this approach appears promising for a 500~GeV center-of-mass collider, extending these concepts to the multi-TeV regime remains an open challenge.



\subsection{Progress over the last two years}



\begin{figure}[htbp]
    \centering
    \includegraphics[width=\textwidth]{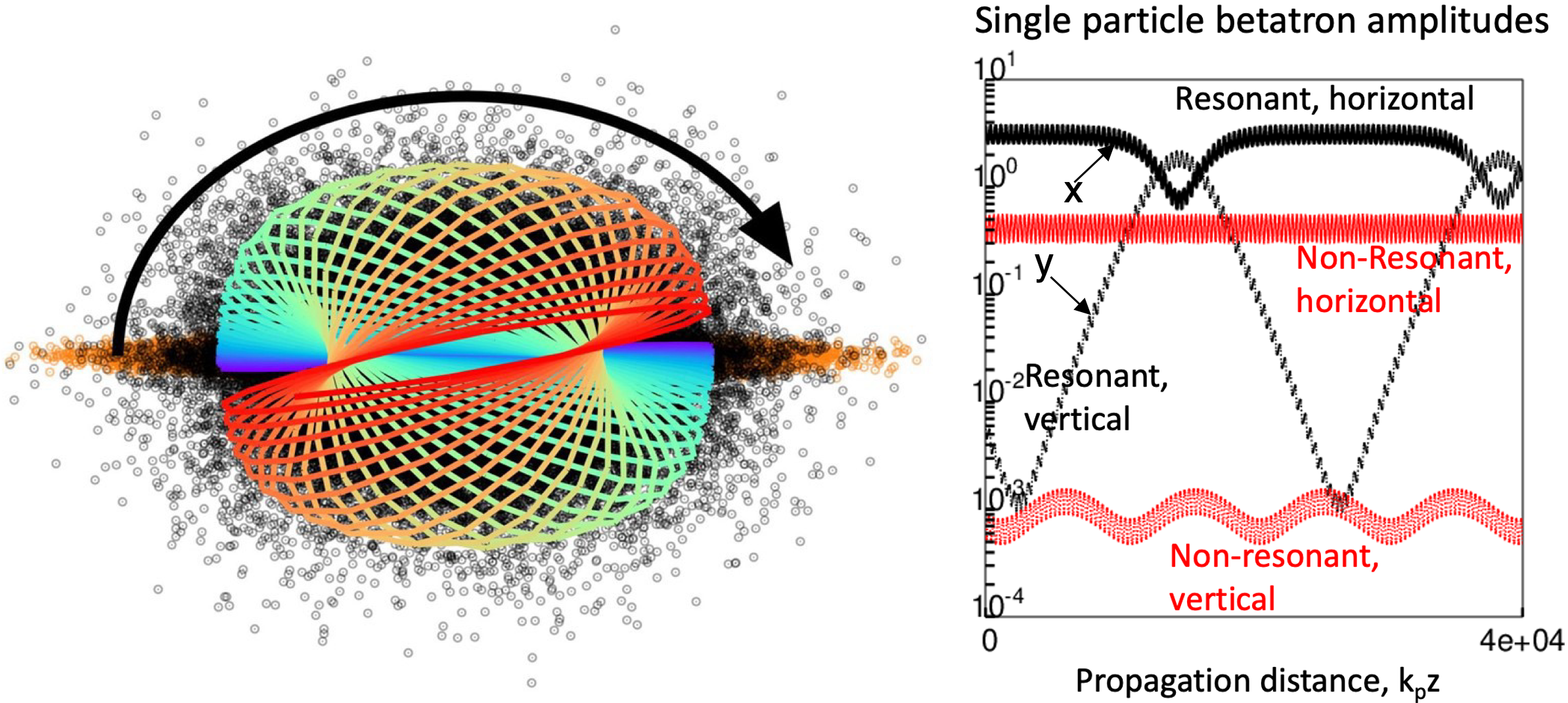}
    \caption[Caption for List of Figures]{\label{fig1} Left: transverse (x,y) trajectory of a resonant particle in plasma wakefield in the presence of field nonlinearity. Right: horizontal (x, solid line) and vertical (y, dashed line) betatron amplitude of a resonant (black) and non-resonant (red) particles. Both particles start with a much large amplitude in x than in y, but this is fully exchanged for the resonant particle around $k_pz\simeq1.4\times16^4$.}
\end{figure}

Over the past two years, we have identified and characterized a previously unrecognized emittance degradation mechanism for flat beams in plasma accelerators. Through analytical modeling and particle-in-cell simulations, we demonstrated that nonlinear wakefield components can induce resonant coupling between the horizontal and vertical betatron oscillations, resulting in emittance transfer from the initially larger horizontal plane to the vertical plane~\cite{diederichs2024resonant}. This is illustrated one Fig.\ref{fig1}. This resonant emittance mixing degrades the beam aspect ratio and compromises the beam quality required for future linear collider applications.

We further established the conditions under which this resonance occurs and proposed practical mitigation strategies. In particular, we showed that, when possible, operating away from resonance significantly suppresses emittance exchange, thereby preserving the flat-beam configuration during plasma acceleration. These findings provide important design constraints for plasma-based linear colliders and serve as the basis for ongoing studies of realistic collider scenarios, including the HALHF concept.

Another work from the community~\cite{manwani2022flat} investigates the use of flat electron beams as drivers for plasma wakefield accelerators, demonstrating that their asymmetric transverse profile generates an elliptical blowout cavity with correspondingly asymmetric focusing forces. Using particle-in-cell simulations, the authors characterize how the cavity ellipticity depends on the drive beam aspect ratio and charge density, and derive the matching conditions required to preserve beam quality during propagation. The study further evaluates experimentally accessible parameter regimes at the Argonne Wakefield Accelerator (AWA) and FACET facilities, establishing the feasibility of producing and exploiting flat-beam-driven plasma wakefields.

\subsection{Time scale for application}

Investigations in the upcoming years will aim at identifying potential solutions for higher energy, up to the multi-TeV range. Although no concrete plan has been proposed, an experimental validation of this mechanism and/or future solutions would be valuable.

\newpage
\section{High-Energy Electron-Positron Beam Collisions with Large-Angle Disruptions}
\label{sec:Grismayer}
\href{https://agenda.infn.it/event/47329/contributions/286131/}{Slides: \url{https://agenda.infn.it/event/47329/contributions/286131/}}
\subsection*{\textit{T. Grismayer\textsuperscript{1}, W. Zhang\textsuperscript{2,3}, L.O. Silva\textsuperscript{1}}}

\noindent \textit{\textsuperscript{1}GoLP/Instituto de Plasmas e Fusão Nuclear, Instituto Superior Técnico, Universidade de Lisboa, Lisboa 1049-001, Portugal}

\noindent \textit{\textsuperscript{2}Jiangxi Province Key Laboratory of Nuclear Physics and Technology, East China University of Technology, Nanchang 330013, China} 

\noindent \textit{\textsuperscript{3}Engineering Research Center of Nuclear Technology Application, Ministry of Education, East China University of Technology, Nanchang 330013, China}

\subsection{Redefining the limits of high-energy lepton colliders}
As the high-energy physics community reaches a critical crossroads in deciding the future of global colliders, the precision of our predictive modeling has never been more vital. Current proposals for next-generation facilities, such as the Future Circular Collider (FCC-ee), the International Linear Collider (ILC), and various plasma-wakefield or dielectric concepts, aim for unprecedented luminosity and center-of-mass energies of several TeVs. To ensure these machines deliver on their scientific promise, it is essential to incorporate a more comprehensive treatment of particle motion than previously required. Our recent research identifies nuanced and complex beam-beam dynamics that emerge as we push these machines toward their operational limits with high disruption and extreme beamstrahlung, suggesting that an upgrade in the modeling standards used in collider design is necessary to maintain high predictive accuracy.

Historically, the interaction of electron and positron beams at the Interaction Point (IP) has been modeled under the assumption that transverse and longitudinal motions are entirely separable~\cite{Chen1988}. In this simplified view, particles are assumed to move at the speed of light along the collision axis regardless of the electromagnetic forces they encounter. However, it has been demonstrated~\cite{Zhang2026} that this assumption fails when the particle disruption angle becomes non-negligible. By introducing a new dimensionless parameter closely related to the disruption angle, we have characterized a regime in which the transverse deflection of particles becomes comparable to their longitudinal motion. 3D fully electromagnetic, particle-in-cell (PIC) simulations reveal that particles do not just "pinch" toward the center; they undergo a significant longitudinal deceleration. In high-disruption scenarios, this can even lead to particle reversal—where the trailing edge of the beam is effectively pushed backward by the fields of the opposing beam~\cite{Zhang2026}.

In our most recent work, we aim to contribute to the collaborative effort to benchmark and modernize the software ecosystem used by the collider community as we approach the new regimes associated with future colliders. Extensive comparisons between the 2D electrostatic codes like GUINEA-PIG~\cite{Schulte1996} and the fully 3D electromagnetic PIC code OSIRIS~\cite{OSIRIS} permitted addressing: the accuracy and limitations of the legacy code GUINEA-PIG, the difference observed in the energy loss and luminosity, the interplay between collective beam dynamics and QED effects, and the prospect of numerically exploring high-energy, high-luminosity advanced colliders based on plasma acceleration.

This research shows that it is critical to adopt more robust simulation standards across the community. As we design colliders that operate with increasingly dense $n \gtrsim 10^{24} \mathrm{cm}^{-3}$ and energetic beams $\mathcal{E} > \mathrm{TeV}$, the collective beam dynamics (coupled with QED processes) will play a vital role in determining the actual physics reach of the facility.  Our study shows that the widely used conventional beam-beam codes cannot explore collisions with considerable disruption angles. More versatile and efficient (MPI-powered) simulation toolkits are needed, and they should be developed in the near future, also integrating these codes in AI-driven workflows. By accurately modeling beam-beam interactions, we can better optimize beam parameters, minimize unwanted backgrounds from pair production, and provide more realistic luminosity forecasts. This ensures that the major decisions being made today about the future of particle physics are built on a foundation of total physical consistency, bridging the gap between theoretical and computational beam and plasma physics and the practical engineering of the world’s next great discovery machines.

\subsection{Bridging plasma physics and high-energy collider design}
The development of future high-energy colliders is increasingly an interdisciplinary, comm-unity-driven, high-investment task, merging the expertise of accelerator physics, particle physics, strong-field QED, and plasma physics. This contribution highlights how simulation tools originally developed for plasma-based acceleration are now essential for understanding the interaction point physics in traditional lepton colliders. They are also completely ideal for studying the advanced plasma-based linear colliders. This collaboration has successfully applied the OSIRIS 3D, electromagnetic, relativistic, massively parallel, PIC code~\cite{OSIRIS}, a staple in plasma research, to the specific problem of lepton beam collisions. We have validated theoretical models, establishing an analytical model for beam-beam interactions that matches simulation data across a broad range of parameters. We have identified novel phenomena such as the anomalous pinch in electron-electron collisions~\cite{Zhang2025}, and analyzed the impact of SF-QED, showing that beamstrahlung and pair production dynamically increase the new dimensionless parameter introduced in this contribution~\cite{Zhang2026}, further amplifying the disruption and beam pinch. This will have an impact on both the collision luminosity and its spectrum with respect to the center-of-mass energies.

The findings emphasize that future collider designs cannot treat beam-beam effects as a secondary technical detail, especially for the dense, ultra-high-energy beams. These activities ensure that the global roadmap for particle physics is supported by the most advanced computational physics available.
\newpage
\section{Plasma Acceleration of Polarized Electrons}
\href{https://agenda.infn.it/event/47329/contributions/281014/}{Slides: \url{https://agenda.infn.it/event/47329/contributions/281014/}}
\label{sec:Reichwein}
\subsection*{\textit{Lars Reichwein\textsuperscript{1,2,3}, Dimitris Sofikitis\textsuperscript{4}, Zheng Gong\textsuperscript{5}, Chuan Zheng\textsuperscript{6}, Liangliang Ji\textsuperscript{2}, Alexander Pukhov\textsuperscript{3}, T. Peter Rakitzis\textsuperscript{7,8}, Markus Büscher\textsuperscript{1,9}}}

\noindent \textit{\textsuperscript{1}Peter Grünberg Institut (PGI-6), Forschungszentrum Jülich, 52425 Jülich, Germany}\\
\noindent \textit{\textsuperscript{2}State Key Laboratory of Ultra-intense Laser Science and Technology, Shanghai Institute of Optics and Fine Mechanics, Chinese Academy of Sciences, Shanghai 201800, People’s Republic of China}\\
\noindent \textit{\textsuperscript{3}Institut für Theoretische Physik I, Heinrich-Heine-Universität Düsseldorf, 40225 Düsseldorf, Germany}\\
\noindent \textit{\textsuperscript{4}Department of Physics, Atomic and Molecular Physics Laboratory, University of Ioannina, University Campus, 45110 Ioannina, Greece}\\
\noindent \textit{\textsuperscript{5}CAS Key Laboratory of Theoretical Physics, Institute of Theoretical Physics, Chinese Academy of Sciences, Beijing 100190, People’s Republic of China}\\
\noindent \textit{\textsuperscript{6}Artemis Targetra GmbH, c/o Collective Incubator, Jülicherstr. 209Q/S, 52070 Aachen, Germany}\\
\noindent \textit{\textsuperscript{7}Institute of Electronic Structure and Lasers, Foundation for Research and Technology-Hellas, 71110 Heraklion-Crete, Greece}\\
\noindent \textit{\textsuperscript{8}Department of Physics, University of Crete, 70013 Heraklion-Crete, Greece}\\
\noindent \textit{\textsuperscript{9}Institut für Laser- und Plasmaphysik, Heinrich-Heine-Universität Düsseldorf, 40225 Düsseldorf, Germany}\\

\subsection{Context}
For future high-energy physics applications of wakefield-based colliders, one important property of the beam is its polarization.
Spin evolution in wakefields has long been neglected in research, and only more recently has gained significant interest.
Plasma-based sources of polarized electrons could potentially provide higher peak currents than conventional sources. Moreover, such sources could be realized in laboratories laid out for wakefield acceleration, without requiring additional infrastructure.
A general overview of the state-of-the-art of plasma-based sources for polarized particle beams is given in Ref.~\cite{Reichwein2025}. 
Applications of polarized electrons can range from surface physics in the low-energy regime \cite{Tusche2024}, over deep-inelastic scattering to probe the nucleon structure~\cite{Glashausser1979}, to research beyond the Standard Model of particle physics on axion-like particles~\cite{Chen2025}.
While a first proof-of-principle experiment on the acceleration of nuclear-polarized helium-3 has recently been conducted~\cite{Zheng2024}, currently no such results exist for polarized electrons.
Nevertheless, experimental realization is planned at DESY in scope of the project ``LEAP''~\cite{Stehr2025}, and more recently at the Shanghai Institute of Optics and Fine Mechanics~\cite{Wu2019_lwfa, Li2022}.

\subsection{Objectives}
Plasma-based sources of polarized electrons either rely on (i) in-situ methods of generating and accelerating polarized beams in a single step, or (ii) on pre-polarized targets, where the target is prepared separately and polarization must be preserved during injection. 
In-situ methods have been proposed by Nie \textit{et al.} in Refs.~\cite{Nie2021, Nie2022}. The injection mechanism is based on the principle of the plasma photocathode~\cite{Hidding2012}, but relies on the ionization of specific orbitals such as $4f^{14}$ of ytterbium, which yields the polarized witness beam. Simulations show that such a setup could potentially generate a 4\,kA electron beam with $\sim 56\%$ polarization.

For the pre-polarized targets, most of the theoretical work relies on hydrogen halide sources, especially HCl. In the following, we highlight the preparation scheme detailed in Ref.~\cite{Sofikitis2025}:

\begin{enumerate}
    \item A first UV laser pulse with circular polarization is used for photodissociation of the HCl molecules and transfer of spin onto the hydrogen component. The hydrogen and halogen fragments exit the irradiated region, leaving behind a hole/channel. 
    \item The process is repeated using a second, wider UV laser pulse, which dissociates molecules in a broader region which forms a reservoir. Since hydrogen is lighter than the halogen atoms, the polarized hydrogen streams back into the hole.
    \item  Afterwards, the hydrogen inside the hole can be ionized and the electrons accelerated. It should be noted that, due to the hyperfine interaction, the polarization is transferred periodically between the hydrogen nucleus and the electrons. Thus, ionization and acceleration need to be timed accordingly.
\end{enumerate}

One crucial challenge of these targets are the restrictive parameter space in terms of density and hole diameter. For example, in order to obtain a pre-polarization of 90\% at a density of $10^{17}$\,cm$^{-3}$, the generated target (``hole'') has a diameter of approx. 1\,{$\mu$}m~\cite{Sofikitis2025}. Lowering the target density allows larger, but nonetheless restrictive, scales.
Most theoretical work so far has simplified the problem of injection by assuming higher densities and/or larger target dimensions. These numerous studies have shown that a variety of injection mechanisms including self-injection \cite{Yin2024}, density down-ramp injection~\cite{Wu2019} and colliding-pulse injection~\cite{Bohlen2023} can be used for pre-polarized targets, provided that feasible target parameters could be achieved.

Spin evolution during injection was shown to be crucial, as the precession during that process is significant.
The precession of the spin vector $\vec{s}$ inside the wakefield can be described according to the T-BMT equation \cite{Bargmann1959}, $\mathrm{d} \vec{s} / \mathrm{d}t = - \vec{\Omega} \times \vec{s}$.
The precession frequency is given as
\begin{align}
    \vec{\Omega} = \frac{e}{mc} \left[  \left(  a + \frac{1}{\gamma}\right) \vec{B} - \frac{a \gamma}{\gamma + 1} \left( \frac{\vec{v}}{c} \cdot \vec{B}\right) \frac{\vec{v}}{c} - \left( a +  \frac{1}{1 + \gamma} \right) \frac{\vec{v}}{c} \times \vec{E} \right] \; ,
\end{align}
where $a \approx 10^{-3}$ denotes the electron's anomalous magnetic moment and $\gamma$ its Lorentz factor. In particular, the precession frequency scales as $\Omega \propto \gamma^{-1}$, meaning that witness beam polarization only changes marginally once the particle is accelerated.

\subsection{Perspectives}
Towards projects like the 10\,TeV pCM collider initiative and research on the high-$\chi$ regime, additional effects could become important: While the Stern-Gerlach force and the Sokolov-Ternov effect are commonly disregarded for wakefield-related studies at comparatively lower energies due to the field strengths and timescales involved~\cite{Thomas2020}, radiative spin-flips due to the witness' betatron oscillations could limit the attainable degree of polarization at higher energies.
\newpage
\section{A Hybrid, Asymmetric, Linear Higgs Factory (HALHF) — Updates and The ‘Repetition Rate’ Question}
\href{https://agenda.infn.it/event/47329/contributions/285244/}{Slides: \url{https://agenda.infn.it/event/47329/contributions/285244/}}
\label{sec:Darcy}
\subsection*{\textit{Richard D'Arcy}}

\noindent \textit{John Adams Institute, University of Oxford, UK}

\subsection{Context and Objective}
Construction of a Higgs factory is the top priority for particle physics in the next decades but the costs are extremely high. Plasma-wakefield accelerators (PWFAs) promise to reduce drastically the footprint and cost of such machines. However, while progress on electron acceleration is rapid, positron acceleration in plasma remains challenging due to the inherent charge asymmetry in plasma. In 2023 we proposed a pragmatic approach to a linear-collider concept that bypasses the positron problem by using a hybrid of classical radio-frequency (RF) accelerators to accelerate positrons and PWFAs to accelerate electrons aka HALHF ~\cite{Foster_2023}.

\subsection{Progress over the last two years}
A new set of baseline parameters, referred to as HALHF 2.0, was produced through a Bayesian optimisation routine ~\cite{Lindstrom_2025}. The optimisation was specifically designed to minimise the full programme cost, including construction, operation, energy consumption, maintenance, and carbon impact. As part of this process, a number of new design constraints were introduced following extensive and carefully considered feedback from the collaboration (most impactful were the need for two separate linacs for the wakefield electron drivers and the colliding positrons). The resulting baseline parameters, together with an outline of future studies, timelines, and projected costs, were subsequently submitted to the ESPPU ~\cite{Foster_2025}.

HALHF has served as an excellent vehicle for consolidating the field’s progress and outlining the next steps required to further the concept. Achieving high luminosity has been highlighted as one of the two central challenges for the programme. Meeting this requirement mandates both a high repetition rate and high average power operation, with the current state of the art still significantly below the levels required for practical application. In particular, it is necessary to demonstrate stable operation at repetition rates in the MHz regime (and beyond), assess the effects on the plasma at these high rates (specifically super-heating of the plasma), and develop novel plasma-cell cooling schemes capable of stabilising the temperature of the system to avoid 

\begin{figure}[htbp]
    \centering
    \includegraphics[width=0.7\textwidth]{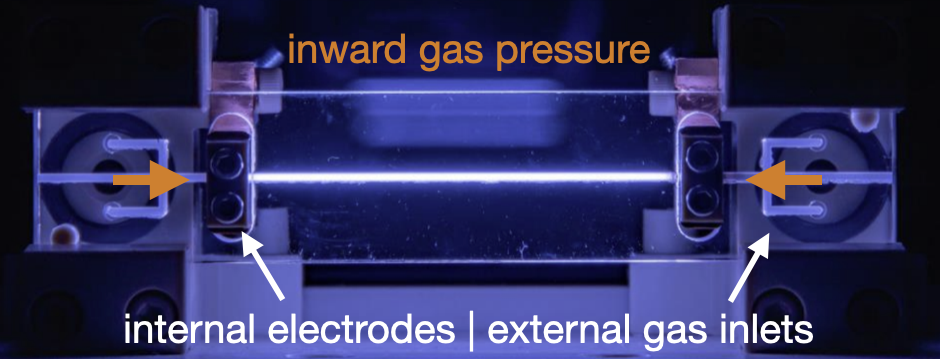}
    \caption[Caption for List of Figures]{\label{YourLabel} The novel plasma cell designed to counteract expulsion and enable rapid plasma regeneration required for MHz plasma-wakefield acceleration.
    \label{darcyfig}
    }
\end{figure}

One of the key challenges associated with high repetition rate operation is maintaining identical plasma properties at microsecond-scale bunch separations. Processes such as plasma expulsion and recombination will inevitably modify the plasma profile if left uncontrolled, potentially degrading acceleration performance. To address this, concepts involving internal electrodes, external gas inlets (Fig.~\ref{darcyfig}), and a MHz “top-up” discharge have been developed. These approaches have already been used to regenerate plasma at MHz repetition rates with sufficiently similar properties for operation with the MHz bunch train at FLASH, enabling studies of acceleration performance under these conditions. Such studies led to the recent world-first demonstration of MHz acceleration at the FLASHForward facility ~\cite{DArcy_2018}, with 0.84\,GV/m gradients  produced in a hydrogen plasma of 7$\times$10$^{15}$\,cm$^{-3}$ electron density used to accelerate FLASH electron bunches with preserved charge (45\,pC) and energy spread (0.19\% FWHM).

Sustained high-repetition-rate (i.e. high-average-power) operation introduces an additional set of challenges defined by plasma heating and plasma-cell cooling. Whilst the recent demonstration of MHz acceleration represents an important first step in this direction, the HALHF parameter regime highlights the need for further investigation, particularly the much higher bunch currents, larger deposited power, and smaller bunch spacing. Thus, several important open questions remain. For example, it is not yet fully understood how the plasma evolves after driving a wakefield with HALHF beam parameters, or whether the acceleration process can be sustained at bunch separations as small as the design parameter (16 ns). Addressing these questions will require the development of a dedicated simulation pipeline, although such simulations will be computationally prohibitive unless judicious physics reductions are made. Furthermore, the effect of sustained power deposition on plasma temperature throughout the bunch train must be understood through direct temperature diagnostics. A novel all-in-one (i.e. plasma-electron, plasma-ion, and plasma-cell) diagnostic suite must be generated. The latter diagnostic in this suite will also determine whether active stabilisation (or other prophylactic measures) will be required to maintain operation of the cell at high powers.

\subsection{Next steps and Timescale for application}
The next step for the HALHF concept is to work towards a self-consistent parameter set required to generate a ‘pre-CDR’ (or Interim/Progress Report) later this year. This will complete the initial list of milestones for the project as laid out by the Lab Directors’ Group in 2023. Future experimental progress will require diversification across accelerator facilities. In the absence of a dedicated green-field facility, multiple complementary facilities will be needed to explore the relevant parameter space. This includes macro-pulse repetition rates of around 100 Hz, micro-pulse repetition rates of approximately 62.5 MHz, driver average powers approaching 500 kW, normalised beam densities of roughly 600, and energy deposition rates of approximately 1 J/m. Complementary experimental programmes at FLASHForward (approximately 10 kW average power), FACET-II (normalised beam density of around 400 and energy deposition rates near 1 J/m), and CLARA (100 Hz operation with micro-pulse rates near 100 MHz) would together enable substantial progress towards the required operating regime. Depending on the progress of this (and other) R\&D, as well as acquiring the missing funding, HALHF could be built within the next 10-15 years (cf. ~\cite{Foster_2025}) for a detailed timeline).
\newpage
\section{Staging concepts for PWFA linacs -- examples from HALHF}
\label{sec:Adli}
\href{https://agenda.infn.it/event/47329/contributions/283656/}{Slides: \url{https://agenda.infn.it/event/47329/contributions/283656/}}

\subsection*{\textit{Erik~Adli and Carl~Lindstr\o m}}

\noindent \textit{Department of Physics, University of Oslo, Oslo, 0316, Norway}
\noindent 

\subsection{Context}

Plasma-wakefield acceleration offers acceleration gradients of several GeV/m, ideal for a next-generation linear collider. However, reaching high energies in a single plasma-accelerator stage is challenging; multiple stages are likely required, which brings its own set of challenges. The beam optics requirements between plasma stages include injection and extraction of drive beams, matching the main-beam beta functions into the next cell, canceling chromaticity, dispersion as well as constraining bunch lengthening. To maintain a high effective acceleration gradient, staging of plasma accelerators must be accomplished in the shortest distance possible. 
\begin{figure}[h!]
	\centering
    \includegraphics[width=\linewidth]{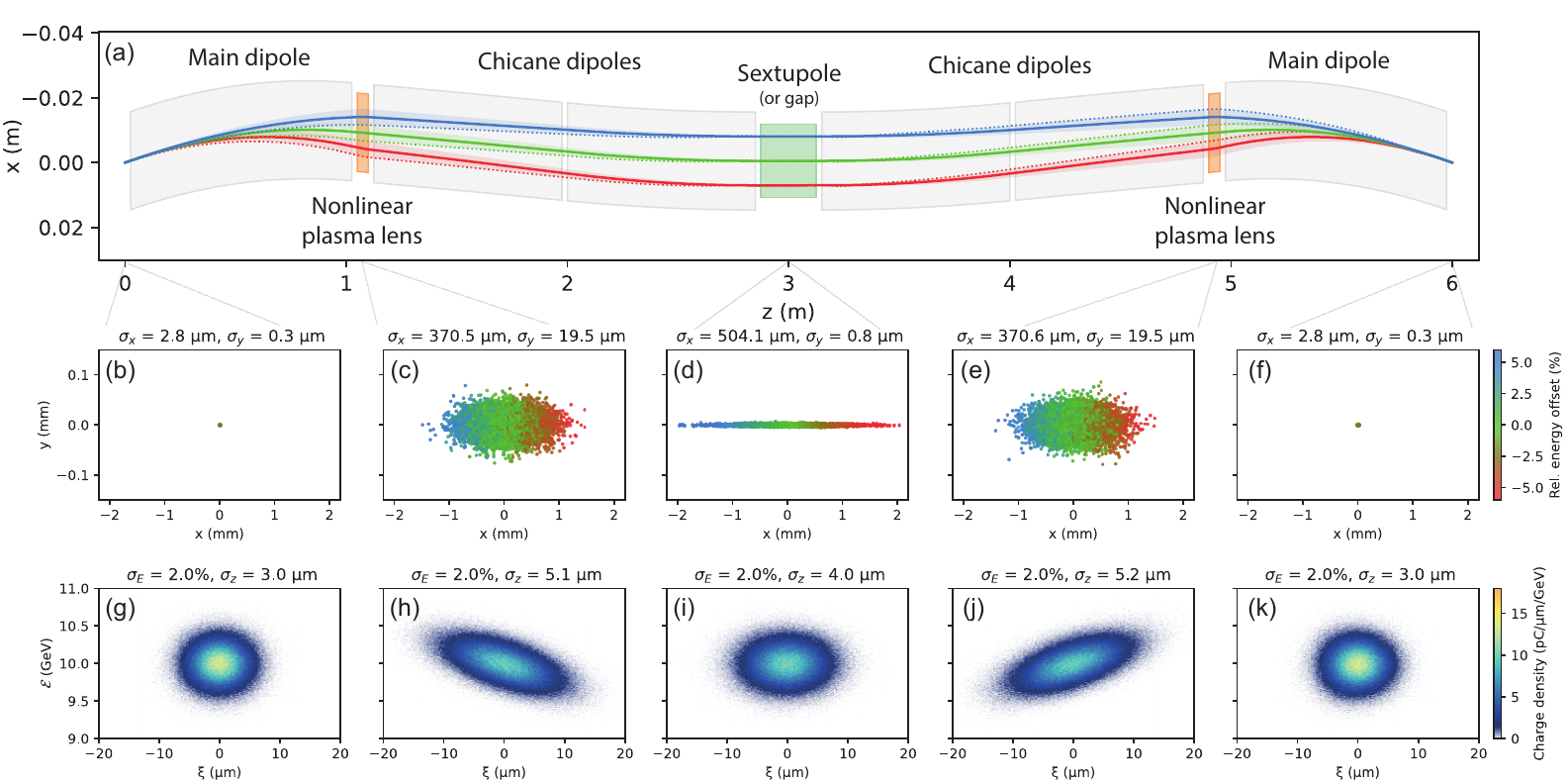}
    \caption{(a) Working principle of the achromatic interstage lattice optics, showing the orbit of three individual energy slices [here using an exaggerated energy spread of $\pm$30\%; low (red), nominal (green) and high energies are shown (blue)]. The beam is initially chromatically dispersed by a dipole, achromatically focused by the first nonlinear plasma lens, then refocused at the midpoint (a sextupole corrects second-order dispersion), then refocused by a second lens, and finally undispersed. (b--f) The evolution of the transverse $x$--$y$ profile is shown along the lattice (the color bar indicates relative energy offset). (g--k) The evolution is also shown in the longitudinal $\xi$--$\mathcal{E}$ phase space. Source: Ref.~\cite{lindstrom2026achromatic}}
    \label{fig:EA:lattice}
\end{figure}

\subsection{Objectives}

Previous plasma-based collider concepts, including those reported in Refs.~\cite{EA:rosenzweig1996, EA:seryi2009, adli_2013_PWFAlinac, EA:schroeder2016}
, did not provide a design, nor clear solutions, for how to stage plasma accelerators.  Following community discussions on how to improve these earlier collider concept, a large effort has been going into finding solutions and designs for the interstage optics, or ``interstage". As thoroughly reviewed in Ref.~\cite{Lindstrom_2021staging}, a number of requirements must be met, such as space efficient in- and out-coupling of the driver, matching of beta functions, chromaticity correction, dispersion cancellation, and ensuring synchronization and alignment to within extremely tight tolerances.

\begin{figure}[h!]
	\centering
    \includegraphics[width=\linewidth]{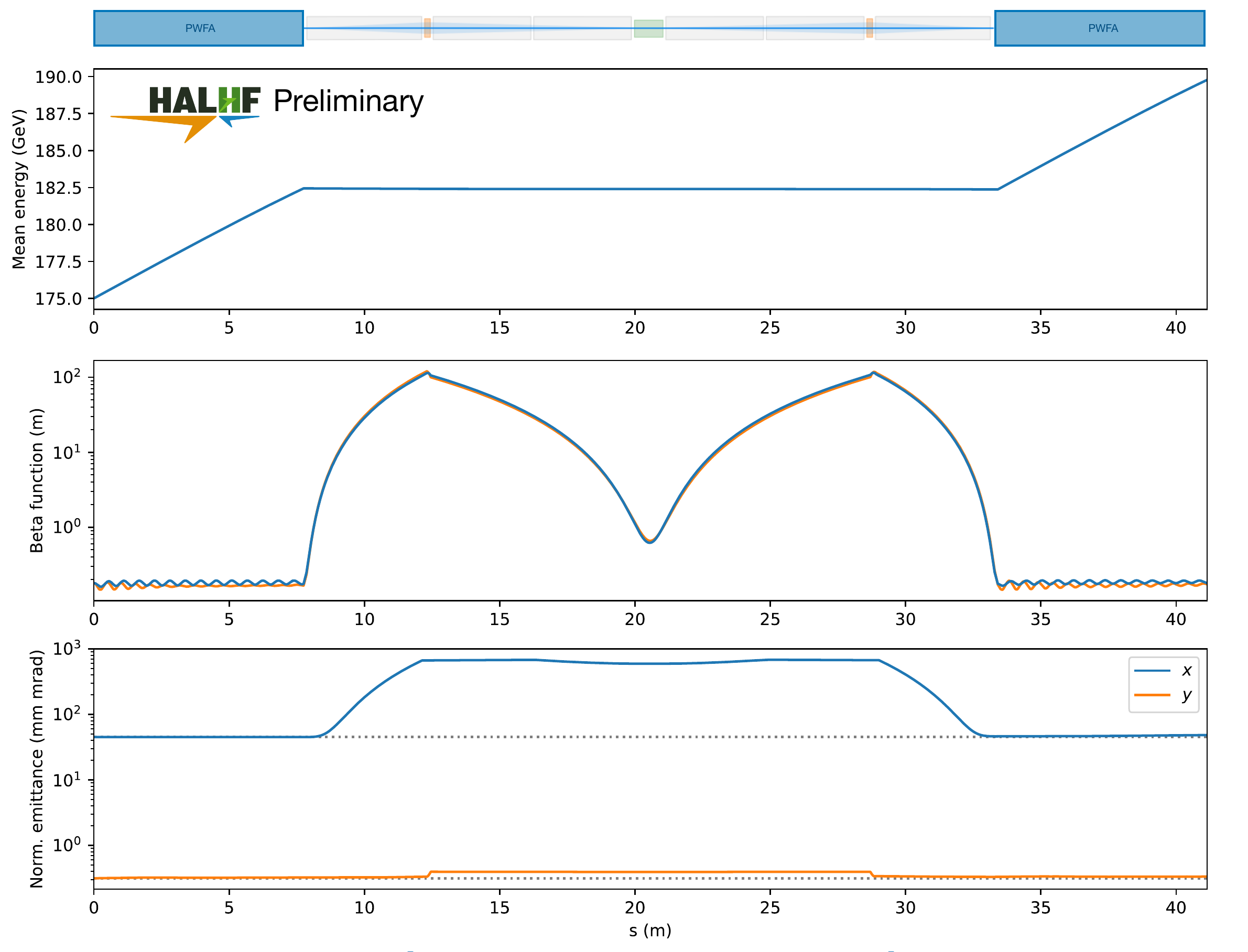}
    \caption{Example of beam-quality-preserving staging from HALHF, using the achromatic lattice. A two-stage simulation is performed in the middle of the HALHF plasma linac (starting at 175 GeV), using HiPACE++ PIC simulation for the plasma stages and ImpactX for the interstage. Here we see that the energy increases by 7.8 GeV per stage (top plot), the beta function increases rapidly when exiting the stage but gets refocused (achromatically) into the next stage (middle plot), and that the emittance is preserved in the plasma stages but temporarily increases due to dispersion and nonlinear effects in the interstage before ultimately being canceled again (bottom plot).}
    \label{fig:EA:HALHFstaging}
\end{figure}
\subsection{Progress over the last two years}
A complete design, fulfilling all requirements, has recently been developed: a compact and achromatic interstage lattice that can transport beams with high divergence and large energy spreads. This design is documented in Ref.~\cite{lindstrom2026achromatic}. The key innovation is the use of a new beam-optics element -- nonlinear plasma lenses \cite{EA:drobniak2025} -- which focuses particles on one side (e.g., left) more strongly than particles on the other side (e.g., right), and can therefore be combined with an energy-dispersion (e.g., higher energies on the left, lower energies on the right) to cancel chromaticity. Figure~\ref{fig:EA:lattice} illustrates this interstage lattice, and key performance aspects.

Being both simpler and shorter than an equivalent quadrupole-based lattice, it also outperforms in terms of transportable energy bandwidth; up to several percent rms. Moreover, the lattice has a tunable $R_{56}$ to allow for longitudinal self-correction in a multistage plasma accelerator, providing intrinsic energy stability and reduced energy spread. The lattice can be scaled to work across 4 orders of magnitude in energy, from \SI{0.5}{GeV} or lower to \SI{5}{TeV} or higher. In short, this achromatic lattice promises to solve the staging problem in plasma acceleration.

\subsection{Time scale for application}

The interstage lattice has been successfully applied to the HALHF collider design study \cite{HALHF_2025}. Figure~\ref{fig:EA:HALHFstaging} shows a two-stage simulation using HALHF parameters, preserving emittance both in the plasma stages and in the interstage. The next step is to show that this works for the full 48-stage plasma linac. 

The practical implementation and development of a working nonlinear plasma lens is being done in the SPARTA project~\cite{EA:lindstrom2025}, with plans is to demonstrate a working interstage lattice experimentally and eventually to implement a medium-scale staging demonstrator facility.

\newpage
\section{High-Energy Physics Applications of the AWAKE Scheme}
\label{sec:Wing}
\href{https://agenda.infn.it/event/47329/contributions/280918/}{Slides: \url{https://agenda.infn.it/event/47329/contributions/280918/}}
\subsection*{\textit{Matthew~Wing, on behalf of the AWAKE collaboration}}

\noindent \textit{University College London, London, UK}

\noindent Proton-driven plasma wakefield acceleration was conceived~\cite{Caldwell:2008oob} as a means to accelerate electrons up to TeV energy scales for applications in high-energy physics.  Given the high proton energy currently available, the acceleration needs only one plasma stage,  in contrast to the laser- and electron-driven schemes.  The AWAKE Collaboration~\cite{AWAKE:2014wjf,Caldwell:2015rkk,AWAKE:2015taz} was established to demonstrate proton-driven plasma wakefield acceleration~\cite{pwfa-AWAKE-acceleration,AWAKE:2018oqk,pwfa-AWAKE-modulation} and show that this is a technology that can be used to produce high-energy electrons.

\subsection{Context}\label{sec:context}

There are many high-energy physics applications for electron beams in the 10\,GeV to multi-TeV energy range~\cite{AWAKE,Wing:2018jjd,Caldwell:2018atq}.  The AWAKE facility could support electrons of a few 10s of GeV, achieved in a few 10s of meters.  The number of electrons that could be provided for experiments would be greater than $10^{15}\,e^-$ per year~\cite{Caldwell:2018atq}.  New searches for dark photons or investigations of strong-field QED can be enabled by such electron beams.  Dark photons~\cite{Okun:1982xi,Galison:1983pa,Holdom:1985ag} are postulated light particles that weakly couple primarily to the electromagnetic current with strength, $\epsilon$.  These particles could be candidates for dark matter and lead to a whole new area of physics.  Searches for dark photons can be performed with electrons hitting a solid target and looking for products where no or very few Standard Model particles would be expected.  The NA64 experiment at CERN performs such searches in the coupling--mass phase space of the dark photon using electrons from the SPS secondary beam~\cite{Andreas:2013lya,Gninenko:2013rka,NA64:2016oww}.  A dedicated experiment in the AWAKE facility could extend the sensitivity to dark photons in the region of $\epsilon \sim 10^{-3} - 10^{-5}$ and masses of $\sim 0.1$\,GeV.
Strong-field QED can be investigated by colliding high-energy electrons with a high-power laser pulse ($\geq$ 10s of TW). No other current or planned strong-field QED experiment has such a high electron beam energy and so an experiment using an AWAKE beam would investigate the QED phase space in a region not otherwise accessible.  Again, this could be performed in the current AWAKE facility with the addition of a high-power laser.

Colliding electrons at 10s of GeV with protons (or ions) from the LHC would  develop an LHeC-like electron--proton/ion collider.  Although the same high energy can be reached, a collider based on the AWAKE scheme would have significantly lower luminosity~\cite{Xia:2014ida,Caldwell:2018atq} than the LHeC design~\cite{LHeCStudyGroup:2012zhm}.  With a luminosity of about $10^{30}$\,cm$^{-2}$\,s$^{-1}$, experiments would focus on investigations of the strong force and have no Higgs physics program.  However, such a collider could be sited at CERN with less new infrastructure compared to LHeC.

Using protons of 7\,TeV from the LHC to drive plasma wakefields, electrons can be accelerated to 3\,TeV.  Colliding with the other proton beam in the LHC would achieve a very-high-energy electron--proton (VHEeP)~\cite{Caldwell:2016cmw} collider with a center-of-mass energy of 9\,TeV, a factor of 30 higher than HERA.  The luminosities would be modest with integrated values in the region of 10--100\,pb$^{-1}$.  However, even with these luminosity values, VHEeP has significant potential to investigate the strong force and QCD in a completely unknown realm and where we expect to see a different kind of fundamental proton structure.

High-energy protons of up to 275\,GeV are also available at BNL and so the AWAKE scheme could be applied there.  A concept~\cite{pwfa-alive-EIC} for an electron injector based on proton-driven plasma wakefield acceleration has been developed for the Electron--Ion Collider (EIC).  Based on the AWAKE scheme, acceleration of electrons of 10\,GeV in bunches of $6 \times 10^{10}$\,electrons is possible with protons of 275 GeV in bunches of $3 \times 10^{11}$\,protons.

\subsection{Objectives}

The remaining objectives~\cite{AWAKE} of the AWAKE experiment are to demonstrate electron acceleration using high, constant gradients and with control of the bunch quality (Run 2c) and to demonstrate the scalability of  plasma sources (Run 2d).  In Run 2c, a bunch of electrons will be externally injected into a 10\,m plasma source and accelerated throughout its full length with gradients above 0.5\,GeV/m.  The electron bunches will have charges of 100s of pC and preserve the beam quality during acceleration with emittances of about 10\,mm\,$\cdot$\,mrad upon exiting the plasma source.  The Run 2c plasma source is based on laser ionization of rubidium vapor and so is not scalable beyond 10s of meters.  Therefore plasma sources are being developed that should be scalable up to kilometers.  These are based on either RF antennas and magnetic field coils (helicon plasma source~\cite{Buttenschon2018,Stollberg:2024jmh,Granetzny:2025hsb}) or with one cathode at high voltage in the middle and a grounded cathode at each end (discharge plasma source~\cite{plasma-torrado-discharge}).  These plasma sources will be developed so that their scalability can be demonstrated in the AWAKE facility, with a plasma source of at least 10\,m in length.

After demonstration of the above, the AWAKE facility could be developed for applications for particle physics experiments.  In principle, the SPS protons can drive wakefields high enough to accelerate electrons up to 200\,GeV~\cite{Lotov:2021nob}, although the current AWAKE facility will only support 10s of GeV without significant modification.

\subsection{Progress over the last two years}

The AWAKE experiment finished Run 2b in mid-2025 to allow for the dismantling of the CNGS target.  Recent progress in AWAKE has been summarized elsewhere~\cite{Gschwendtner:2025dcv}.  Key aspects for Run 2c and 2d are the increased electron energy gain when introducing a step in the plasma density~\cite{Pannell:2025ore} and the developments for scalable plasma sources; a ten-meter-long discharge plasma source was already installed and used in the AWAKE facility.

Many of the applications discussed in Section~\ref{sec:context} were proposed during the development of the case for AWAKE Run 2.  The EIC injector, also proposed around that time~\cite{Chappell:2019ovd}, has recently been significantly developed~\cite{pwfa-alive-EIC} into a realistic scenario.

\subsection{Time scale for application}

Upon completion of AWAKE Run 2c and 2d, proton-driven plasma wakefield acceleration should have been sufficiently demonstrated such that first particle physics applications could be realized.  In particular, experiments to search for dark photons and to measure strong-field QED should be possible, as well as an electron injector for the EIC.  Run 2c and Run 2d should require a combined 4\,years of data taking which sets the timescale for application.  In the original AWAKE plan, this would mean application to particle physics in the mid-2030s. 
\newpage
\section{Update on ALiVE}
\label{sec:Caldwell}
\href{https://agenda.infn.it/event/47329/contributions/281387/}{Slides: \url{https://agenda.infn.it/event/47329/contributions/281387/}}

\subsection*{\textit{Allen Caldwell\textsuperscript{1}, John Farmer\textsuperscript{1}, Nelson Lopes\textsuperscript{2}, Alexander Pukhov\textsuperscript{3}}, Ferdinand Willeke\textsuperscript{4}, Thomas Wilson\textsuperscript{3}}

\noindent \textit{\textsuperscript{1}Max Planck Institute for Physics, Garching, Germany}\\
\noindent \textit{\textsuperscript{2}Instituto Superior Tecnico, Lisbon, Portugal}\\
\noindent \textit{\textsuperscript{3}Heinrich-Heine-Universit\"at, D\"usseldorf, Germany}\\
\noindent \textit{\textsuperscript{4}Brookhaven National Laboratory, Brookhaven, USA}

\subsection{Context}

Acceleration in plasma generates interest due to the very strong accelerating gradients it allows. Initially, laser driven plasma wakefield acceleration was considered in the literature~\cite{PhysRevLett.43.267} and it was later recognized that the plasma could also be excited by an electron bunch~\cite{PhysRevLett.54.693}.  More recently, proton-driven plasma wakefield acceleration was also introduced~\cite{pwfa-caldwell-protondriven}.

The advantage of a proton driver is that the energy of the driver, using today's technology, is sufficient to reach accelerated bunch particle energies required, e.g., for a Higgs Factory~\cite{pwfa-ALiVE-EPPSU}.  This is achieved without staging, greatly simplifying the accelerator complex. A limitation of the proton-driven scheme has been the repetition rate of the driver, which limits the achievable luminosity.  With the FFA concept of F. Willeke~\cite{accelerator-willeke-FFA}, high luminosities are also achievable thereby making the proton-driven PWFA  (PDPWA) case extremely attractive.

\begin{figure*}[htb] 
    \centering
\includegraphics[width=0.5\textwidth]{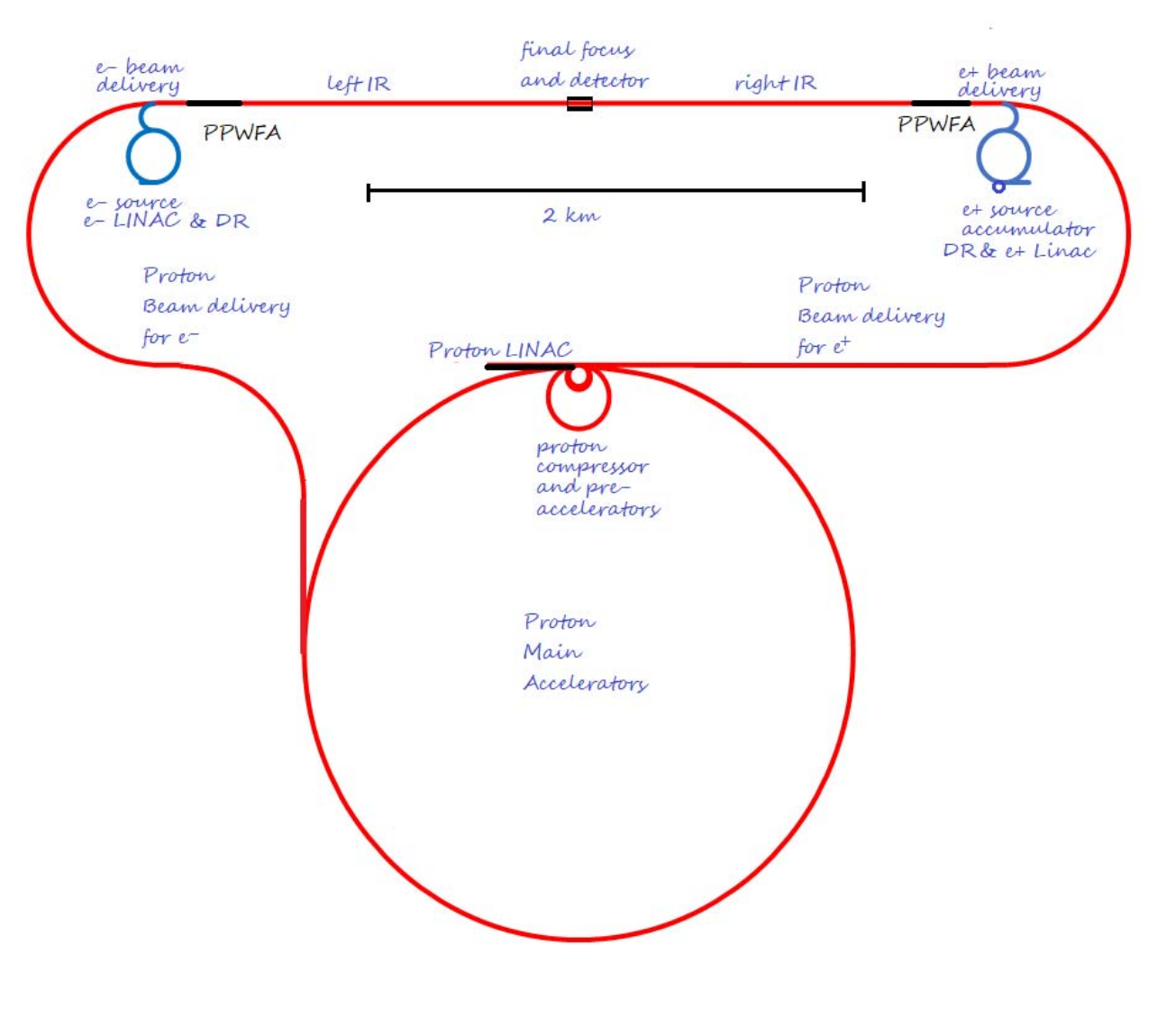}
\caption{\it  Footprint of a PDPWFA Higgs factory facility driven by an FFA proton accelerator. Figure taken from~\cite{pwfa-ALiVE-EPPSU}.}
\label{fig:footprint} 
\end{figure*}

An FFA scheme for a PDPWA collider is shown in Fig.~\ref{fig:footprint}.  A train of proton bunches are brought to the necessary energy in a rapid-cycling ring and then extracted one bunch at a time.  These are then injected into the plasma acceleration sections, trailed by the lepton bunches to be accelerated.  The accelerated bunches are then extracted and brought to the interaction point.

\subsection{Status of FFA study}

We summarize the main findings of the FFA concept to date:
\begin{itemize}

\item An FFA for acceleration of a continuous stream of proton bunches for plasma
wakefield acceleration of a lepton collider beam appears to be possible.

\item
A large amount of work needs to be done to evaluate all the details and to come
up with a complete FFA optical design.
The FFA has interesting beam physics, including collective effects and magnetic non-linearities.

\item Exponential magnets are beyond present magnet capability but extrapolating
recent progress forward will likely provide the superconducting technogy to make
the FFA magnet possible.

\end{itemize}

\subsection{Status of LEP-3 study}

Our initial focus was the study of a
Higgs Factory based on PDPWA~\cite{pwfa-farmer-higgs}, but wakefield acceleration of low-emittance positron bunches remains an unsolved challenge. We consider now only accelerating electron beams, a
HALHF-like scheme~\cite{pwfa-HALHF}, for first projects.  A very attractive option at this time is to develop a high-energy electron beam concept to increase the LEP-3 energy reach~\cite{HEP-LEP3-EPPSU} to the $t\bar{t}$ center-of-mass energy.  In our scheme, one high charge positron bunch would circulate in the LEP-3 tunnel and would be collided with high-energy bunches of electrons accelerated by proton bunches produced in an FFA installed in the SPS tunnel and extracted at $11.5~kHz$.  The parameters assumed and the luminosity resulting from our studies are given in Table~\ref{tab:LEP3}.  We find that interesting luminosities would be reachable.  The ability to increase the LEP-3 energy reach to the  $t\bar{t}$ production threshold make this an exciting option for a future project at CERN should the FCC-ee not be realized.

\begin{table}[hbpt]
    \centering
    \begin{tabular}{|l|c|c|}
    \hline
         & LEP3 Baseline & LEP3-ALIVE \\
         \hline
         electron beam energy & $115$~GeV & $290$~GeV \\
        positron beam energy & $115$~GeV & $115$~GeV \\
        CoM Energy  & $230$~GeV &  $365$~GeV\\
        electron bunch population & $2.5\cdot 10^{11}$&$ 2\cdot 10^{10}$ \\
        positron bunch population &$2.5\cdot 10^{11}$ & $5\cdot 10^{12}$\\
        circulating electron bunches & 20 & 0\\
        circulating positron bunches & 20 & 1 \\

       Luminosity  & $1.6\cdot 10^{34}$~cm$^{-2}$s$^{-1}$ & $1.8\cdot 10^{33}$~cm$^{-2}$s$^{-1}$ \\
              top 1~\% Luminosity  & $1.6\cdot 10^{34}$~cm$^{-2}$s$^{-1}$ & $1.3\cdot 10^{33}$~cm$^{-2}$s$^{-1}$ \\
\hline
    \end{tabular}
    \caption{LEP-3 parameters assumed and simulation results}
    \label{tab:LEP3}
\end{table}

\subsection{Summary}
We find that, with the development of the FFA concept, a path has been found for a high-luminosity, high energy lepton collider based on proton-driven plasma wakefield acceleration.  The reduced footprint and reuse of existing facilities make this a very interesting option for a future collider facility.

\newpage
\section{Plasma Wakefield Injector for EIC}
\label{sec:Farmer}
\href{https://agenda.infn.it/event/47329/contributions/281387/}{Slides: \url{https://agenda.infn.it/event/47329/contributions/281387/}}


\subsection*{\textit{
        J.~P.~Farmer\textsuperscript{1},
        H.~Jaworska\textsuperscript{2},
        A.~Caldwell\textsuperscript{1}, 
		N.~Lopes\textsuperscript{3}, 
        A.~Pukhov\textsuperscript{2},
        L.~Reichwein\textsuperscript{2,4},
        F.~Willeke\textsuperscript{5}, 
        M.~Wing\textsuperscript{6,7}}}
        
\noindent \textit{\textsuperscript{1}Max Planck Institute for Physics, Garching, Germany}\\
\noindent \textit{\textsuperscript{2}Heinrich-Heine-Universit\"at, D\"usseldorf, Germany}\\
\noindent \textit{\textsuperscript{3}Instituto Superior Tecnico, Lisbon, Portugal}\\
\noindent \textit{\textsuperscript{4}Forschungszentrum Jülich, Germany}\\
\noindent \textit{\textsuperscript{5}Brookhaven National Laboratory, Brookhaven, NY, USA}\\
\noindent \textit{\textsuperscript{6}UCL, London, UK}\\
\noindent \textit{\textsuperscript{7}DESY, Hamburg, Germany}


\subsection{Context}
The Electron Ion Collider (EIC) is planned to be constructed at Brookhaven National Laboratory~\cite{accel-willeke-eic_cdr}, making use of the existing infrastructure of the Relativistic Heavy Ion Collider (RHIC).  The existing RHIC Yellow-Ring will be upgraded and repurposed as a 275\,GeV Hadron Storage Ring (HSR), and an Electron Storage Ring (ESR) supporting energies of up to 18\,GeV will be installed.

The EIC will support an ambitious scientific program, deepening our understanding of fundamental science including nuclear structure, gluon saturation and proton spin.  This program imposes demanding technical requirements on the accelerator, which must be met at an affordable cost.

The existing RHIC Blue-Ring, which accelerates protons in the same direction as the EIC ESR, is not currently considered necessary for the EIC.  We propose to make use of the Blue-Ring to generate proton drive beams for a plasma wakefield accelerator, which has the potential to accelerate electron bunches for the ESR in a relatively compact and cost-effective manner.

\subsection{Objectives}
The requirements for the electron beam are challenging for a plasma-based accelerator due to the high bunch charge.  This is especially true for 10 GeV operation, where the required beam charge is 10x higher than for 18~GeV operation.  1160 polarized electron bunches of 28~nC must be accelerated and stored, with the beam storage time limited by the proton beam lifetime of ten hours.

We propose a novel top-up scheme to meet the requirement of high bunch charge~\cite{pwfa-alive-EIC}, in which plasma-accelerated electron bunches are accelerated and accumulated in the ESR until the full beam charge is accumulated.  The repetition rate for the plasma accelerator is limited by the fill time of the Blue-Ring, imposing an average repetition rate of $\sim1$~Hz.  This places a minimum requirement of 1~nC electron charge accelerated per proton drive beam.

\begin{figure}
    \centering
    \includegraphics[width=0.9\textwidth]{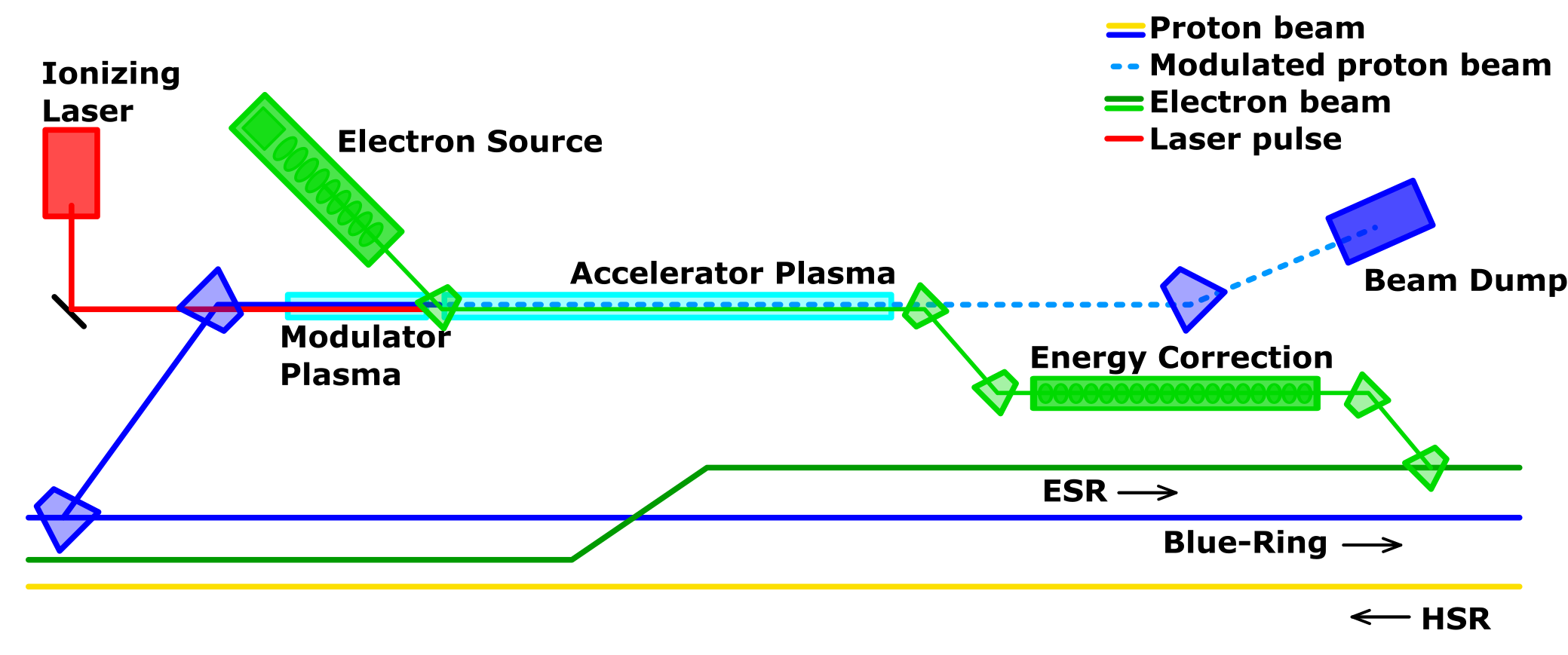}
    \caption[Caption for List of Figures]{\label{fig:EICinjectorlayout} Conceptual layout of an electron injector for the EIC based on plasma wakefield acceleration~\cite{pwfa-alive-EIC}.  Proton bunches are extracted from the Blue-Ring and modulated in a short plasma section.  The modulated bunch is then used to accelerate a witness bunch of electrons.  The spent proton beam is dumped, and the electrons are sent to an RF energy correction stage before being injected and accumulated in the ESR.  The full injector is expected to fit within a length of $< 200$~m, equivalent to one of the straight sections of the EIC.}
\end{figure}

Many of the challenges in realizing a proton-driven wakefield accelerator have already been addressed by the AWAKE experiment at CERN~\cite{pwfa-AWAKE-symmetry}.  A long proton beam, such as those accelerated in the Blue-Ring, can be self-modulated in a plasma~\cite{pwfa-lotov-smi,pwfa-AWAKE-modulation}, with the resulting train of microbunches used to accelerate a witness bunch to high energy~\cite{pwfa-AWAKE-acceleration}.  A schematic layout of the proposed EIC injector is shown in Fig.~\ref{fig:EICinjectorlayout}, and follows general scheme proposed by AWAKE Run~2c and d.  However, the requirements for an EIC injector differ from the applications proposed for the AWAKE experiment, requiring the acceleration of high electron bunch charge with a sub-percent level energy spread.

Upgrading the Blue-Ring to 275~GeV and increasing the bunch charge would allow proton drive bunches similar to those already used in AWAKE.  Particle-in-cell simulations have demonstrated that such drive bunches would allow 1\,nC electron bunches to be accelerated from 150\,MeV to 10~GeV in 50\,m of plasma~\cite{pwfa-alive-EIC}.  Such plasma sources could be achieved using a discharge plasma, again building on experience with scalable plasma sources from AWAKE~\cite{plasma-torrado-discharge}.

The resulting plasma-based injector is discussed in detail in reference~\cite{pwfa-alive-EIC} and offers performance close to the EIC baseline.  Improvements in performance can be achieved by increasing the electron charge accelerated per drive bunch.  This is expected to be possible through moderate optimization, for example through shaping the electron bunch~\cite{pwfa-meer-beamloading,pwfa-lindstrom-energyspread} or by injecting multiple electron bunches per drive bunch~\cite{pwfa-farmer-multibunch}.

\subsection{Outlook}
These studies have demonstrated that an electron injector for the EIC based on plasma wakefield acceleration could deliver performance close to the EIC baseline~\cite{pwfa-alive-EIC}, at a potentially much reduced cost.  Detailed studies are required for some components, and improvements in performance are expected to be possible through moderate optimization.  These results support the continued development of this scheme towards a full conceptual design.

\newpage
\section{Wakefield Accelerators for Future 10\,TeV Colliders \& Near-Term Applications}
\label{sec:Osterhoff}
\href{https://agenda.infn.it/event/47329/contributions/284252/}{Slides: \url{https://agenda.infn.it/event/47329/contributions/284252/}}

\subsection*{\textit{Jens Osterhoff} on behalf of the 10 TeV Wakefield Collider Collaboration}

\noindent \textit{Lawrence Berkeley National Laboratory, Berkeley, California 94720, USA}

\subsection{Introduction}
Wakefield technology is maturing rapidly toward the performance required for first accelerator applications. In the US, this development has largely been driven by R\&D aimed at delivering the accelerator technology for future particle colliders. That effort, together with targeted US and worldwide work on non-collider use cases, has markedly improved wakefield technology in energy, beam quality, stability, and control. These systems are approaching the performance needed in early applications such as particle and radiation sources that drive free-electron lasers (FELs), synchrotron injectors, and beam manipulation in the context of Basic Energy Sciences. High Energy Physics early applications could include injectors, for electron and muon test beams, and in Axion search and strong-field quantum electrodynamics studies. Compact wakefield technology can also enable novel medical therapy and imaging modalities, and advanced industrial and security applications. All of these develop technology towards potential future colliders.

\subsection{Technology and vision}
Four GV/m wakefield technologies are under development: laser- and beam-driven schemes that can each excite strong accelerating fields in either structure- or plasma-based media, with each of these four variations offering distinct characteristics. Plasma-based technologies, which deliver the highest gradients up to the 100 GV/m level, are the largest area of international effort with laser drivers also offering a credible path to ultra-compact and potentially mobile sources with broad near-term applicability.\\
These near-term applications can act as an accelerant, catalyzing the development of a mature, high-gradient, affordable technology for an energy upgrade of a future linear-collider Higgs factory in the LCVision context \cite{LCVision} or for a compact green-field machine. To qualify as a particle collider technology, the field must deliver three advances simultaneously: 1) high energy, through staging or very high-energy drivers; 2) high luminosity, through high repetition rate, stability, ultra-low emittance, and high charge; and 3) high efficiency in transferring energy from the power grid into the colliding beams. Dedicated wakefield collider R\&D is needed, as these goals substantially exceed the demands of near-term applications.

\subsection{Wakefield accelerators are maturing}
The community has made notable progress in recent years. Laser-driven plasma experiments reached the 10 GeV level over a 30 cm acceleration stage \cite{PhysRevLett.133.255001}, an energy considered viable for a TeV-scale multi-staged plasma linac built from 10 GeV modules, or for near term XFELs. Beam-driven experiments simultaneously demonstrated micron-level normalized emittance preservation and sub-percent energy spread in plasma \cite{Lindstrom:2024zbo}. Teams worldwide produced beams from laser- and beam-driven plasmas of sufficient quality to drive FELs \cite{BARBER:2025,WANG:2021,LABAT:2023,pompili2022_pwfa_fel}. Control and stability improved markedly through passive and active methods, including machine-learning-based control: energy spread and shot-to-shot jitter below 0.1\% were shown \cite{winkler}, and a laser-plasma FEL ran over tens of thousands of shots across days of operation \cite{BARBER:2025}. Beam-driven experiments recently reached 60\% transfer efficiency from driver to wakefield and 42\% from wakefield to the accelerated beam \cite{pwfa-lindstrom-energyspread}, and preliminary results indicate the potential for acceleration at MHz repetition rates in wakefields \cite{darcy}.\\
These advances are prompting the community to integrate plasma-based subsystems into \$100M-scale projects, as FELs (EuPRAXIA in Europe, C2FEL in China) and full-energy synchrotron injectors (PETRA IV in Germany). Meanwhile, companies are beginning to invest in the technology for near-term applications, and the broad particle-physics community is deeply involved in designing a 10 TeV collider based on wakefield concepts \cite{10TeVStudy}.

\subsection{The 10 TeV wakefield collider design study}
This community study was initiated in 2024 in response to the latest P5 Report \cite{P5:2023wyd}, aiming to develop a unified 10 TeV parton-center-of-mass wakefield collider design by 2028. Nearing the end of a first assessment phase, it explores design options, metrics, and technology trade-offs, funded in the US entirely through LDRDs, to first answer whether compelling wakefield collider options exist at low cost. The study necessarily began at the interaction point, where a new linear-collider paradigm emerges. At the 10 TeV scale, vector boson fusion complements traditional s-channel annihilation in the total particle production rate, motivating exploration of $e^+e^-$, $e^-e^-$, and $\gamma\gamma$ collision modes. Beamstrahlung must also be revisited in the quantum regime, since its strong, unavoidable effect on the luminosity spectrum requires considering round beam collisions alongside traditional flat beam shapes.\\
We have modeled the IP with new (WARPX) and legacy (GUINEA-PIG, CAIN) codes, using conceptual beam-parameter sets developed for the most recent Snowmass process. Owing to strong beamstrahlung, all considered collider types exhibit broad luminosity spectra, with beams effectively interacting as composite initial states. Together with theorists and detector specialists, we modeled particle production rates for the three colliders ($e^+e^-$, $e^-e^-$, and $\gamma\gamma$) with round and flat beams across production channels; the results point to round-beam $e^+e^-$ or $\gamma\gamma$ colliders as interesting candidates for a future machine \cite{chigusa.arxiv.2025}.\\
Originating in the United States under LDRD funding that ends in September 2026, the effort has grown into a global initiative through the LCVision and ALEGRO frameworks. With indications that US funding in the near future may emphasize near term applications while work on future colliders will be constrained, international partnership will be needed to carry the study forward.

\subsection{Conclusion}
Wakefield technology is approaching the performance required for near-term applications, with strong synergies across scientific sectors, medicine, industry, and security. Its rapid development also reinforces work on long-term collider applications, with the ultimate goal of a future linear energy-frontier machine at 10 TeV. The 10 TeV wakefield collider design study is off to a strong start, and the study will need to be carried by the global community, possibly coordinated through ALEGRO.

\newpage
\section{The Physics Case for a 10\,TeV Wakefield Collider: Turning Beam Physics into New Physics}
\label{sec:Opferkuch}
\href{https://agenda.infn.it/event/47329/contributions/281984/}{Slides: \url{https://agenda.infn.it/event/47329/contributions/281984/}}
\subsection*{\textit{Toby Opferkuch}}
\noindent \textit{SISSA International School for Advanced Studies, Via Bonomea 265, 34136 Trieste, Italy}\\
\noindent \textit{INFN Sezione di Trieste, Via Bonomea 265, 34136 Trieste, Italy}

\subsection{Context}
Reaching a parton-level center-of-mass energy of about \SI{10}{\TeV} is widely regarded as the next major step at the energy frontier and is a central element of the strategies set out in the 2020 European Strategy and the 2023 P5 Report~\cite{Adolphsen:2022ibf,P5:2023wyd}. Three accelerator paths are being pursued: a $\sim$\SI{100}{\TeV} proton-proton collider, a \SI{10}{\TeV} muon collider, and a \SI{10}{\TeV} plasma-wakefield linear collider (WFC)~\cite{Gessner:2025acq}. None are construction-ready, and each faces distinct technical challenges. In particular, a WFC must achieve efficient \emph{positron} acceleration, multi-stage operation, and tolerable levels of beam-beam radiation (beamstrahlung) at the interaction point. While the physics case for a \SI{10}{\TeV} muon collider is by now well established, that of a wakefield electron-based collider has received comparatively little theoretical attention. The work summarized here aims to address this imbalance by examining two questions: (i)~whether a compelling new-physics case exists for $e^-e^-$ and $\gamma\gamma$ machines, which circumvent the positron-acceleration problem; and (ii)~whether the strong beamstrahlung characteristic of a wakefield collider should be regarded as a limitation or as an asset. Quantitative answers to these questions can directly inform R\&D priorities (positron source, staging, beam shape) for the wakefield community.

\subsection{Objectives}
The overarching objective is to translate realistic, simulation-driven WFC beam parameters into quantitative new-physics reach, so that machine and detector R\&D can be guided by the physics targets that become accessible. To this end, we incorporate luminosity spectra obtained from full beam-beam simulations~\cite{schroeder.jinst.2023} into matrix-element calculations of new-physics signals, treating beamstrahlung-induced photons and secondary positrons as ``initial-state partons'' in analogy with hadronic PDFs. Five WFC configurations are benchmarked: $e^+e^-$ and $e^-e^-$, each with round and flat beams, plus a $\gamma\gamma$ option. The same framework is applied to a representative set of targets: heavy kinetically mixed $Z^\prime$ resonances~\cite{Cipressi:2026WFC}, pair production of electroweak multiplets including the thermal Higgsino~\cite{chigusa.arxiv.2025}, and ongoing studies of the triple Higgs coupling, lepton-flavor violation, and single-particle production in vector boson fusion. Projections are compared against the HL-LHC, FCC-hh, and a \SI{10}{\TeV} muon collider under matched assumptions.

\begin{figure}[t]
    \centering
    \includegraphics[width=\textwidth]{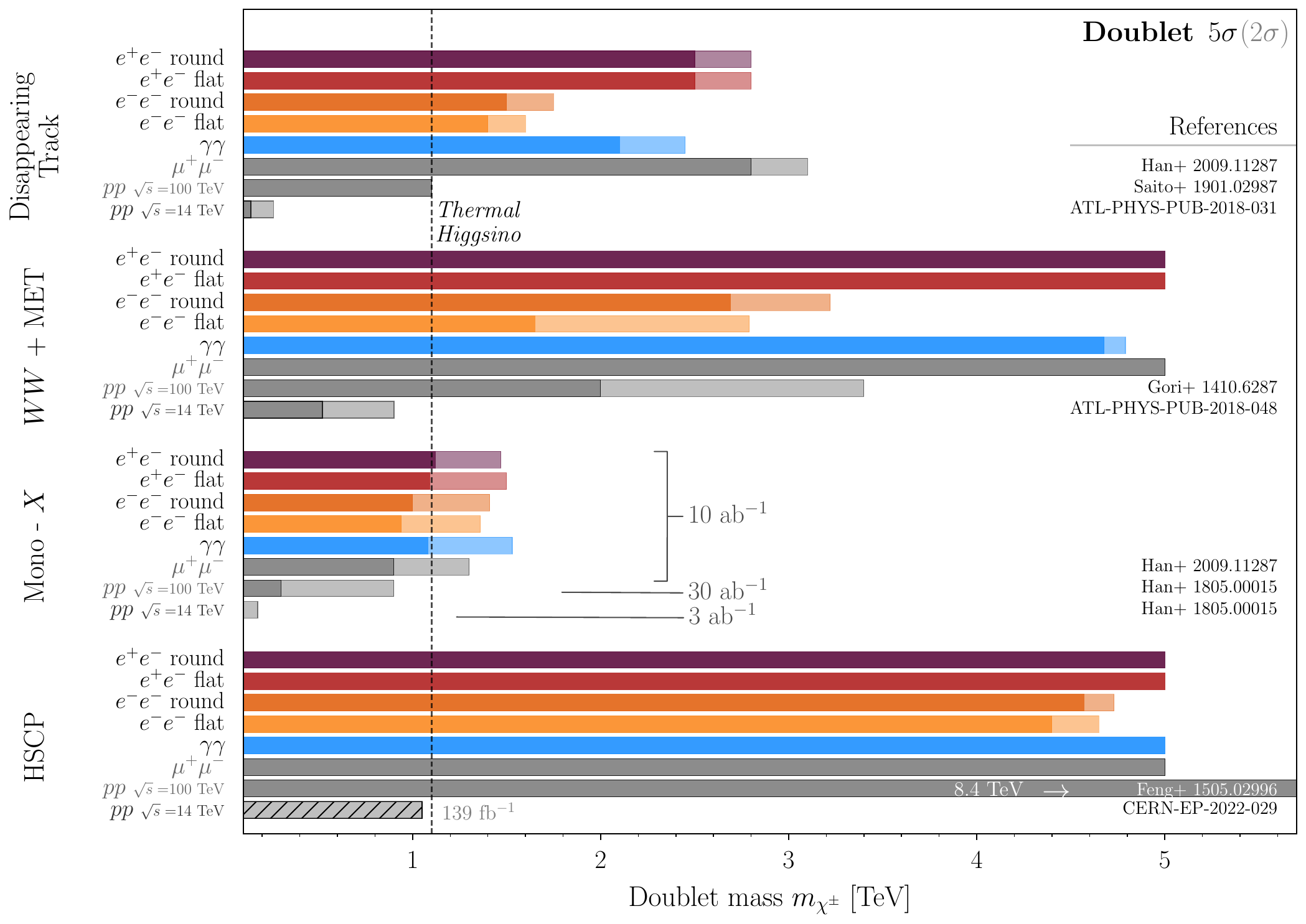}
    \caption[Higgsino mass reach across collider concepts]{\label{fig:Higgsino} $5\sigma$ discovery (solid bars) and $2\sigma$ exclusion (light extensions) reach in the doublet (Higgsino) mass $m_{\chi^\pm}$ for four complementary search strategies, comparing the five WFC configurations against a \SI{10}{\TeV} muon collider and $pp$ machines at $\sqrt{s}=14$ and \SI{100}{\TeV}. WFC and $\mu^+\mu^-$ assume 10\,ab$^{-1}$, FCC-hh 30\,ab$^{-1}$, HL-LHC 3\,ab$^{-1}$. The dashed vertical line marks the thermal Higgsino target. From~\cite{chigusa.arxiv.2025}.}
\end{figure}

\subsection{Progress over the last two years}
Three principal findings have emerged. \emph{Beamstrahlung is an asset, not a liability.} The broad luminosity spectrum it induces effectively scans a wide range of sub-nominal energies, enhancing narrow-resonance production by orders of magnitude relative to initial-state radiation alone~\cite{Cipressi:2026WFC}. For a kinetically mixed $Z^\prime$ at $M_{Z^\prime}\!\sim\!1$\,TeV, an $e^+e^-$ WFC at $\sqrt{s}=\SI{10}{\TeV}$ improves the sensitivity to the kinetic-mixing parameter by more than an order of magnitude relative to a muon collider at the same energy and geometric luminosity, reaching into theoretically motivated, one-loop generated regions; the advantage diminishes as $M_{Z^\prime}$ approaches $\sqrt{s}$. \emph{Round beams outperform flat beams despite increased radiation.} Although round beams emit more beamstrahlung, their larger geometric luminosity yields a higher signal yield for essentially every target studied. \emph{Photon-photon colliders are highly competitive.} For thermal-mass electroweak multiplets, a $\gamma\gamma$ collider with \SI{10}{ab^{-1}} performs comparably to a $\mu^+\mu^-$ collider at the same luminosity and exceeds the reach of a \SI{100}{\TeV} $pp$ machine; combined with disappearing-track and heavy stable charged particle searches, it covers the thermal Higgsino target~\cite{chigusa.arxiv.2025} (Fig.~\ref{fig:Higgsino}). Crucially, \emph{secondary} photons and positrons produced via beamstrahlung give electron-only configurations meaningful access to channels otherwise unavailable, partially mitigating the positron-acceleration bottleneck.

\subsection{Time scale for application}
The studies above use realistic but pre-design beam parameters; their near-term role (2--3 years) is to provide quantitative physics input to the \SI{10}{\TeV} Wakefield Collider Design Study~\cite{Gessner:2025acq}. On a five-year horizon, sustained interaction with accelerator and detector teams should yield self-consistent benchmark scenarios against which R\&D milestones can be calibrated. Particle-physics experiments at a wakefield collider remain a 2040+ prospect, contingent on energy-gain scaling with plasma length and on resolving the positron and staging challenges. The present work demonstrates, however, that several of the most compelling physics targets do not require their simultaneous resolution.
\newpage
\section{High-Energy Gamma-Gamma Collider}
\label{sec:Barklow}
\href{https://agenda.infn.it/event/47329/contributions/281990/}{Slides: \url{https://agenda.infn.it/event/47329/contributions/281990/}}

\subsection*{\textit{T.~Barklow\textsuperscript{1} and
             S.~S.~Bulanov\textsuperscript{2}}}

\noindent \textit{\textsuperscript{1}SLAC National Accelerator Laboratory, Menlo Park, CA, USA}\\
\noindent \textit{\textsuperscript{2}Lawrence Berkeley National Laboratory, Berkeley, CA, USA}

\subsection{Context}
The 2023 P5 report~\cite{P5:2023report} identified the 10~TeV parton center-of-mass energy frontier as a priority for particle physics.
A \textit{standalone} $\gamma\gamma$ collider at this scale, in which a portion of the electrons in a  5~TeV electron beam
are converted into $\sim 5$~TeV photons through Compton scattering with laser beams,
offers a compelling positron-free path to beyond Standard Model (BSM) discovery. 
Unlike $e^+e^-$ colliders, a $\gamma\gamma$ collider requires only
electron beams, sidestepping the technological burden of positron production and acceleration.
The $\gamma\gamma$ initial state is 
uniquely powerful: the inclusive
$\gamma\gamma\!\to\!WW$ cross section ($\sim\!90$~pb) exceeds
$e^+e^-\!\to\!WW$ by three orders of magnitude~\cite{tesla.ijmpa.2004}, suggesting similarly enhanced cross-sections for 
vector boson scattering $\gamma\gamma\rightarrow WWX$.
Furthermore, the rate of $e^+e^-$ pair-production of heavy BSM particles at a $\gamma\gamma$ collider is enhanced relative to that of an $e^-e^-$ collider 
due to  positron production in
the Compton interaction regions through $\gamma_\textrm{ laser}\gamma\rightarrow e^+e^-$.

Conventional $\gamma\gamma$ collider designs stipulate that the scattering laser 
wavelength satisfy $\lambda[\mu\text{m}]>4.2\,E_0[\text{TeV}]$ to suppress
photon conversion into $e^+e^-$ pairs inside the laser pulse.
At 10~TeV this requires $>\!25\;\mu$m mid-infrared lasers — a severe
technological challenge that our work addresses directly.

\subsection{Objectives}

Our goal is to demonstrate that a 10~TeV $\gamma\gamma$ collider can be
realized with lasers spanning optical to X-ray wavelengths,
operating in the high-$x$ regime ($x\!\gg\!4.8$, where
$x=4E_0\omega_0/m_e^2$, with $E_0$ and $\omega_0$ the electron and laser photon energies, respectively, and $m_e$ the electron mass) even in the presence of prolific
$e^+e^-$ pair production during photon generation~\cite{barklow.jinst.2023a,barklow.jinst.2023b}.
The envisioned machine uses two 5~TeV electron beams from wakefield
accelerators~\cite{schroeder.jinst.2023,10TeV} that each Compton-scatter off a
moderate-intensity ($a_0\!=\!0.3$) laser pulse just upstream of the interaction
point (IP).
By scanning the laser photon energy ($\omega_0 = 1.2$~eV to $1$~keV) and
pulse length, we seek to maximize the partial $\gamma\gamma$ luminosities
$L_{\gamma\gamma}^{5\%}$ and $L_{\gamma\gamma}^{20\%}$—the luminosities in
$\gamma\gamma$ collisions with center-of-mass energy exceeding 95\% and 80\%
of the maximum value, respectively.


\label{fig:lumi pol}

\subsection{Progress over the last two years}

\textbf{Analytical model.}
A 1D rate-equation model shows that the competing Compton photon production and
Breit-Wheeler absorption processes reach a photon-number maximum of
$n_\gamma^{\max}\!=\!n_0\,\kappa^{\kappa/(1-\kappa)}$, where $n_\gamma$ is the number of high energy photons,
$\kappa\!=\!W_{BW}/W_C$, and $W_C$ ($W_{BW}$) are the rates of the Compton (Breit-Wheeler) processes,
 yielding 0.25--0.45 high-energy photons per initial
electron for $x$ ranging from 4.8 to $\infty$.

\begin{figure*}[!ht]
\centering
\includegraphics[width=5.7cm]{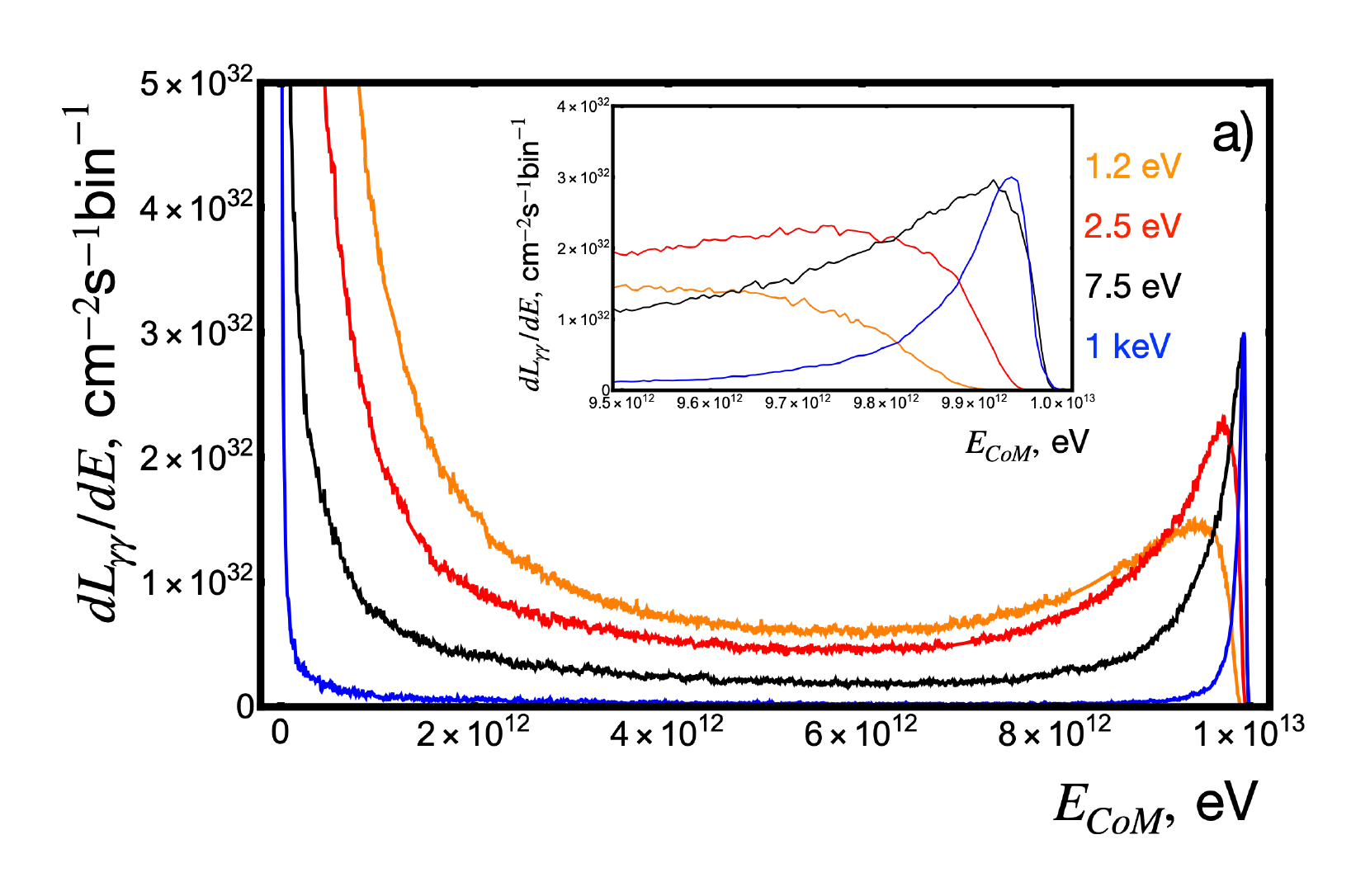}
\includegraphics[width=5.7cm]{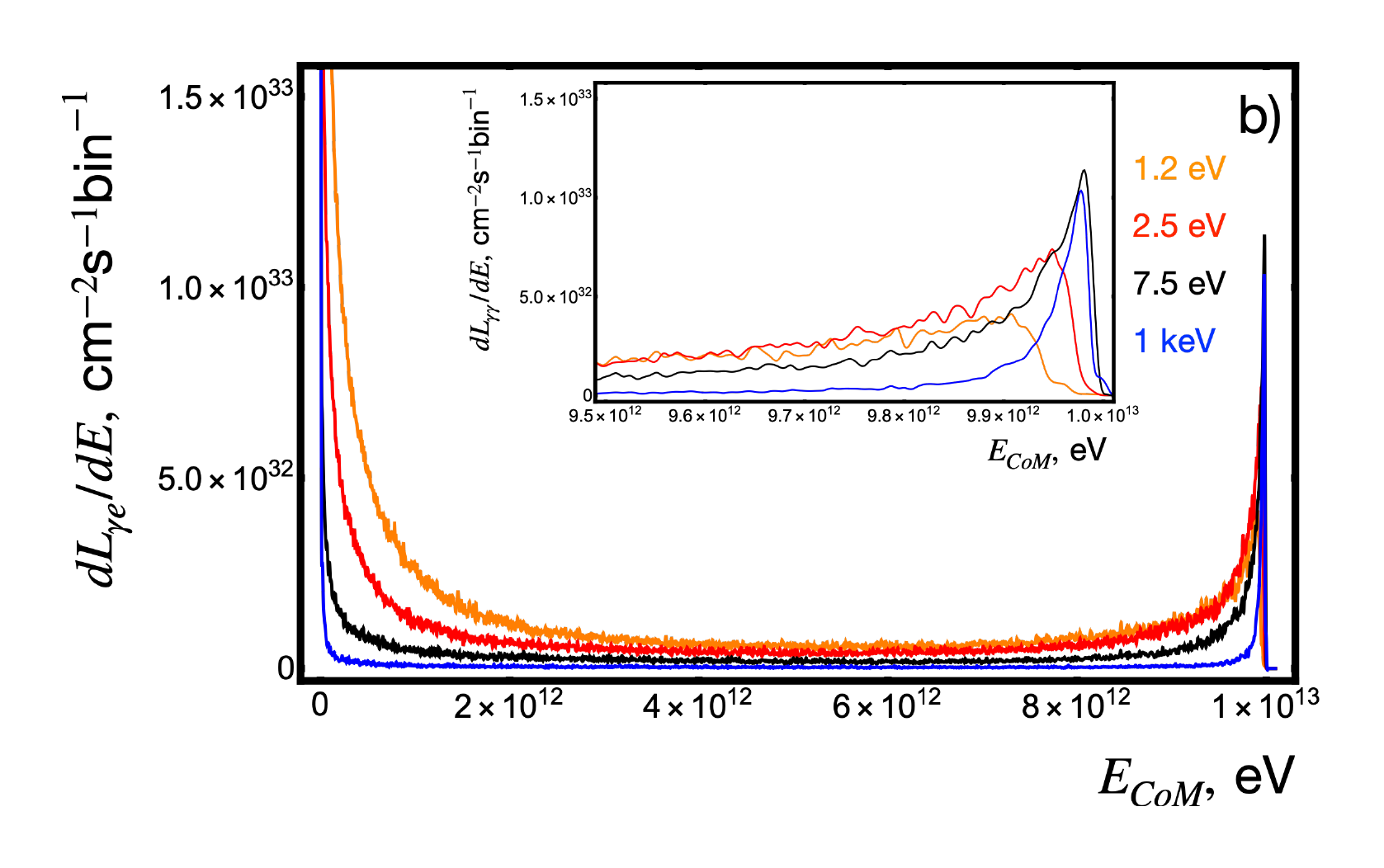}
\includegraphics[width=5.7cm]{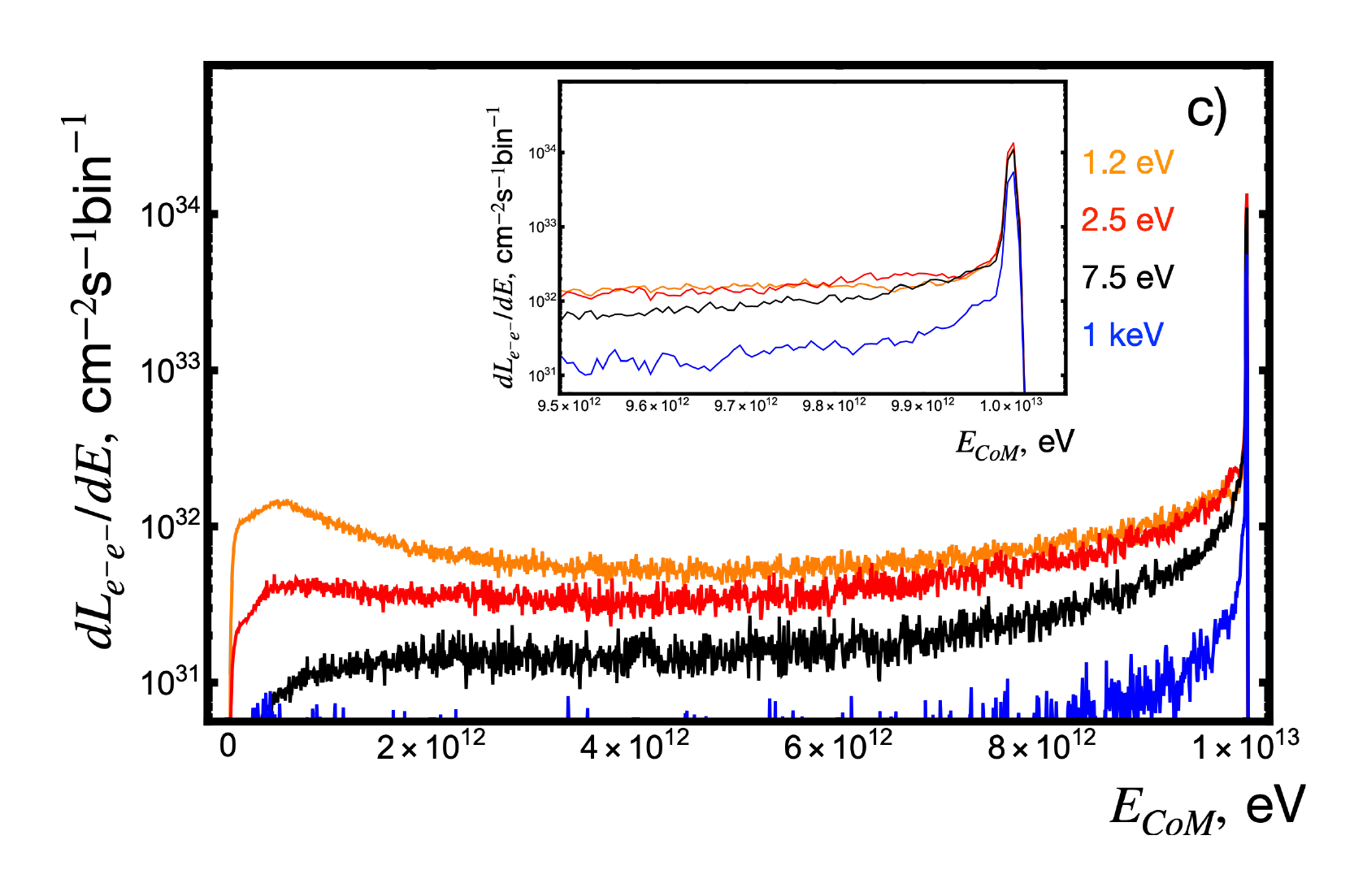}
\includegraphics[width=5.7cm]{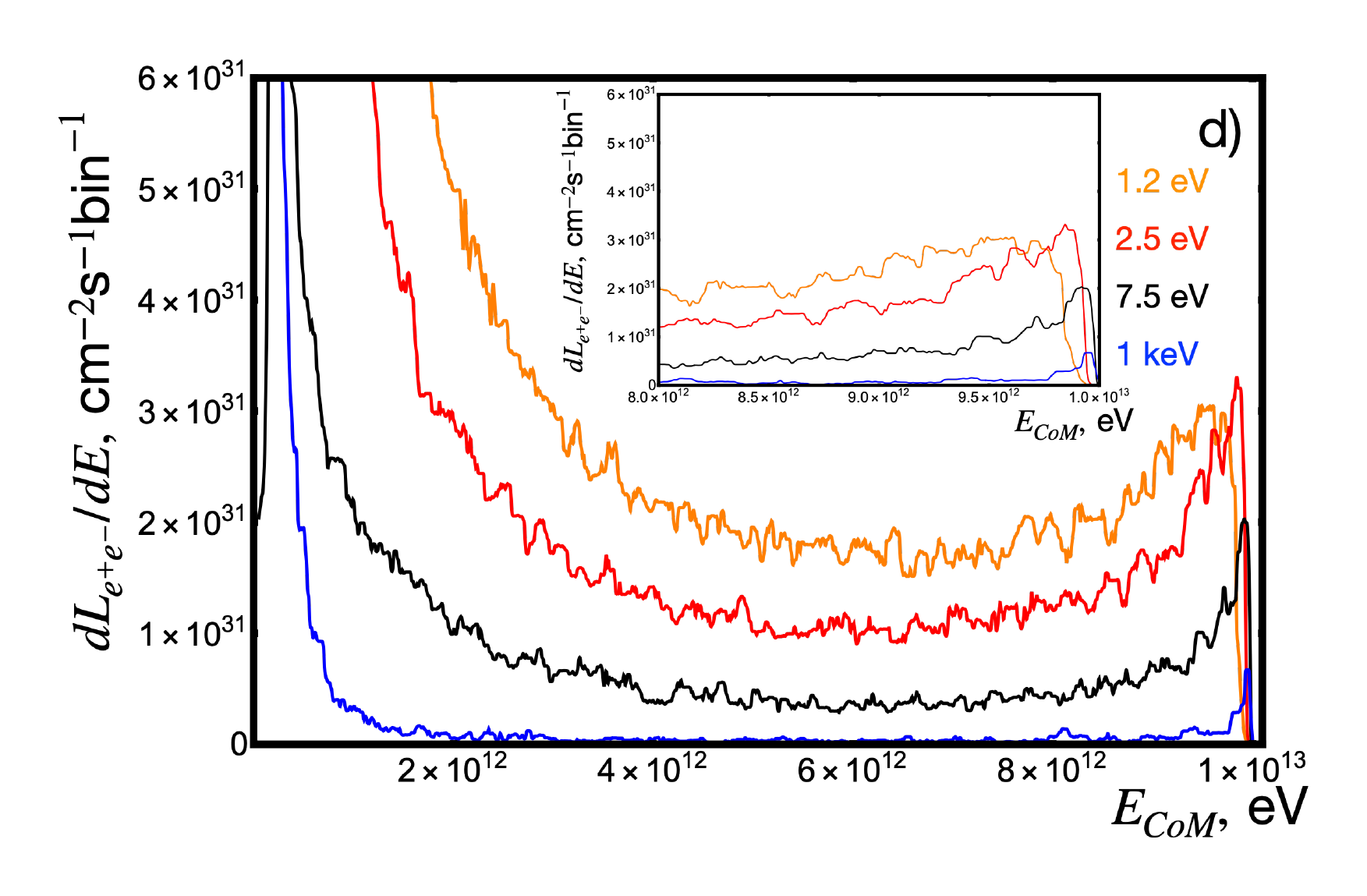}
\caption{The $\gamma\gamma$ (a), $\gamma e$ (b), $e^-e^-$ (c), and $e^+e^-$ (d) luminosity spectra for different laser frequencies and pulse lengths: $\omega_0=$ 1.2 eV, $c\tau=8$ mm (orange), 2.5 eV, $c\tau=6$ mm (red), 7.5 eV, $c\tau=5.3$ mm (black), 1 keV, $c\tau=80$ $\mu$m (blue). The inserts show the luminosity spectra near the 10 TeV peak.}
\label{fig:lumi}
\end{figure*}

\textbf{CAIN simulations.}
Full two-stage simulations with the CAIN Monte Carlo code~\cite{CAIN}, covering
both the conversion and beam–beam collision stages, confirm and refine the
analytical picture.  
Typical $\gamma\gamma$, $\gamma e^-$, and $e^-e^-$ luminosity spectra for different scattering laser wavelengths assuming a bunch population of 
$N_e\!=\!2.1\!\times\!10^9$ and  an IP beta function of
$\beta\!=\!0.6$~mm are shown in Fig.~\ref{fig:lumi}.
Scanning over laser frequency and duration  reveals \textit{two near-optimal
operating points}: 

\begin{itemize}
  \item \textbf{Near-optical} ($\omega_0\!=\!5$~eV, $c\tau\!=\!6$~mm,
        $W_L\!=\!125$~J): broad luminosity spectrum with
        $L_{\gamma\gamma}^{5\%}\!=\!1.3\!\times\!10^{34}$~cm$^{-2}$s$^{-1}$.
  \item \textbf{X-ray} ($\omega_0\!=\!1$~keV, $c\tau\!=\!80\;\mu$m,
        $W_L\!=\!1.7$~J): sharply peaked spectrum
        ($\Delta E_{\rm CoM}\!=\!75$~GeV FWHM at 9.95~TeV) with
        $L_{\gamma\gamma}^{5\%}\!=\!5.3\!\times\!10^{33}$~cm$^{-2}$s$^{-1}$.
\end{itemize}

A recent physics study~\cite{chigusa.arxiv.2025} confirms that broad luminosity
spectra do not undermine electroweak discovery reach at multi-TeV energies,
validating both operating points for BSM searches.

\textbf{Polarization.}  Anti-parallel electron–laser helicities
($2\lambda_e P_c\!=\!-0.9$) enhance $L_{\gamma\gamma}^{5\%}$ by a factor of
6 relative to the parallel case, underscoring the importance of polarized beams.

\textbf{High-luminosity configuration.}  Increasing the bunch charge to
$N_e\!=\!6.3\!\times\!10^9$ and tightening the IP beta function to
$\beta\!=\!0.34$~mm raises $L_{e^-e^-}^{geo}$ by a factor of~11, boosting
the 1~keV $L_{\gamma\gamma}^{1.5\%}$ from $0.035$ to
$0.33\!\times\!10^{35}$~cm$^{-2}$s$^{-1}$ — sufficient for detailed studies of narrow
$J\!=\!0,2$ resonances.

\subsection{Time scale for application}

The $\gamma\gamma$ collider development is tightly coupled to the wakefield
accelerator roadmap~\cite{10TeV}: energy-frontier-quality beams
in the 100~GeV range are foreseen in the early 2030s, TeV-scale staging in
the 2040s, and a $\sim\!10$~TeV facility decision in the following decade.
Near-term priorities include: (i)~experimental validation of high-$x$
photon-beam generation; (ii)~laser technology development — the 5~eV operating
point requires $\sim\!125$~J pulses at $\sim\!50$~kHz, while the X-ray point
needs $\sim\!1.7$~J at 1~keV, motivating XFEL-based schemes~\cite{barklow.jinst.2023b};
(iii)~implementation of the ``flying focus'' technique to extend effective
conversion lengths beyond the Rayleigh limit; and (iv)~detailed physics-reach
comparisons of $\gamma\gamma$, $e^+e^-$, and $e^-e^-$ scenarios at
10~TeV~\cite{chigusa.arxiv.2025}.
Our results establish that a 10~TeV $\gamma\gamma$ collider is a scientifically
compelling and technically achievable goal using a wide range of scattering
laser technologies, opening a rich BSM discovery program at the energy frontier.
\newpage
\section{Advances in SWFA R\&D for Integration in Linear Colliders}
\label{sec:Power}
\href{https://agenda.infn.it/event/47329/contributions/287322/}{Slides: \url{https://agenda.infn.it/event/47329/contributions/287322/}}
\subsection*{\textit{Gongxiaohui~Chen and John~Power}}

\noindent \textit{Argonne National Laboratory, Argonne, Illinois 60439, USA}\\

\subsection{Context}
Structure wakefield acceleration (SWFA) is an advanced accelerator concept based on the excitation of electromagnetic wakefields in engineered metallic or dielectric structures by a relativistic drive beam~\cite{gai2012short}. In contrast to plasma-based approaches, SWFA operates in precisely defined RF structures with well-controlled boundary conditions, enabling reproducible fields, compatibility with both electrons and positrons, and a modular architecture naturally suited to staging. Two main implementations are under development: collinear wakefield acceleration (CWA), in which drive and main beams share the same structure, and two-beam acceleration (TBA), in which a drive beam generates high-power RF pulses in a power extraction and transfer structure (PETS) that are coupled to a separate accelerating structure for the main beam. For linear collider applications, TBA is presently the more mature approach due to its decoupling of drive and main beam dynamics and its similarity to conventional RF linac architectures.

The scientific motivation for SWFA in the collider context is the requirement of real-estate accelerating gradients well beyond those achieved in conventional long-pulse RF systems. A $10~\mathrm{TeV}$ center-of-mass collider with $5~\mathrm{TeV}$ per beam and a total length of order $10$--$20~\mathrm{km}$ requires effective gradients approaching or exceeding $500~\mathrm{MeV/m}$. Earlier SWFA studies targeting multi-TeV energies operated near $200~\mathrm{MeV/m}$ real-estate gradient. The increase to the $10~\mathrm{TeV}$ scale therefore demands operation in a regime that significantly exceeds conventional normal-conducting limits. The key scientific enabler is operation in the short-pulse RF regime, with pulse durations on the order of $5$--$10~\mathrm{ns}$. Empirical breakdown scaling derived from long-pulse operation indicates a strong dependence of breakdown rate on both field and pulse length. Experimental results have demonstrated that nanosecond-scale RF pulses substantially suppress breakdown probability, allowing access to surface fields approaching the sub-GV/m regime. This short-pulse regime is intrinsic to TBA, where the RF pulse duration is determined by the drive bunch train structure rather than by external klystrons.

Over the past decade, the SWFA program has established several experimental milestones that directly inform collider-scale extrapolation. High-power RF generation has progressed from early dielectric structures to X-band metallic and metamaterial PETS devices, culminating in the generation of $565~\mathrm{MW}$ of X-band RF power using a metamaterial PETS~\cite{picard2022generation}, with ongoing progress toward the gigawatt level. In parallel, accelerating structures operating in the short-pulse regime have achieved gradients near $300~\mathrm{MV/m}$~\cite{shao2018recent}, corresponding to surface fields around $500~\mathrm{MV/m}$. In short-pulse X-band RF photoinjectors, surface fields approaching $0.6~\mathrm{GV/m}$ have been reached with low dark current and acceptable breakdown rates~\cite{tan2022demonstration}. Multi-stage TBA acceleration has been experimentally demonstrated~\cite{jing2018electron}, establishing the feasibility of modular energy gain. In collinear wakefield acceleration, drive-beam shaping techniques have achieved transformer ratios exceeding five~\cite{gao2018observation}, improving the efficiency of energy transfer from drive to main beam. These results provide a coherent experimental foundation for scaling studies. In addition, SWFA concepts have been explored for compact light-source applications~\cite{piot2023development,zholents2018conceptual}, illustrating the breadth of the underlying accelerator science.

The present $10~\mathrm{TeV}$ SWFA linac concept is based on a modular TBA architecture. Each module consists of a PETS that converts drive-beam kinetic energy into a short RF pulse, followed by an accelerating structure that boosts the main beam. The central parameters are the RF frequency, RF pulse length, drive and main bunch formats, and the achievable loaded gradient. X-band technology at $11.7~\mathrm{GHz}$ serves as a near-term baseline. Existing X-band PETS routinely generate several hundred megawatts of RF power~\cite{picard2022generation}, with pulse lengths of a few nanoseconds. X-band accelerating structures operating with these pulses achieve unloaded gradients near $200$--$300~\mathrm{MV/m}$~\cite{shao2018recent}. However, reaching $500~\mathrm{MeV/m}$ real-estate gradient at X-band is a major challenge and would require multi-gigawatt RF power per structure and pushes aperture and breakdown limits.

\subsection{Objectives}
Frequency scaling provides a promising path forward. Increasing frequency by a factor of three to Ka-band ($\approx 35~\mathrm{GHz}$) increases shunt impedance and reduces structure dimensions, enabling higher gradients for a given RF power. First-pass Ka-band studies indicate that peak gradients approaching $1~\mathrm{GV/m}$ are feasible in short structures with nanosecond-scale pulses. For a $30~\mathrm{cm}$ accelerating structure, loaded energy gains on the order of $150$--$190~\mathrm{MeV}$ correspond to real-estate gradients of approximately $500$--$700~\mathrm{MeV/m}$, depending on fill factor and beam loading. Sector-level extrapolations suggest that a sequence of such modules could achieve $5~\mathrm{TeV}$ per beam within roughly $10~\mathrm{km}$ of active length. These estimates assume RF pulse durations of $\sim 6~\mathrm{ns}$, main bunch spacing of order $0.3$--$0.4~\mathrm{ns}$, and repetition rates around $100~\mathrm{Hz}$.

Luminosity considerations impose additional constraints. In short-pulse SWFA, the number of main bunches per RF pulse is limited by the pulse duration, reducing the number of bunches per train relative to long-pulse systems. To achieve geometric luminosities above $10^{35}~\mathrm{cm^{-2}\,s^{-1}}$, the bunch charge, repetition rate, and transverse beam sizes must be carefully optimized. Preliminary parameter studies indicate that with nanometer-scale beam spot sizes, main bunch charges of order $1$--$2~\mathrm{nC}$, and approximately $10$--$20$ bunches per RF pulse, luminosities in the $10^{35}~\mathrm{cm^{-2}\,s^{-1}}$ range are attainable. The resulting drive-to-main beam energy transfer efficiency in initial Ka-band sector estimates is on the order of $20$--$25\%$, comparable to other advanced accelerator proposals. These studies remain first-pass and require refinement with integrated beam dynamics simulations.

\subsection{Next steps}
Several scientific challenges must be addressed before a $10~\mathrm{TeV}$ SWFA linac can be considered viable. On the high-gradient side, operation near $1~\mathrm{GV/m}$ peak field at Ka-band must be validated experimentally, including systematic measurements of breakdown rate, dark current, and conditioning behavior in the short-pulse regime. Thermal management and pulsed heating become increasingly important as frequency rises and apertures shrink. On the beam dynamics side, control of beam breakup (BBU) and multi-bunch instabilities in both PETS and accelerating structures is essential, particularly given sub-nanosecond bunch spacing. Theoretical and simulation studies of drive-beam stability in dielectric wakefield accelerators~\cite{baturin2018stability,tan2022simulation}, along with experimental demonstrations of transverse stability and wakefield control in structured geometries~\cite{lynn2023demonstration,lynn2024observation}, provide important groundwork but must be extended to collider-relevant parameters. Emittance preservation at nanometer scales requires precise alignment, wakefield control, and potentially active feedback systems. On the drive-beam side, production of high-charge, high-repetition-rate bunch trains compatible with Ka-band spacing represents a major systems challenge. Efficiency optimization across the full chain---drive beam to RF to main beam---remains a critical area of study.

Despite these challenges, SWFA offers several intrinsic scientific strengths. It supports both electron and positron acceleration within conventional vacuum structures. Its modular TBA architecture allows independent optimization of drive and main beamlines and straightforward staging. Operation in the short-pulse regime provides access to gradients beyond those practical in long-pulse RF systems. In addition, it is worth noting that SWFA shares physics synergies with plasma wakefield acceleration, including drive-beam shaping and phase-space manipulation, opening avenues for hybrid concepts and cross-disciplinary collaboration.

In summary, SWFA has progressed from conceptual proposals to experimentally validated high-power, high-field operation in the short-pulse regime. Scaling studies to Ka-band frequencies indicate a plausible path toward real-estate gradients of $500~\mathrm{MeV/m}$ and beyond, meeting the central requirement for a $10~\mathrm{TeV}$ linear collider. While substantial R\&D is required in high-gradient physics, beam stability, drive-beam generation, and system efficiency, the underlying accelerator science is sound and increasingly supported by experimental results~\cite{gai2012short,picard2022generation,shao2018recent,tan2022demonstration,jing2018electron,gao2018observation,piot2023development,zholents2018conceptual,baturin2018stability,tan2022simulation,lynn2023demonstration,lynn2024observation}. %
Continued focused investigation of short-pulse TBA modules, frequency scaling, and integrated beam dynamics will determine whether SWFA can fulfill its potential as a collider technology for the multi-TeV frontier.

\newpage
\section{LWFA Linac for 10 TeV Collider}
\label{sec:Benedetti}
\href{https://agenda.infn.it/event/47329/contributions/280917/}{Slides: \url{https://agenda.infn.it/event/47329/contributions/280917/}}

\subsection*{\textit{Francesco Massimo\textsuperscript{1} and Carlo Benedetti\textsuperscript{2}}}

\noindent \textit{\textsuperscript{1}LPGP, CNRS, Univ. Paris-Saclay, Paris, France}\\
\noindent \textit{\textsuperscript{2}Lawrence Berkeley National Laboratory, Berkeley, CA, USA}

\subsection{Context and objectives of LWFA LINAC WG}
The scope of the Laser Wakefield Acceleration (LWFA) Linear Accelerator (LINAC) Working Group (WG) is to accelerate particle (electron and positron) beams  from the source to the beam delivery system at an energy of 5\,TeV, in a compact and cost-effective manner using staged LWFAs.

The WG was tasked with identifying key challenges (e.g., quality-preserving staging, acceleration of positrons, stability, etc.) and perform reviews of different technologies and design options (i.e., LWFA regime, inter-stage beam transport, etc.) with the goal of identifying a LINAC design that maximizes the luminosity-per-power ($> 10^{32}$/cm$^2$/s/MW), and minimizes the LINAC footprint (geometric gradient $>500$\,MV/m). 




\subsection{Progress over the last two years}

Over the last two years, the WG focused on the following activities:
\begin{itemize}[itemsep=1em]
    \item \emph{Review of LWFA LINAC challenges.} A comprehensive list of LWFA LINAC challenges was discussed during the initial phase of the 10\,TeV collider design study initiative. Owing to the limited resources available, it was agreed during the LWFA LINAC WG kick-off meeting (February 2025) and subsequent discussions at the ALEGRO2025 Workshop to focus the design effort on a subset of these challenges related to the electron side of the collider, relevant to electron–electron or gamma–gamma collider concepts, since no suitable LWFA-based scheme capable of providing collider-relevant positron beams had yet been identified~\cite{PhysRevAccelBeams.27.034801}.
    
    \item \emph{Design of a LWFA stage suitable for collider applications.} A design was developed for an LWFA stage capable of accelerating 330\,pC electron bunches by 6.2\,GeV over a distance of 25\,cm using a 10\,J, 1\,$\mu$m laser driver with a FWHM pulse duration of 42\,fs and a spot size at focus of 41 $\mu$m. The laser is focused at the entrance of a plasma target with a density of $8.9\times 10^{16}$\,cm$^{-3}$. The longitudinal plasma profile is nonlinearly tapered to reduce dephasing and increase the laser-to-beam efficiency. The transverse plasma profile is parabolic with a matched radius of 41\,$\mu$m. The laser-to-wake and wake-to-bunch energy transfer efficiencies are 51\%, and 40\%, respectively, resulting in a  laser-to-beam efficiency of 20.4\%. Assuming a wall-to-laser energy transfer efficiency of 40\%, one obtains a wall-to-beam efficiency of 8.2\%, comparable to that of CLIC.
    By properly tailoring the bunch current profile (the total bunch length is 18\,$\mu$m, and the peak current is 5.5\,kA) a relative rms energy spread of 0.4\%  can be achieved.
    This stage provides stable acceleration owing to the background ion motion triggered by the intense electron bunch~\cite{PhysRevLett.121.264802}. Emittance growth induced by ion motion~\cite{PhysRevAccelBeams.20.111301} can be mitigated by employing an electron beam with a suitable longitudinal tapering of its size, which can be obtained through an adiabatic matching procedure~\cite{10.1063/5.0043847}. Owing to the nonlinearity of the transverse wakefield, round electron beams are considered, since flat beams are prone to resonant emittance mixing~\cite{diederichs2024resonant}. Finally, owing to the strong focusing provided by the background ions, emittance growth due to Coulomb collisions is negligible (i.e., $\ll 1$\,nm) for beam energies up to several TeV~\cite{Schroeder:2022}. 

\item \emph{Design considerations for inter-stage bunch transport.} In a staged LWFA setup, laser drivers can be in-coupled at a short distance ($\sim 0.15$\,m for the laser parameters considered above) by means of plasma mirrors~\cite{Steinke2016}. However, inter-stage distance is generally determined by the distance required to transport the electron beam. This value depends on the gradient of the focusing system (usually fixed to the max value allowed by the chosen transport technology) and scales with bunch energy as $\sim E_{bunch}^{1/2}$, hence the accelerating stages are further apart the higher the energy. For instance, assuming the transport is performed with an active plasma lens~\cite{PhysRevLett.115.184802} with a focusing gradient of 2000\,T/m, the bunch transport distance becomes several meters for TeV energies. More specifically, for the bunch parameters discussed above, this results in an average LINAC length of 0.45\,km for 1\,TeV beams (geometric accelerating gradient of 2.1\,GV/m, average in-coupling distance of 2.7\,m), and a length of 5.1\,km for 5\,TeV beams (geometric accelerating gradient of 1.0\,GV/m, average in-coupling distance of 6.0\,m). For bunch energy spreads of $(0.1-1)\%$, and a multi-meter average inter-stage distance, chromatic effects during drifts become the main source of emittance degradation~\cite{10.1063/1.4740456, PhysRevSTAB.16.011302, PhysRevAccelBeams.24.104602}. Development of transport optics with a large chromatic acceptance in order to preserve beam quality is required~\cite{Lindstrom_2021staging}.

\item \emph{Optimization of the LWFA stage design using multifidelity optimization techniques.} Self-consistent Particle in Cell (PIC) simulations of this kind of acceleration stage are resource-consuming, even using an optimized code for LWFA like INF$\&$RNO~\cite{Benedetti2010,Benedetti2012,Benedetti2013,Benedetti2018}. The INF$\&$RNO simulations that found the described design used a combination of cylindrical symmetry, quasi-static approximation~\cite{Mora1996} and time-averaged ponderomotive approximation~\cite{Terzani2021,Massimo2025} to run quicker simulations. A typical run required $\approx 30-40$  hours on 1 cpu.
To further optimize a regime identified through INF$\&$RNO, a multi-fidelity optimization strategy is being tested, where a quick tracking code like Wake-T~\cite{Ferran_Pousa2019_WAKET,} is used to perform Bayesian Optimization~\cite{Jalas2021} using the library~\cite{Discoveri}. When a suitable working point is found through these lower-fidelity simulations, it is verified by the higher-fidelity model of INF$\&$RNO. This strategy presents some challenges, such as identifying an objective function suited for the design~\cite{Jalas2021,Irshad2023}, and the choice of parameters to optimize, e.g. the plasma density, the delay of the particle beam after the laser pulse.

\item \emph{Interaction with WG Laser Drivers and review of laser quality effects on acceleration.}
The identification of a suitable working point benefited from sustained interactions with the WG Laser Drivers. These exchanges were initiated to ensure that the WG LWFA LINAC accounts for anticipated developments in high-intensity laser technology, particularly for drivers of high-energy LWFA stages. In parallel, this collaboration aimed to inform the WG Laser Drivers about the laser parameters and characteristics most favorable for defining an optimal working point.

A key challenge identified in this context is the requirement for a high degree of symmetry in the transverse laser intensity distribution. To better understand the impact of deviations from ideal symmetry, a literature review was conducted focusing on studies of transverse asymmetries in the LWFA process.

Over the last decade, the progress in phase retrieval techniques to reconstruct the transverse distribution of the laser electric field~\cite{Beaurepaire2015,Ferri2016,Zemzemi2020,Moulanier2023JOSAB,Massimo2025} for high-fidelity PIC simulations allowed to study the effects of a realistic laser model in injector stages. Several studies showed that, depending on the working point, the accelerated charge can be smaller~\cite{Ferri2016,Dickson2022,Moulanier2023}, or greater~\cite{Massimo2026} than the one in simulations using an idealized, rotationally symmetric laser, with the energy spectra significantly changing from one model to the other~\cite{Ferri2016,Zemzemi2023,Moulanier2023,Massimo2026}. Asymmetric transverse laser fluence distributions can cause deviations of the electron beam trajectory from the laser propagation direction~\cite{Dickson2022}, or oscillations of the electron beam centroid around the laser propagation direction~\cite{Moulanier2025}. 

Despite these advances, there remains a lack of comprehensive studies addressing the impact of transverse laser asymmetries in LWFA regimes relevant to collider applications.

\end{itemize}

\newpage
\section{PWFA Linac for 10 TeV Collider}
\label{sec:Storey}
\href{https://agenda.infn.it/event/47329/contributions/280916/}{Slides: \url{https://agenda.infn.it/event/47329/contributions/280916/}}

\subsection*{\textit{D. Storey\textsuperscript{1}, A. Knetsch\textsuperscript{1}, L. Verra\textsuperscript{2}}}

\noindent \textit{\textsuperscript{1}SLAC National Accelerator Laboratory, Menlo Park, CA, USA}\\
\noindent \textit{\textsuperscript{2}INFN, Laboratori Nazionali di Frascati, Frascati, Italy}

\subsection{Overview}

Beam-driven plasma wakefield acceleration (PWFA) can sustain accelerating gradients many orders of magnitude higher than conventional RF structures, offering a path to multi‑TeV energy colliders. Candidate configurations include e$^+$ e$^-$, e$^-$ e$^-$, and $\gamma$-$\gamma$ colliders, each requiring on the order of 5 TeV per arm of the collider.
While a single plasma stage can provide GV/m‑scale fields, an operational collider may, depending on driver species and energy, require multiple stages of acceleration and interstage elements to reach full colliding beam energy. Accordingly, early design work focuses on the choice of driver species and implications for efficiency, staging, and technical readiness, the effect of synchrotron radiation and emittance growth in the interstages between plasmas, and strategies and challenges for accelerating flat and/or polarized beams in PWFA.

\subsection{Driver species considerations}

A range of PWFA-based collider concepts have been proposed, from early-stage conceptual studies \cite{rosenzweig1998, adli_2013_PWFAlinac} to more fully optimized designs such as HALHF \cite{Foster_2023, HALHF_2025}. Most adopt electron-driven PWFA with multiple stages, each on the order of tens of GeV, to reach the target beam energy.

Proton driven concepts are also under active study\,\cite{pwfa-farmer-higgs}, leveraging multi-TeV proton beams to drive a single, long plasma stage to reach the target energy eliminating the need for staging. Realizing such an approach would require several technical advances 
including: the acceleration of short proton bunches with lengths comparable to the plasma wavelength; tailored proton-beam current and emittance profiles for stable propagation, efficient wake excitation, and high transformer ratio long and uniform plasma channels on the order of >100m; and high repetition rate and average power proton accelerators. Progress towards proton-driven PWFA is discussed further elsewhere in this workshop summary.

Staged, electron-driven PWFA benefits from mature, efficient RF linac technology for producing high-quality drive beams, with potential paths to high repetition rate and energy efficiency. However, many stages are needed to reach full beam energy, bringing challenges in interstage optics and beam quality preservation,
synchronization and transverse alignment, spent-driver removal, and more. 

\subsection{Staging considerations}
Each plasma module must be linked by an interstage that extracts the spent driver bunch, injects a fresh driver, and preserves the witness beam’s emittance, energy spread, and polarization while matching it into the next plasma \cite{Lindstrom_2021staging}. In practice, this involves a chicane and septum system
together with high‑gradient optics to capture and rematch the witness bunch. Synchrotron radiation generated in these bends can significantly reduce the effective accelerating gradient and dilute emittance.
While coherent synchrotron radiation becomes less prominent with higher energy, incoherent synchrotron scales strongly with beam energy and inversely with bend radius. Balancing chicane length, field strength, and the use of compact matching optics is therefore critical to preserving beam quality and maintaining a high effective gradient\,\cite{verra2026effect}. 
First experimental results towards a laser-gated two-stage PWFA were discussed\,\cite{knetsch2023high} which could test the prediction of a self-correcting energy spread \cite{lindstrom2021self} or reduce accumulated magnetic field.

\subsection{PWFA with flat and polarized beams}


Flat colliding bunches can reduce beamstrahlung at the interaction point and lead to a more favorable luminosity spectrum in the case of a e$^+$ e$^-$ or e$^-$ e$^-$ collider \cite{Barklow:2023iav}. 
Hence the acceleration of flat beams in plasma wakefields is being studied intensely by the community with theory and experiments in preparation phase. It was found that in a nonlinear plasma wakefield, betatron oscillation of the trailing beam in x and y planes resonantly mix, leading to transfer of emittance from the larger- to the smaller-emittance plane, thereby degrading overall beam quality\,\cite{diederichs2024resonant}. Possible mitigations under study include avoiding deleterious resonant conditions by detuning the x/y betatron tunes, by using flat drivers to generate an intentionally asymmetric wake, and employing controlled ion motion to break symmetry. These solutions come with challenges of their own. For instance, a plasma wake driven by an asymmetric electron beam in absence of ion motion or beam loading has different gradients in the transverse focusing fields between the transverse planes, and the accelerating field strength shows a transverse dependence \cite{manwani2025analysis}. 

When accelerating polarized beam by PWFA, several effects must be taken under consideration to ensure the preservation of sufficient polarization \cite{Reichwein2025}. For ultrarelativistic beams, spin precession is expected to be reduced in the plasma’s accelerating and focusing fields, but cumulative precession in interstage magnets should still be evaluated. In staged PWFA machines, the energy scaling and cumulative impact of Stern–Gerlach and Sokolov–Ternov effects across many plasma stages and chicanes should be quantified to determine their overall significance and mitigation if required.

\subsection{Next steps}
Ongoing studies and interaction across the community working groups will continue to  improve the understanding of requirements and advantages of PWFA as a collider technology. 
Further studies on luminosity spectra would inform if collision of flat beams is in fact the more desirable configuration. 
Flat beams could be delivered either by accelerating flat beams through a multi-stage PWFA LINAC or by developing a final focusing design. 
Since drive-beams need to be coupled into the PWFA stages, transverse asymmetries in the magnetic lattice seem unavoidable. 
Such an asymmetry, combined with the effect of synchrotron-radiation emission, will influence energy spread, emittance asymmetry, and polarization. While the HALFH interstage design can be adapted to keep the effect of SR on energy spread and emittance withing target parameters even at acceleration up to 5 TeV \cite{lindstrom2026achromatic}, at this point it is unclear whether an interstage design would favor a polarized or flat beam.

\newpage
\section{Beam Sources for 10 TeV Wakefield Collider}
\label{sec:Chubenko}
\href{https://agenda.infn.it/event/47329/contributions/285160/}{Slides: \url{https://agenda.infn.it/event/47329/contributions/285160/}}

\subsection*{\textit{Oksana Chubenko\textsuperscript{1}, Siddharth Karkare\textsuperscript{2}, Joe Grames\textsuperscript{3}, Matthias Fuchs\textsuperscript{4}}}

\noindent \textit{\textsuperscript{1}Department of Physics, Northern Illinois University, DeKalb, IL 60115, USA}\\
\noindent \textit{\textsuperscript{2}Department of Physics, Arizona State University, Tempe, AZ 85287, USA}\\
\noindent \textit{\textsuperscript{3}Center for Injector and Sources, Thomas Jefferson National Accelerator facility, Newport News, VA 23606, USA}\\
\noindent \textit{\textsuperscript{4}Institute for Beam Physics and Technology, Karlsruhe Institute of Technology, Germany}

\subsection{Context}
Wakefield particle acceleration has been suggested as a promising approach toward a 10 TeV collider~\cite{Gessner_2025}. Several concepts, including Laser Wakefield Acceleration (LWFA), Plasma Wakefield Acceleration (PWFA), and Structure Wakefield Acceleration (SWFA), are being actively investigated in parallel as potential pathways to a 10 TeV collider. Each of these approaches requires particle sources (for the witness beam or for both drive and witness beams) with specific parameter sets to enable efficient wakefield acceleration. A comprehensive and systematic evaluation of existing and emerging particle generation technologies is therefore essential for the successful implementation of these schemes.

\subsection{Objectives}
The Particle Beam Sources Working Group of the 10 TeV Wakefield Accelerator Design Study was established in late 2024 - early 2025. It is dedicated to the assessment of capabilities of a wide range of particle generation techniques in the context of specific collider design requirements, with a particular focus on achievable brightness and projected cost. The group’s overall goal is to help with determining the optimal configuration of the 10 TeV wakefield collider.

\subsection{Progress over the last two years}

\begin{figure}[!h]
  \centering
  \begin{subfigure}[b]{0.9\textwidth}
    \centering
    \includegraphics[width=\textwidth]{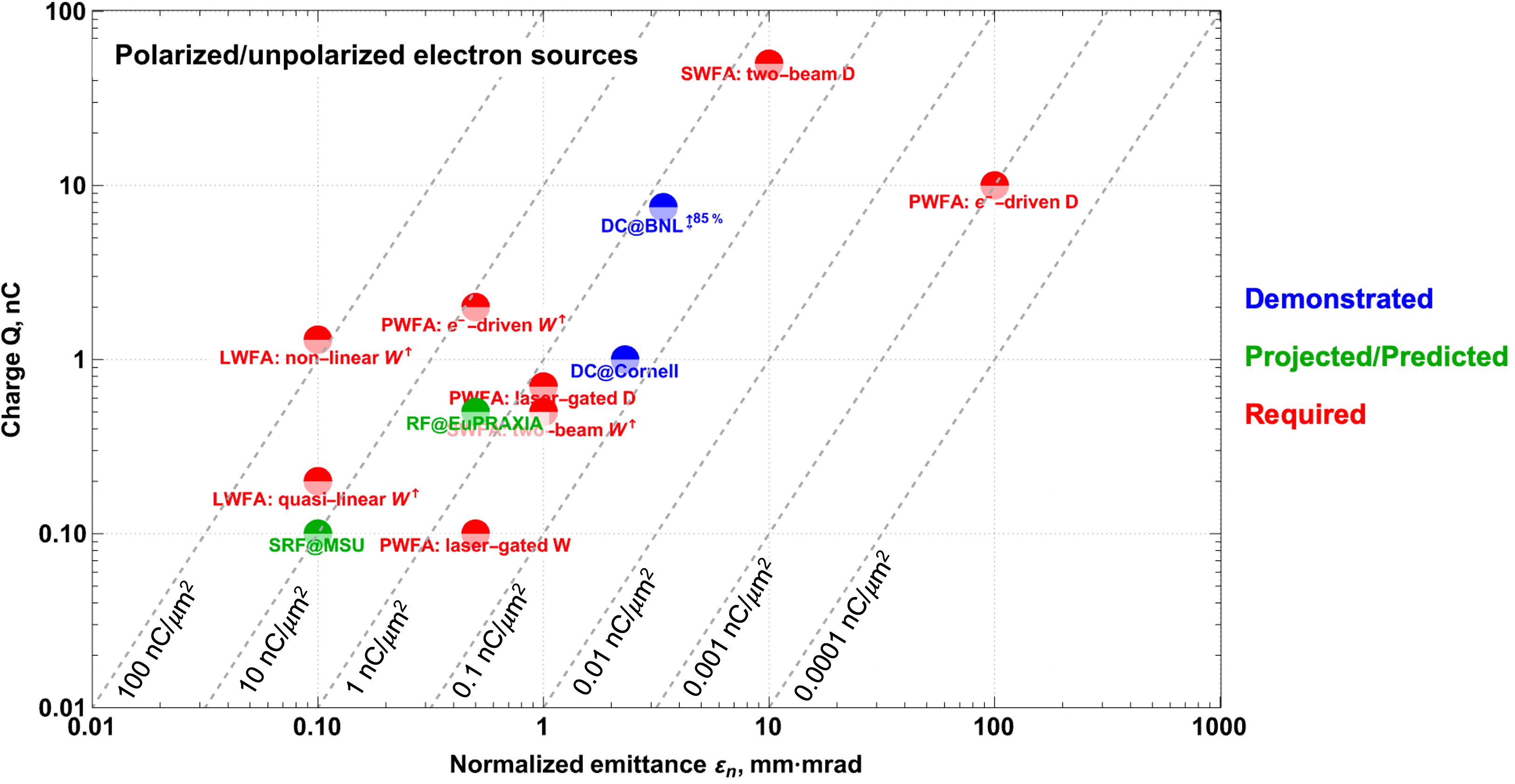}
    \label{electrons}
  \end{subfigure}
  \hfill
  \begin{subfigure}[b]{0.9\textwidth}
    \centering
    \includegraphics[width=\textwidth]{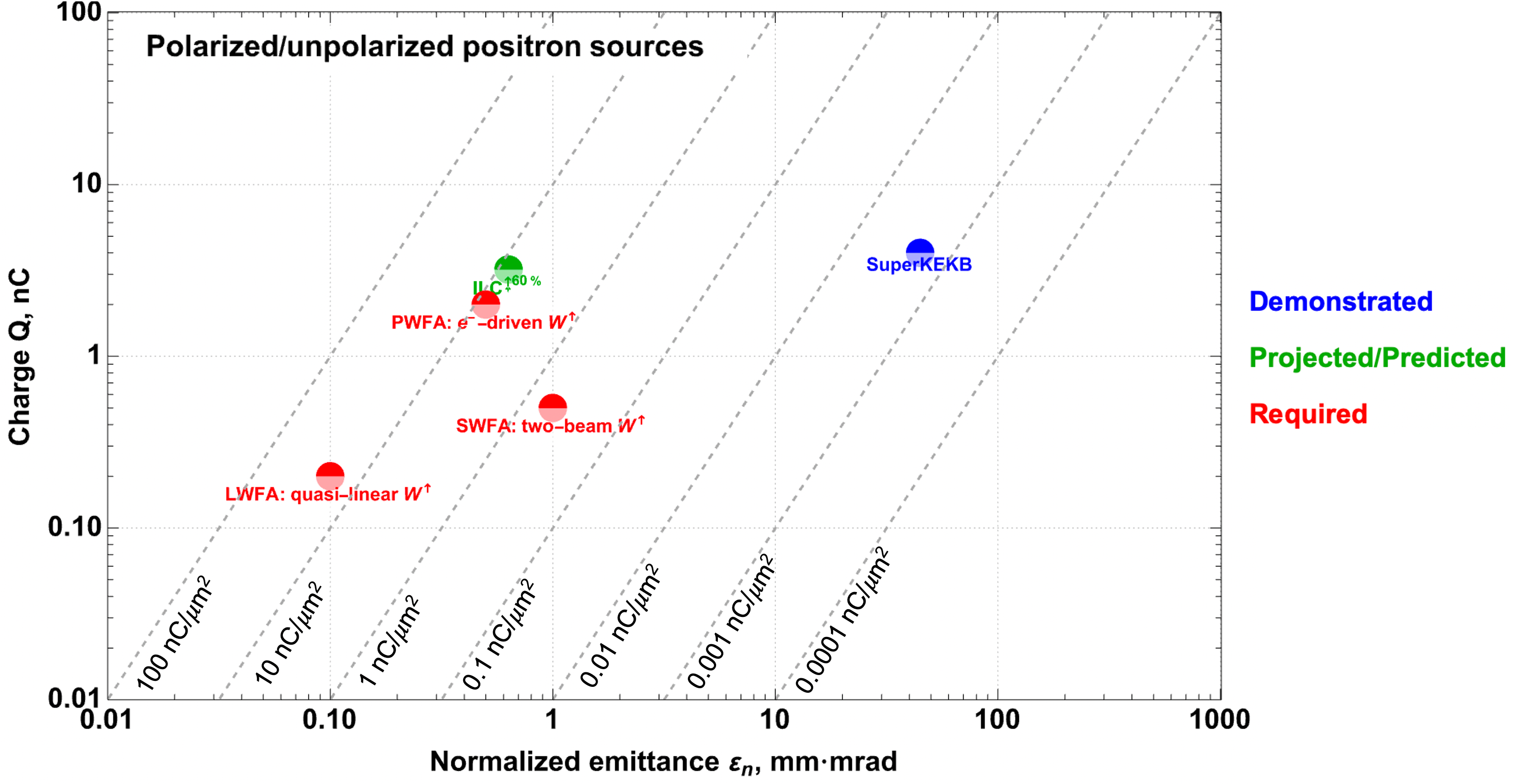}
    \label{positrons}
  \end{subfigure}
  \caption{Comparison of the electron (top) and positron (bottom) sources required to enable different wakefield acceleration schemes with some operational and projected state-of-the-art particle generation technologies.}
  \label{comparison}
\end{figure}

We reviewed a range of operational particle generation technologies, as well as emerging facilities under development~\cite{Nie2021, Wang_2022, Litvinenko_2026, Bartnik_2015, Silvi_2024, Del_Dotto_2025, Yin_2025, Miller_2023, Alharthi_2025, Zhu_2016, Satoh_2016, Adolphsen_2013}. A comparison of their capabilities in terms of normalized emittance $\epsilon_n$ and bunch charge $Q$ is presented in Fig.~\ref{comparison} for electrons (top) and positrons (bottom), and is discussed in greater detail in Ref.~\cite{Chubenko_2026}. Operational facilities are shown in blue, while projected or simulated capabilities are shown in green. The required source parameters for the proposed wakefield acceleration schemes are indicated in red and represent preliminary target parameters provided by the corresponding wakefield working group leaders. Spin-polarization capability is indicated by the upward arrow, $\uparrow$. The sloped lines represent contours of constant $Q/\epsilon_n^2$, as labeled, providing an indication of the achievable 4D beam brightness.

\subsection{Summary and next steps}
Preliminary analysis indicates that some schemes, in their present form, require combinations of parameters whose feasibility remains uncertain. The group will continue this effort over the next several years in collaboration with other working groups, with increased focus on additional required beam parameters and projected cost.

\section*{Acknowledgements}
This material is based upon work supported by the U.S. Department of Energy, Office of Science, Office of Nuclear Physics under contract DE-AC05-06OR23177. The authors thank the Center for Bright Beams, NSF award PHY-1549132.

\newpage
\section{Beam-Beam Interactions at 10\,TeV Center-of-Mass Energy}
\label{sec:Gessner}
\href{https://agenda.infn.it/event/47329/contributions/287333/}{Slides: \url{https://agenda.infn.it/event/47329/contributions/287333/}}
\subsection*{\textit{Spencer Gessner\textsuperscript{1} and Arianna Formenti\textsuperscript{2}}}

\noindent \textit{\textsuperscript{1}SLAC National Accelerator Laboratory}\\
\noindent \textit{\textsuperscript{2}Lawrence Berkeley National Laboratory}
\noindent

\subsection{Context}
The Beam-Beam Working Group, co-convened by Spencer Gessner and Thomas Grismayer (IST Lisbon), has met 14 times~\cite{Indico10TeV} in the context of the 10 TeV Wakefield Collider Design Study~\cite{10TeVStudy}. The Beam-Beam Working Group includes accelerator physicists, detector physicists, and particle theorists, representing a ``full-stack" of HEP expertise. 

Many of the Beam-Beam Working Group members contribute to the development of PIC codes for beam-beam interactions. These codes include VLPL, led by Alexander Pukhov at Heinrich Heine University, OSIRIS led by Thomas Grismayer at IST Lisbon, and WarpX led by Arianna Formenti at Lawrence Berkeley National Laboratory.

At the ALEGRO 2026 meeting, we presented an overview of these meetings which covered topics related to particle-in-cell (PIC) modeling of beam-beam interactions, analytic theory, final focus systems, detector modeling, and physics studies at 10 TeV center-of-mass energy. We proceeded to highlight three topics led by the authors of this summary: 1) development of an analytic theory for beamstrahlung at 10 TeV, 2) addition of physics processes and validation for beam-beam interactions in WarpX at 10 TeV, and 3) applications of WarpX beam-beam modeling for the FCC-ee collider.

\subsection{An analytic theory for beamstrahlung at large energy}
Beamstrahlung is the process by which colliding particles emit radiation while traversing the intense electromagnetic field of the opposing bunch~\cite{YokoyaChen}. Beamstrahlung has two principal consequences for a future lepton collider: it broadens the luminosity spectrum away from the nominal center-of-mass energy, and it establishes the initial conditions for electron-positron pair production. The key dimensionless parameter characterizing the interaction strength is $\Upsilon$, defined as the ratio of the particle energy to the Schwinger critical field energy. For $\Upsilon \leq 1$, the interaction is in the classical synchrotron-radiation regime with incoherent pair creation; for $\Upsilon \geq 1$, one enters the strong-field quantum regime with coherent pair creation. At 10 TeV center-of-mass, $\Upsilon \gg 1$ for all realistic beam parameter sets.

In our efforts to model beamstrahlung for different collision energies with legacy PIC code GUINEA-PIG~\cite{Schulte}, we made a striking observation. When the center-of-mass energy is varied while holding $\epsilon$, $N$, $\sigma_z$, and $\beta^*$ fixed, the resulting luminosity spectra are self-similar. We derived scaling laws to provide an intuitive explanation of this effect. Starting from the Yokoya-Chen model~\cite{YokoyaChen}, we derived an approximation for the luminosity spectra in the limit $\Upsilon \rightarrow \infty$ and showed that the shape of the spectra depends only on the number of radiated beamstrahlung photons $N_\gamma$~\cite{He}.

\subsection{Development and validation of the WarpX beam-beam code}
The legacy codes GUINEA-PIG~\cite{Schulte} and CAIN~\cite{CAIN}, which rely on 2D Poisson solvers, were used extensively during the recent Snowmass process to simulate 10 TeV beam-beam interactions~\cite{Barklow:2023iav}. Several limitations became apparent: both codes are single-threaded, leading to prohibitively long run times; the large number of pairs produced in high-$\Upsilon$ collisions caused memory overflow; the codes are no longer regularly maintained; and documentation is sparse. These difficulties prompted our team to develop the high performance computing code WarpX~\cite{WarpX} as a next-generation platform for beam-beam studies.

\begin{figure}
    \centering
    \includegraphics[width=0.5\linewidth]{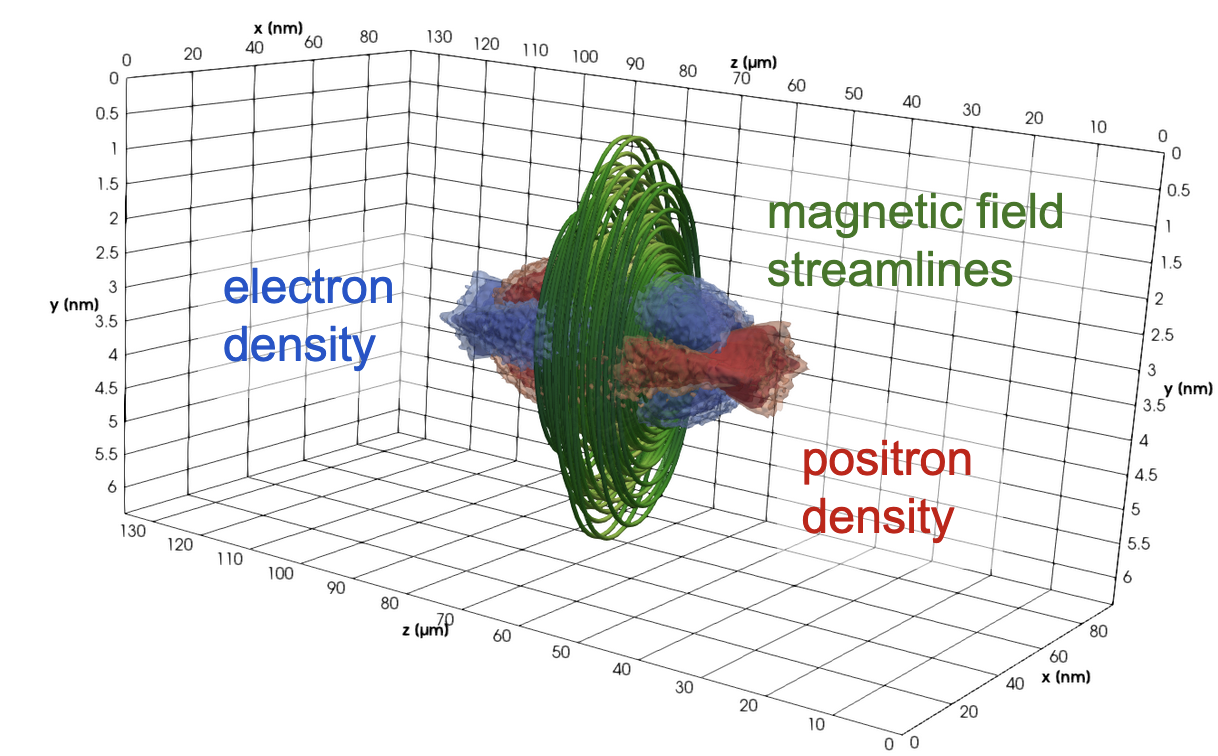}
    \caption{Particle beam distributions and magnetic field lines during a high energy collision modeled in WarpX.}
    \label{fig:WarpX}
\end{figure}

WarpX was missing many of the physics models needed to model high energy beam-beam collisions, specifically models related to pair creation. The LBNL-SLAC team implemented elastic Bhabha scattering and radiative Bhabha scattering as well as the Breit-Wheeler, Bethe-Heitler, and Landau-Lifshitz pair production processes. Extensive work went into validating WarpX in well-benchmarked regimes. Figure~\ref{fig:WarpX} shows an example of particle beam distributions and magnetic field lines during a high energy collision modeled in WarpX.

\subsection{Application of WarpX for the FCC-ee}

In the effort to develop and apply WarpX for beam-beam collisions at 10 TeV center-of-mass energy, it quickly became apparent that we needed to validate the code in well-studied regimes. We used the ILC~\cite{LCVision} as our initial target for validation, as it had been well-studied with the GUINEA-PIG and CAIN codes. At the same time, the FCC-ee~\cite{FCC} was gaining momentum as the next global collider project. 

Recently, the LBNL-SLAC team has devoted significant effort to apply WarpX to the FCC-ee design parameters. Specific challenges include: crab-waist interactions, detector solenoid effects, particle hand-off for turn-by-turn modeling, and particle hand-off of incoherent pairs for detector background modeling. The goal of this work is to provide a new, high-performance code for beam-beam interactions to the HEP community to replace the legacy GUINEA-PIG and CAIN codes. We are planning to roll out the code to the HEP community with a series of tutorials starting with the upcoming US FCC Workshop at SLAC in September, 2026~\cite{USFCC}.
\newpage
\section{Detectors for 10 TeV Collider}
\label{sec:Rastogi}
\href{https://agenda.infn.it/event/47329/contributions/282155/}{Slides: \url{https://agenda.infn.it/event/47329/contributions/282155/}}

\subsection*{\textit{Angira Rastogi and Simone Pagan Griso}}

\noindent \textit{Lawrence Berkeley National Laboratory, Berkeley, CA, USA}
\\ 

\subsection{Context}
The key design parameters of an effective detector depend strongly on the properties of the beam particles, the beam configuration (round vs. flat), the machine-detector interface, and the targeted physics program. In developing a detector concept for a 10 TeV wakefield-accelerated $e^+e^-$ or $e^-e^-$ collider, it is essential to understand the backgrounds arising from beam–beam interactions at each bunch crossing. High-energy photons radiated in the intense electromagnetic fields of the opposing beams generate substantial beam-induced backgrounds (BIB), primarily through $e^+e^-$ pair production and hadronic photoproduction $\gamma\gamma\rightarrow qq$. In addition, coherent $e^+e^-$ pair production can occur via interactions of particles in one beam with the collective field of the other (the non-linear Breit-Wheeler process), alongside incoherent pair production resulting from interactions between individual particles in opposing beams (e.g., Landau-Lifshitz, Bethe-Heitler, and Breit-Wheeler processes).

\subsection{Progress}
We have developed a Key4Hep-based~\cite{Key4hep:2023rka} software workflow for end-to-end studies, spanning signal event generation, detector-level simulation, and physics object reconstruction. Within this framework, the detector geometry is defined using the \textit{DD4hep} toolkit~\cite{Frank_2014}, which also provides the necessary interfaces to Geant4~\cite{Allison:2016nima835} for simulating particle–matter interactions via the \textit{ddsim}~\cite{Petrič_2017} application. The \textit{DDPlanarDigi} processor emulates the detector front-end electronics response in a simplified manner. It takes simulated hit information, such as position, time, and energy deposition, in detector layers and converts it into digitized signals by applying Gaussian smearing, thereby incorporating the effects of spatial and timing resolution. The output of each stage is stored in the EDM4hep~\cite{refId0} format, the standard event data model used within Key4Hep.

We use beam-beam simulations from WarpX~\cite{Vay_2018}, converted to EDM4hep format, to characterize the BIB environment in the vicinity of the interaction region. The number of particles entering the detector per bunch crossing (BX), after fiducial selections i.e. transverse momentum $p_T > 20$ MeV, $10^\circ < \theta < 170^\circ$ for $e^\pm$, and $0.7^\circ < \theta < 179^\circ$ for photons are summarized in Table~\ref{tab:beams}. Additionally, there is a contribution from machine-induced backgrounds arising from particles that scatter back from the forward instrumentation (back-splash pairs). These particles are out of time with respect to the primary BX, as they originate several meters downstream ($\sim$6 m) and arrive with a delay of order 20 ns. As a result, their impact can be effectively mitigated using modest timing cuts.

\begin{table}[h!]
\centering
\begin{tabular}{|c|c|}
\hline
\bf{Beam-beam interaction} & \bf{Number of particles / BX} \\
\hline
Coherent pair production & $\num{3e8}\times 2$ \\
\hline
Incoherent pair production & $\num{1e6}\times 2$ \\
\hline
Beamstrahlung photons & $\num{7e10}\times 2$ \\
\hline
Disrupted primary beam & $\num{3e6}\times 2$ \\
\hline
\end{tabular}
\caption{Number of particles per bunch crossing (BX) from various interactions with detector fiducial selections i.e. $p_{T}$ $>$ 20 MeV, $10^{\circ}< \theta < 170^{\circ}$ for $e^+/e^-$ and $0.7^{\circ} < \theta < 179^{\circ}$ for photons in $e^+e^-$ round beam collision.}
\label{tab:beams}
\end{table}

\begin{figure}[h!]
    \centering
    \includegraphics[width=1\textwidth]{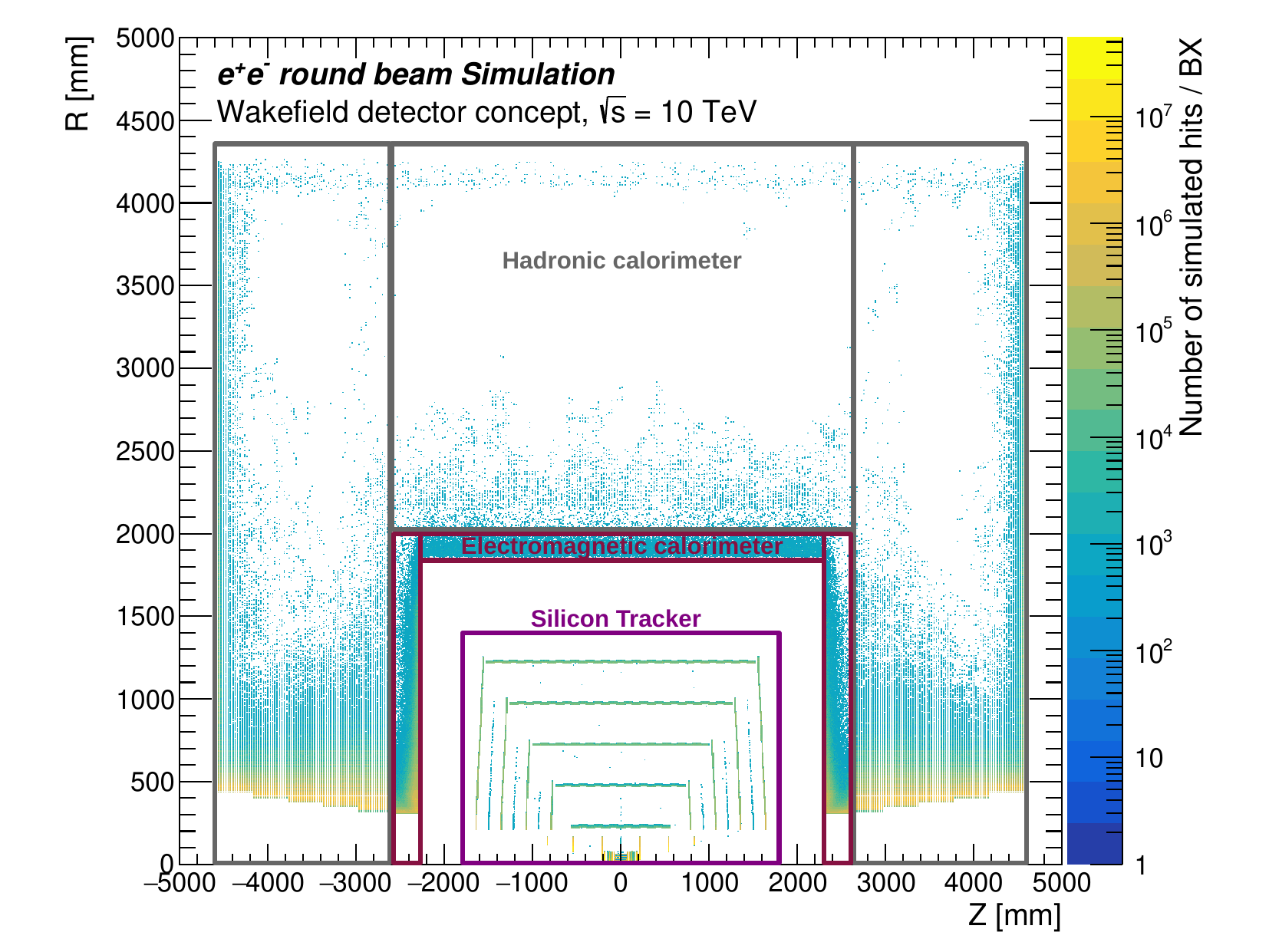} 
    \caption{Simulated hits multiplicity per BX from all BIB particles in the initial 10 TeV Wakefield detector geometry.}
    \label{fig:simrz}
\end{figure}

Guided by this, we propose an initial detector concept for a 10 TeV collider, inspired by existing designs for future high-energy machines operating in comparable radiation environments. The tracking system adopts a full-silicon approach, similar to the ILC's Silicon Detector (SiD)~\cite{breidenbach2025updatingsiddetectorconcept}, which enables high-precision measurements through excellent secondary-vertex reconstruction and jet flavor tagging. Silicon sensors provide fine spatial resolution ($\sim$20 $\mu$m pitch), timing resolution at the level of tens of picoseconds, and strong radiation tolerance. A 5 T solenoidal magnetic field, combined with a large tracking volume and low material budget, ensures good momentum resolution while efficiently sweeping low-momentum $e^+e^-$ pairs from incoherent production out of the central detector region. Given the higher flux of $e^+e^-$ pairs from coherent production in the forward region, the solenoid is positioned upstream of the calorimeters to mitigate long-term radiation damage. The calorimeter system is designed to meet multiple performance goals: fine granularity for photon-hadron discrimination, sufficient depth (high $\lambda_0$) for improved jet-mass resolution and $W/Z$ separation, and extended forward coverage for high-energy electrons from vector-boson-fusion processes. Accordingly, we adopt silicon-tungsten electromagnetic and iron-scintillator hadronic calorimeters, with timing resolutions of approximately 50 ps and 1 ns, respectively, along with a dedicated muon system for efficient tagging of $Z'$ bosons in $ZH$ production, an important channel for the Higgs physics program, following the design philosophy of the MAIA 10 TeV muon collider detector concept~\cite{bell2025maianewdetectorconcept}. The simulated hits multiplicity per BX from all the BIB particles in this detector geometry is shown in Figure~\ref{fig:simrz}.

\subsection{Next Steps}
The next steps are to evaluate the viability of the proposed detector design by quantifying realistic detector occupancies under continuous, real-time readout conditions and assessing the performance of physics object reconstruction. We will also investigate the impact of $\gamma\gamma \rightarrow q\bar{q}$ backgrounds at 10 TeV, once implemented in WarpX, with particular emphasis on jet energy resolution and flavor tagging, and explore pileup mitigation strategies similar to those developed at the CERN Large Hadron Collider. In parallel, we plan to study benchmark physics processes to further optimize the detector concept, refine the design parameters of individual subdetector systems, and help validate assumptions of key physics projections. Finally, we will compare alternative beam-beam configurations including $e^-e^-$ and $\gamma\gamma$ collisions, as well as round versus flat beams, to identify the most feasible and efficient path towards realizing a 10 TeV wakefield collider.

\newpage
\section{AI for Accelerators: Current and Future Trends}
\label{sec:Kain}
\href{https://agenda.infn.it/event/47329/contributions/287383/}{Slides: \url{https://agenda.infn.it/event/47329/contributions/287383/}}

\subsection*{\textit{Verena Kain}}

\noindent \textit{Accelerator and Beam Physics group (ABP), Beams Department, CERN, Geneva, Switzerland}\\
\noindent \textit{Integrated Climate Impact group (ICI), Energy Climate and Environment programme (ECE), IIASA, Laxenburg, Austria}

\subsection{Context}

Artificial intelligence for high energy physics and particle accelerators is causing a paradigm shift with direct parallels to the revolution in numerical weather prediction (NWP). Neural network concepts were introduced to weather forecasting as early as the 1990s, but early models could not scale to replace full dynamical simulations. The breakthrough came in 2023 with Pangu-Weather~\cite{Bi:2023}, an artificial neural network that replaced the numerical solver and outperformed state-of-the-art global forecasting systems. AI models now dominate leaderboards, with conventional numerical methods relegated to the role of ground truth. A similar transition is taking place in theoretical particle physics, where AI is beginning to replace or augment analytical methods for quantities such as scattering amplitudes. Alongside these technical advances, the environmental footprint of AI compute is increasingly under scrutiny~\cite{Sevilla:2022}. AI models are however very competitive in comparison to simulations on HPC clusters. For example an AI 10-day weather forecast requires typically a factor 1000 less energy than NWP.  

\subsection{Key themes}

This invited talk surveys the state of AI for particle accelerators from three angles: (i) lessons from the broader AI revolution and what they imply for accelerator physics; (ii) the current status and trends in machine learning for accelerators based on recent community workshops; and (iii) a vision for how AI should be embedded throughout the full lifecycle of next-generation machines, with the FCC-ee as a concrete case study.

\subsection{Status and trends}

Optimisation and control of particle beams or accelerator components are the dominant ML/AI applications to date, with Bayesian Optimisation the most mainstream technique, see for example recent workshops~\cite{MLworkshop:2025}. The community has developed impressive expertise, with state-of-the-art algorithms deployed at multiple facilities. A notable recent extension is Bayesian Algorithm eXecution (BAX), which formulates virtual objectives---for example minimum-emittance tuning---as information-theoretic acquisition problems. Virtual emittance scans are performed on posterior samples of a Gaussian process surrogate and used to choose measurement points that simultaneously maximise information gain and minimise emittance~\cite{BAX:2024}. The method has been demonstrated at the Linac Coherent Light Source (LCLS) and is being adopted more widely. Reinforcement learning is also gaining traction, with a strong programme at Jefferson Laboratory developing new algorithms tailored for online accelerator control including constraints. 

Another major trend is the adoption of differentiable simulation codes. Differentiability allows gradients to be propagated through the simulator, enabling powerful new workflows: 6D phase-space reconstruction from as few as 20 measurements~\cite{Roussel:2024}, gradient-based lattice design, and rapid surrogate model fitting. The Integrable Optics Test Accelerator (IOTA) at Fermilab has a leading example of a fully differentiable accelerator model. New or upgraded simulation codes should be written to be differentiable by default.

On the operational side, large language models (LLMs) are beginning to find a role in accelerator control rooms. Several laboratories---including EuXFEL and Berkeley Lab---are exploring LLM-assisted interfaces that let operators prompt high-level operational concepts rather than issue detailed machine commands \cite{Sulc:2025}. The General AI Assistant for Intelligent Accelerator Operations (GAIA) at DESY combines LLMs with the ReAct (Reasoning and Acting) framework to create an agent that can reason about machine state and invoke control-system tools. At present these systems are proof-of-principle, but they point towards a future in which static GUIs and paper logbooks are replaced by conversational agents.

\subsection{Vision for future machines}

The FCC-ee requires a completely new accelerator complex with a new injector chain, offering a greenfield opportunity to rethink the human--machine interface and data infrastructure from first principles. Rather than digitising the workflows inherited from analogue-era machines, future facilities should be designed from the outset for AI-driven operation: machine-readable data stores, differentiable simulation models integrated into the control system, and agent-based interfaces that promote high-level operational goals~\cite{FCC}. The overall trend identified across the community is a drive towards full automation of routine machine operation.

\subsection{Conclusions}

AI is changing how science is done, bringing democratisation, faster onboarding, and qualitatively better results across disciplines. For particle accelerators, the community is still in transition, but the tools available today are already sufficient to address most operational challenges. The recommendation is to be more ambitious: remove scientific silos, share data openly across laboratories, and embrace differentiable and AI-native design practices for new machines. Those who engage now can still shape the direction of the field.
\newpage
\section{Beam Delivery System for 10 TeV Collider}
\label{sec:Downham}
\href{https://agenda.infn.it/event/47329/contributions/280919/}{Slides: \url{https://agenda.infn.it/event/47329/contributions/280919/}}

\subsection*{\textit{Keegan Downham}}

\noindent \textit{University of California, Santa Barbara, CA, USA}\\
\noindent \textit{SLAC National Accelerator Laboratory, Menlo Park, CA, USA}

\subsection{Context}
To reach the high luminosity goals at smaller repetition rate than 
circular colliders, high energy plasma wakefield colliders must 
compensate by providing small beam spot sizes at the interaction point. 
This is achieved by the beam delivery system (BDS), comprised primarily of a collimation 
section for machine protection and a final focusing system (FFS) for demagnification of the 
beams. Linear colliders typically utilize flat 
beams ($\sigma_{x}^{*} >> \sigma_{y}^{*}$) to minimize the effects of beamstrahlung, 
however plasma wakefield acceleration (PWFA) with flat 
beams provides additional challenges, making it 
more favorable to use round beams ($\sigma_{x}^{*} = \sigma_{y}^{*}$) in the 
accelerating stages. 
\subsection{Objectives}
A flat-beam BDS for a 10 TeV wakefield collider aims to deliver a luminosity of $\mathcal{L} \approx 34.1 \times 10^{34}$ $\rm{cm}^{-2}$ $\rm{s}^{-1}$~\cite{Barklow:2023iav} while being 
compatible with plasma acceleration stages and maintaining a reasonable 
length footprint. Additional considerations 
include correction of chromaticity, mitigation of synchrotron radiation effects, 
cancellation of geometric aberrations, and the 
characterization of beamstrahlung. 
\subsection{Progress over the last two years}
Over the past year, efforts have been made to 
re-purpose the FFS design for CLIC 7 TeV~\cite{lewis} 
to be compatible with the increased beam energy of a 10 TeV wakefield collider. A first design 
for a flat-beam FFS at 10 TeV was recently 
achieved~\cite{Downham:2026zwc}, with a luminosity 
$\mathcal{L} = 10.8 \times 10^{34}$ $\rm{cm}^{2}$ 
$\rm{s}^{-1}$. The compatibility of the chosen 
parameters used in the 10 TeV FFS design with 
achieved values from PWFA experiments was also 
considered, as shown in Table~\ref{tab:accel:compatibility}; the 
bunch lengths and bunch energy spreads assumed 
in the 10 TeV design have been achieved, although 
the assumed emittances and bunch charges are 
still beyond current capabilities.  


\begin{table}[hbtp!]
    \centering
      \scalebox{0.95}{
      \begin{tabular}{lccccccc@{}}
  \hline\hline
      Metric & Achieved Values & 10 TeV FFS & Compatible? \\
      \hline \hline
      Bunch energy spread & 0.12\%~\cite{Lindstrom:2024zbo} & 0.3\% & \textcolor{green}{Yes} \\
      \hline
      Bunch length & 10-20 $\mu$m~\cite{osti_3016925} & 44 $\mu$m & \textcolor{green}{Yes} \\
      \hline
      Bunch charge & $\mathcal{O}$(10 pC)~\cite{Lindstrom:2024zbo,osti_3016925} & 596 pC & \textcolor{red}{No} \\
      \hline
      Emittance & $\mathcal{O}$(1 $\mu$m)$^{*}$~\cite{Lindstrom:2024zbo,osti_3016925} & 0.66 $\mu$m (x) / 0.02 $\mu$m (y) & \textcolor{red}{No} \\
\hline\hline
  \end{tabular}}
    \caption{Comparison of achieved parameter values from PWFA experiments with the values assumed for the 10 TeV FFS design~\cite{Downham:2026zwc}. $^{*}$ Emittances are (approximately) equal 
    in both horizontal and vertical directions. }
    \label{tab:accel:compatibility}
\end{table}

\subsection{Time scale for application}
The first design of a flat-beam FFS for a 10 TeV 
wakefield collider is a step in the right 
direction, however an improved design is needed to 
achieve the desired luminosity and establish 
compatibility with machine parameter goals. 
Additionally, a number of other considerations 
must be addressed before the 10 TeV BDS can be 
seriously considered for a collider application. 
To comply with round-beam plasma acceleration stages, 
two options are possible. The first of which is 
the addition of a matching section to convert 
the round beams from the plasma acceleration 
stages to the flat beams required at the 
beginning of the current FFS design. The second 
option is the design of a round-beam FFS at 
10 TeV, which requires a substantial effort and 
careful design to meet the luminosity goals. 
Additionally, the collimation system for a 
10 TeV collider has not yet been considered and 
will be required before a mature BDS design can 
be proposed. Further work would also include the 
study of magnet misalignment tolerances and 
beam jitter to understand the effect of errors 
on the FFS performance. Altogether, assuming that the emittance and bunch charge targets are met from PWFA experiments, the timeline  to have a full collider-ready BDS 
at 10 TeV is likely in excess of two years, 
depending on whether a round-beam BDS is needed 
and whether the collimation section length at 
10 TeV is acceptable. 

\newpage
\section{Operating High-Power Laser Facilities at Scale: Lessons from EPAC and
Vulcan 20-20}
\label{sec:Oliveira}
\href{https://agenda.infn.it/event/47329/contributions/281478/}{Slides: \url{https://agenda.infn.it/event/47329/contributions/281478/}}

\subsection*{\textit{Pedro Oliveira}}

\noindent \textit{Central Laser Facility, Science and Technology Facilities Council, Rutherford Appleton Laboratory, Harwell Campus, Didcot, UK}

\subsection{Operational challenge}
The value of a user facility is determined not only by peak laser performance, but also by the fraction of scheduled time for which a
qualified beam and experimental environment are available, the stability of the system and the diagnostic coverage provided to users.
Historically, the CLF's high-power systems have operated in extended daytime and evening shifts rather than continuously. Moving
beyond this model in new systems will require time and operational experience, because high-power lasers combine complex optical
chains, micron- and microradian-scale alignment tolerances, high-voltage and vacuum systems, large diagnostic suites and optics
that may degrade abruptly under high fluence. A single failed component can stop an entire experimental run.
Operational design must therefore be treated as an all-encompassing systems-engineering problem from the start. Simplicity and
compactness reduce failure modes, while standardized components across facilities improve maintainability. Automated turn-on,
alignment, adaptive-optics control, diagnostic analysis and damage detection reduce routine workload and make performance less
dependent on individual intervention. These functions must be integrated with timing, configuration control and experimental
diagnostics rather than deployed as isolated tools.

\subsection{Maintainability and operational readiness}

\begin{figure}[htbp]
    \centering
    \includegraphics[width=0.7\textwidth]{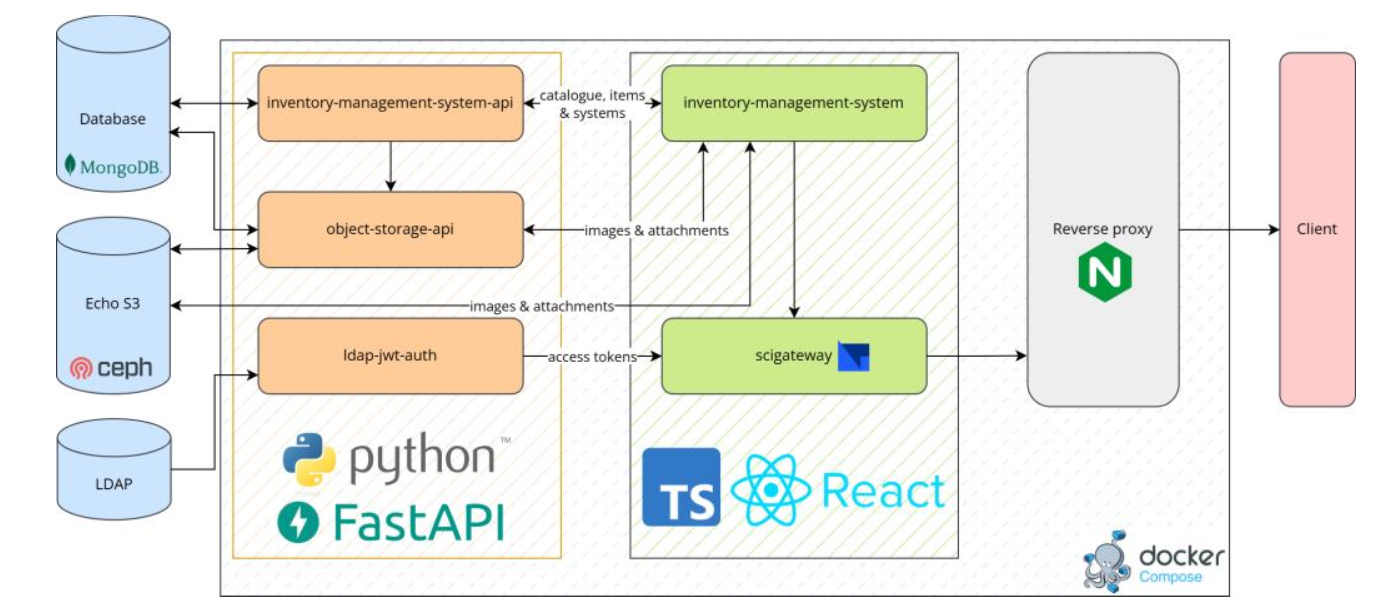}
    \caption[Caption for List of Figures]{\label{YourLabel} CLF spare-parts database architecture
    \label{fig:Oliveira1}
    }
\end{figure}

The spares strategy is a critical part of operations management and is the ultimate practical limit on availability. The CLF adopts a
two-pronged approach: retaining critical stock on site and shortening the replacement route where rapid manufacturing or
experiment-specific solutions are possible. This is normally achieved by keeping blank substrates in stock, either in-house or with
trusted suppliers, which substantially reduces the wait time for new optics.
In addition, stock requirements are assessed against component lifetime, delivery time, number in service and system criticality. A
structured database links systems, component catalogues, installed items and storage locations. For example, if an optic has a lifetime
$L$ and a delivery timescale $T$, the margin factor $M$ is:
\begin{equation}
    M = \frac{NT}{L}
\end{equation}
where $N$ is the number of that optic in use. In this case, the minimum number of optics needed in stock ($MNS$) at any one time is:
\begin{equation}
    MNS = M + 2\sqrt{M}
\end{equation}
Criticality can then be defined as the actual stock minus the minimum required stock. To make this effective across complex systems,
we have developed a spare-parts database that includes a catalogue of all items that can be used in the facilities and a hierarchical
structure of systems and subsystems, from the facility level to the position of each individual optic. This information is combined to
determine the resilience of the facility. Figure~\ref{fig:Oliveira1} shows the architecture of such a database.
Technical resilience also requires explicit ownership. Competence matrices identify the first, second and third points of contact for
each subsystem, reducing dependence on informal knowledge. An emergency replacement matrix also allows us to replace critical
elements, normally lasers, with others of similar characteristics.
Finally, an important part of optimizing facility time is to prepare and test experiments initially in a virtual environment; this supports
design, commissioning and user preparation. For EPAC, the vEPAC environment is intended to connect laser, plasma, particle-
transport, radiation-generation and application simulations. 
 It can be used to test sensitivity to misalignment, drift and jitter, optimize
operating points and assess proposed experiments before changes are made to the facility.

\subsection{Controls, automation, protection and data}

\begin{figure}[htbp]
    \centering
    \includegraphics[width=0.7\textwidth]{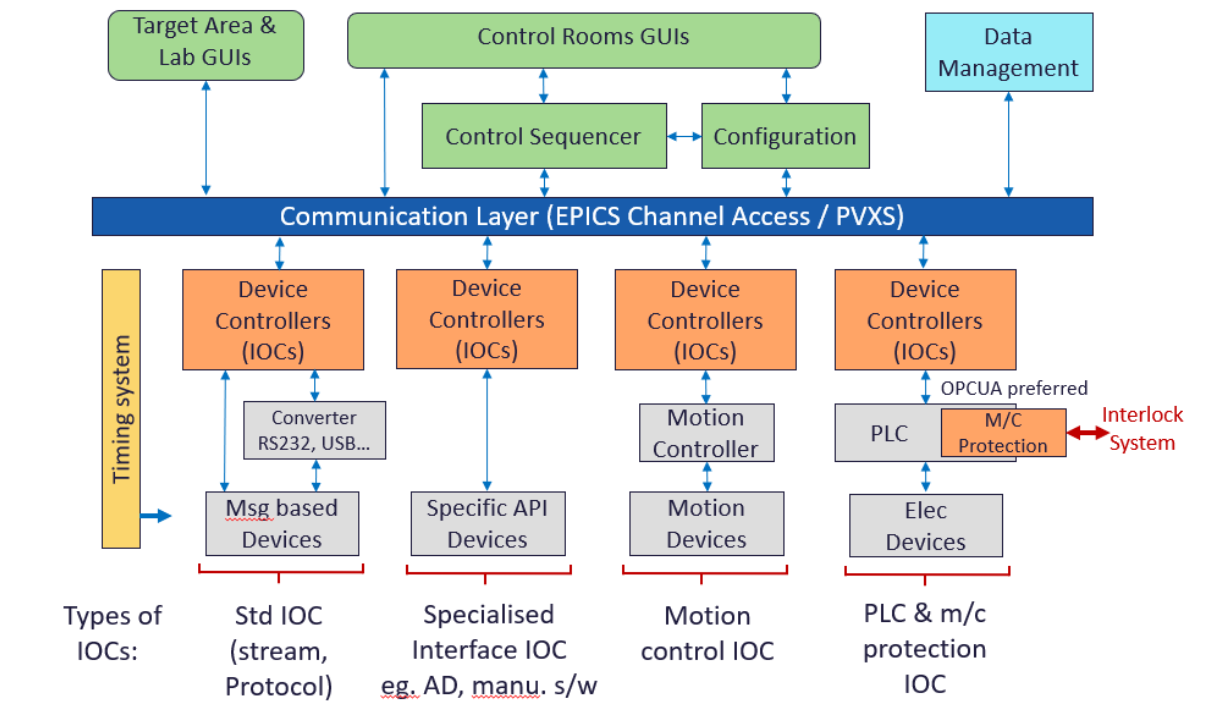}
    \caption[Caption for List of Figures]{\label{YourLabel} Layered EPICS control architecture linking devices, timing, sequencing, interlock systems and data management.
    \label{fig:Oliveira2}
    }
\end{figure}

The main way to achieve this in increasingly complex and demanding facilities is to rely on automation. Automation is the
intervention most likely to reduce or stabilise resource requirements throughout the life of the facilities. The high-power facilities
being built at the CLF are therefore developing automatic alignment, adaptive-optics control, remote turn-on, integrated plasma
diagnostics, immediate diagnostic analysis, damage detection and a data pipeline and storage system for later review. A single control
system coordinates these activities (Fig.~\ref{fig:Oliveira2}).
The new CLF facilities use a distributed control architecture based on the Experimental Physics and Industrial Control System
(EPICS). Device controllers provide standard interfaces for message-based equipment, motion systems, programmable logic
controllers and devices with specialist application programming interfaces. Above this layer, control-room interfaces, configuration
services, sequencers and timing systems provide coordinated operation. This structure allows local subsystem development while
maintaining a facility-wide operational model.

Machine protection is being trialled in EPAC, with logic hosted in programmable controllers and monitored through EPICS via an
OPC UA server. A soft stop protects against degradation that would become damaging if repeated, while a hard stop responds to
immediate hazards. This distinction is important at high repetition rate, where a modest fault can otherwise become a rapid
sequence of damaging events.
Experiment orchestration and data provenance are treated as part of the control problem. Bluesky provides programmable shot
sequences and controlled access to hardware and diagnostics. Data from EPICS controllers are passed through a Kafka pipeline for
display and live analysis before being written to HDF5 files. Users can access previous data and analyses through managed
computing services. The same approach supports remote expert participation and provides the traceability needed to compare
performance across long campaigns.

\subsection{Facility upgrades}

EPAC and Vulcan 20-20 illustrate how the same operational principles apply to different upgrade paths. EPAC is a new 10 Hz
petawatt-class user facility based on an optical parametric chirped-pulse-amplification front end, a gas-cooled Ti:sapphire power
amplifier and a 10\,Hz diode-pumped solid-state pump laser; its first experimental area is designed for laser-wakefield acceleration,
and user access is planned after commissioning. At 10 Hz, diagnostics may generate data rates approaching 5\,GBs$^{-1}$ and annual
volumes of about 1-2\,PB, making automated shot orchestration, online analysis and reliable metadata association essential. Vulcan
20-20 targets a lower-repetition-rate but higher-energy regime, combining a 20\,PW short-pulse beam, specified at 400\,J in 20\,fs with
a five-minute shot interval, with multi-kilojoule long-pulse capability and auxiliary beams. This places the emphasis on broadband
OPCPA, large-aperture compression and transport, flashlamp-pumped Nd:glass amplification, multiple target-area configurations,
safe sequencing, fault recovery and condition monitoring.

\subsection{Conclusions}

Longer high-power-laser operations cannot be achieved sustainably by staffing alone. Availability must be designed into the facility
through automation, standardization, maintainability, protected operating states and rapid access to trustworthy data. EPAC and
Vulcan 20-20 demonstrate that these principles apply across very different operating regimes: one prioritizes 10 Hz throughput and
data volumes, while the other coordinates extreme pulse energy, large optics and multiple beam configurations. The common
objective is a facility that can repeatedly deliver characterized beam conditions, recover efficiently from faults and give users a
coherent experimental and data environment. This systems approach is the practical route towards higher-value, longer-duration
operation of future high-power laser facilities.
\newpage
\section{Operation and Management of a 24/7 FEL User Facility}
\label{sec:Giannessi}
\href{https://agenda.infn.it/event/47329/contributions/286130/}{Slides: \url{https://agenda.infn.it/event/47329/contributions/286130/}}

\subsection*{\textit{Luca Giannessi\textsuperscript{1,2}, Filippo Bencivenga\textsuperscript{1}, Claudio Masciovecchio\textsuperscript{1}, Giuseppe Penco\textsuperscript{1}, Mauro Trovò\textsuperscript{1} }}

\noindent \textit{\textsuperscript{1}Elettra-Sincrotrone Trieste S.C.p.A.  Strada Statale 14 - km 163,5 in AREA Science Park 34149 Basovizza (Trieste) ITALY}\\
\noindent \textit{\textsuperscript{2}INFN Laboratori Nazionali di Frascati Via E. FERMI 51, 00044 Frascati (Roma) ITALY}
\noindent \textit{\textsuperscript{3}}

\subsection{Context}
The operation of a modern Free-Electron Laser (FEL) user facility entails a highly complex integration of advanced technologies, scientific expertise, and coordinated infrastructure, extending well beyond the traditional notion of accelerator operation. Facilities such as FERMI (the FEL facility at the Elettra Sincrotrone Trieste; see Fig. \ref{Elettraandfermi}) exemplify this paradigm, functioning simultaneously as precision photon sources, large-scale research infrastructures, and service laboratories for an international user community. For this reason, FERMI  has been adopted as a reference case to illustrate the various aspects involved in the operation of such a facility.

Seeded FELs offer distinctive performance characteristics compared to SASE-based sources, notably in terms of spectral purity, stability, and control of the emitted radiation. Through external seeding schemes such as High Gain Harmonic Generation \cite{Yu:1991}, the coherence properties of the driving laser are transferred to the FEL output, enabling narrow bandwidth, high wavelength stability \cite{Allaria:2012a,Allaria:2013a}, and laser-like photon statistics \cite{Gorobtsov:2018}. At FERMI, two FEL lines (FEL-1 and FEL-2) operate using harmonic generation techniques, extending coherent emission across a broad spectral range while preserving these properties \cite{Allaria:2015}. Typical performance includes high spectral coherence, wavelength stability at the level of $10^{-5}$ rms, and pulse energy fluctuations of a few percent under optimized conditions, with femtosecond-level temporal jitter \cite{Danailov:2014}.

The operation of such a facility is inherently user-driven and highly competitive. Access is granted through periodic calls for proposals evaluated by international review panels, which assess scientific merit. Only a fraction of submitted proposals can be accommodated, leading to significant oversubscription ($\sim\,3.5)$.
A defining feature of FEL facilities is the necessity for simultaneous and precise coordination of multiple subsystems. The FEL process itself relies on the exponential amplification of radiation through a beam–radiation instability, which is highly sensitive to electron beam parameters such as current, emittance, energy spread, and alignment \cite{Kondratenko:1980,Dattoli:1981,Bonifacio:1984}. Maintaining optimal conditions requires continuous monitoring and feedback. In seeded FELs, this complexity is further increased by the critical role of the seed laser, which directly participates in the gain mechanism and must be synchronized with the electron beam at the tens of femtosecond level.

The scientific production of a Free-Electron Laser (FEL) is inherently lower in quantitative terms than that of synchrotron facilities; operationally, FELs differ substantially from synchrotron light sources. While synchrotrons typically serve multiple beamlines in parallel, FEL facilities often operate on few or just one experiment at a time, with beam delivery and machine parameters optimized sequentially for each experiment. FELs naturally focus on delivering high-impact, unique experiments, prioritizing quality over quantity. This experiment-driven mode of operation demands dedicated machine tuning and close interaction between accelerator physicists, beamline scientists, and users. 

The successful execution of experiments relies on the coordinated effort of multidisciplinary teams, including experts in accelerator physics, laser systems, diagnostics, controls, and data infrastructure. Preparation for each experiment involves detailed planning, often initiated months in advance, to ensure that all technical requirements are met. Diagnostics and real-time feedback systems are essential components of operation, as they enable stabilization of both electron and photon beams. In addition, the large data volumes generated by experiments carried out on a "shot-to-shot" basis necessitate robust data acquisition, storage, and processing capabilities.
A key strength of FEL facilities lies in their versatility. Advanced operational modes—such as multi-color pulse generation \cite{Ferrari:2016,Prince:2016}, pulse shaping \cite{Maroju:2020}, attosecond synchronization \cite{Maroju:2023}, polarization control and exotic polarization states \cite{Allaria:2014, RebernikRibic:2017a} —can be implemented to meet specific experimental requirements. These capabilities are often developed through close collaboration between facility staff and users, contributing to continuous innovation and expansion of the scientific reach of the facility.
Scientific output provides an important metric of facility performance. Experiments conducted at FEL facilities contribute to a wide range of disciplines, including condensed matter physics \cite{Lam:2018,Bencivenga:2019,Yao:2022,Capotondi:2025}, ultrafast magnetism \cite{KorffSchmising:2014, Fanciulli:2022, Leveille:2022},  atomic and molecular physics and ultrafast science \cite{Pathak:2020,Nandi:2022,Maroju:2023}. The path from experiment to publication typically spans several years, reflecting the complexity of data analysis and interpretation, but results in a steady production of high-impact scientific work.

\subsection{Next Steps}
Future developments aim to further enhance performance and extend capabilities, reducing the radiation pulse duration and extending toward higher photon energies the operation spectral range. 

Planned upgrades at FERMI include increasing the accessible photon energy range toward soft X-ray regimes, reducing pulse durations into the few-femtosecond regime \cite{Mirian:2021}, and implementing advanced seeding schemes such as Echo-Enabled Harmonic Generation (EEHG)\cite{Ribic:2019}.  Recent experiment carried out at FERMI have shown the possibility of extending the range of emission to the water window and to the L-edges of transition metals by implementing the higher order harmonic emission of the FEL. Importantly these harmonic still preserve the coherence properties of the driving source \cite{Penco:2024,Spezzani:2024}. These developments are designed to preserve the key advantages of seeded FELs—coherence, stability, and control—while pushing the frontiers of ultrafast and high-resolution photon science.
The FERMI FELs have been the first operational seeded FEL facility, providing fully coherent radiation and setting a benchmark in the field. Today, however, the landscape of seeded FELs is rapidly evolving: new and upgraded facilities such as Dalian Coherent Light Source \cite{Yu:2019DCLS}, Shanghai Soft X-ray Free Electron Laser \cite{Feng:2022}, the FLASH1 as part of the FLASH 2020+ program in DESY \cite{Beye:2023}, and the Athos beamline at SwissFEL \cite{Abela:2019} are significantly expanding the global availability of fully coherent FEL radiation, enriching the scientific opportunities worldwide.

Ultimately, the performance and evolution of an FEL facility depend critically on the expertise and coordination of the people involved. The combination of advanced instrumentation and highly skilled teams enables not only reliable operation but also continuous innovation, ensuring that such facilities remain at the forefront of scientific research.

\begin{figure}[htbp]
    \centering
    \includegraphics[width=0.7\textwidth]{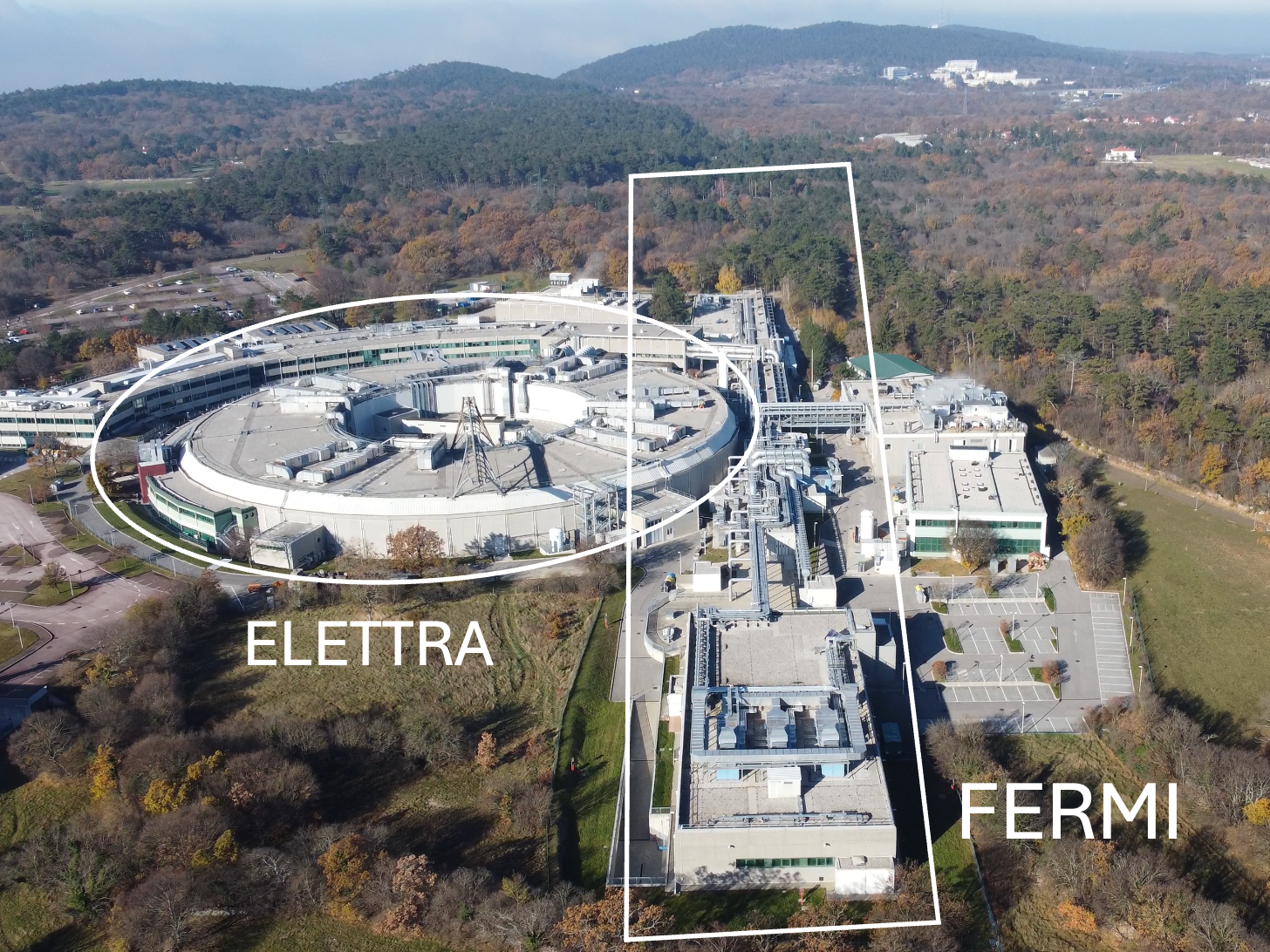}
    \caption[Caption for List of Figures]{\label{YourLabel} An aerial view of the Elettra Sincrotrone Trieste laboratory. The FERMI free-electron laser is the structure appearing on the right of the Elettra storage ring.}
    \label{elettraandfermi}
\end{figure}
\newpage
\section{Direct Laser Acceleration - a Path Towards High-Charge Electron Bunches for High-Field QED Applications}
\label{sec:Vranic}
\href{https://agenda.infn.it/event/47329/contributions/285159/}{Slides: \url{https://agenda.infn.it/event/47329/contributions/285159/}}
\subsection*{\textit{Marija Vranic\textsuperscript{1} and Robert Babjak\textsuperscript{1,2}}}
\noindent \textit{\textsuperscript{1}GoLP/IPFN, Instituto Superior Técnico - Universidade de Lisboa, 1049-001 Lisbon, Portugal }\\
\noindent \textit{\textsuperscript{2}CEA, DAM, DIF, F-91297 Arpajon, France}

\subsection{Context}
Direct Laser Acceleration (DLA) in low-density plasmas can provide high-charge ($\sim~\mu$C) electron beams with multi-GeV energies. These beams can be used to generate radiation or other particles (e.g. positrons and muons), and DLA mechanism can also be used to accelerate positive charges like positrons. Most research to date on DLA has been focused on near-critical density materials (to obtain high-brilliance X-ray radiation) where energy gain is limited, or using moderate intensity picosecond laser pulses in gaseous targets. Combining the new class of intense lasers (multi-PW, ~100 fs duration) with low-density gas targets, it is possible to take DLA electron energy cutoff  to $\sim$ 10 GeV,  which opens up a variety of possibilities for applications, most notably for high-field QED experiments. 

\subsection{Objectives}
One of the scientific objectives is to obtain a pair plasma in the lab, which would serve as a testbed for studying the collective effects in electron-positron plasmas. The requirement is to create a high enough density and size to contain at least several plasma skin depths. The lepton plasmas are expected to have unique properties due to their mass ratio of 1. The DLA electrons could be converted to pair plasma either through electron-laser scattering (Breit-Wheeler process) or by using a high-Z, thick converter target (Bethe-Heitler process). Another option for creating an electron-positron plasmas is using these beams in combination with intense lasers to seed QED cascades in laboratory. 

Apart from creation, the acceleration of positrons in plasmas is a grand challenge by itself. Previous experience from electron acceleration schemes cannot be readily ported to positron acceleration. DLA-accelerated electrons provide a traveling guiding structure for a positron beam that can be accelerated also through DLA. The obtained beam is, however, broadband. 

Another objective is to reach high-chi ($\chi\gg1$) regime of QED using laser reflection on a solid target ($\chi$ is defined as a ratio of the electric field in the electron rest frame to the Schwinger critical limit $E_C\simeq 1.3\times 10^{18}$ V/m). The laser pulse front gets etched in the acceleration stage, and then gets reflected with a fast-rising envelope, which allows electrons to immediately access the intense field, without  first getting slowed down in the low-intensity section of the laser. 

The fourth objective is generation of muons. The DLA beams are broadband, but they have a lot of charge (experimentally obtained $\gg$ 100 nC). Increasing the average energy of the beam by using low density plasma channels for acceleration allows taking lots of charge to several GeV energies, which then can be converted to muons in the interaction with a solid target. 
\subsection{Progress over the last two years}
Several advances have occurred in the last two years. A recent experiment showed the importance of sheath field for electron acceleration \cite{Tang:2025}, which adds to the effect of density gradient on acceleration identified by Babjak et al \cite{Babjak:2024}. Properties of radiation emitted in the DLA acceleration stage were calculated \cite{Babjak:2025}, and requirements for reaching the QED regime in the laboratory with DLA beams identified \cite{Babjak:2026}. In addition, a configuration for acceleration of Bethe-Heitler positrons was identified \cite{Martinez:2025}.

\subsection{Time scale for application}
Some of the mentioned applications can be expected with lasers of Extreme Light Infrastructure in the next few years. Reaching the QED dominated regime is possible with that technology, but producing sufficient pairs to obtain electron-positron plasma depends on control of multiple parameters, and may eventually require a laser power over 10\,PW. The timeline is then tied to the timeline of next laser upgrades, which is uncertain at the moment.  
\newpage
\section{EuPRAXIA@SPARC\_LAB - Final Layout}
\label{sec:Ferrario}
\href{https://agenda.infn.it/event/47329/contributions/281012/}{Slides: \url{https://agenda.infn.it/event/47329/contributions/281012/}}
\subsection*{\textit{Massimo Ferrario 
\\[1mm]
\normalfont (on behalf of the EuPRAXIA@SPARC\_LAB Collaboration)}}

\noindent \textit{INFN -- Laboratori Nazionali di Frascati, Italy}

\subsection{Context}
EuPRAXIA (European Plasma Research Accelerator with eXcellence In Applications) is the first plasma accelerator infrastructure included in the ESFRI Roadmap and represents one of the major European initiatives toward next-generation accelerator technologies ~\cite{assmann2020eupraxia}. Its objective is to demonstrate that plasma-based accelerators can evolve from laboratory-scale experiments into reliable, user-oriented research infrastructures for photon science, advanced radiation sources, and future high-energy physics applications.
EuPRAXIA@SPARC\_LAB, hosted at INFN-LNF in Frascati, is the beam-driven plasma wakefield acceleration (PWFA) implementation of the EuPRAXIA infrastructure ~\cite{ferrario2018eupraxia}. The facility combines compact high-gradient X-band radio-frequency technology with plasma acceleration, aiming to realize the first Free-Electron Laser (FEL) driven by a plasma-accelerated electron beam operating as a user facility.

The project builds upon more than two decades of experience at SPARC\_LAB in high-brightness electron beams, FEL physics, plasma acceleration, and advanced diagnostics ~\cite{ferrario2013sparc_lab, Pompili2024ApplSciSPARCLAB}. In particular, SPARC\_LAB demonstrated the first FEL lasing driven by a plasma-accelerated electron beam, providing a key proof-of-principle for the EuPRAXIA concept ~\cite{pompili2022_pwfa_fel}.

The facility is designed as a compact and sustainable accelerator infrastructure capable of delivering soft X-ray FEL radiation in the water-window region for applications in biology, materials science, ultrafast photon science, and advanced accelerator research. At the same time, EuPRAXIA@SPARC\_LAB represents an important technological stepping stone toward future plasma-based linear colliders and compact accelerator-driven infrastructures ~\cite{tdr2026}.

\subsection{Objectives}
The primary objective of EuPRAXIA@SPARC\_LAB is to demonstrate stable operation of a high-brightness FEL driven by a beam produced through plasma wakefield acceleration. The facility will operate with a 1 GeV electron accelerator in two complementary configurations: a full RF mode for conventional operation and a hybrid RF+PWFA mode in which approximately half of the energy gain is provided by the plasma stage.

As shown in Fgure 1, the baseline layout consists of:

\begin{itemize}
    \item an S-band high-brightness photoinjector,
    \item a compact X-band linac,
    \item a 60 cm plasma accelerating module,
    \item the FEL undulator lines,
    \item dedicated photon beamlines for users.
\end{itemize}

The main scientific and technological goals include:

\begin{itemize}
    \item demonstrating FEL-quality electron beams from plasma acceleration,
    \item validating compact high-gradient accelerator technologies,
    \item achieving sub-percent-level beam stability compatible with user operation,
    \item developing advanced synchronization and single-shot diagnostics,
    \item providing a user platform for photon science and accelerator R\&D.
\end{itemize}

A central aspect of the project is sustainability. By replacing a substantial fraction of conventional RF acceleration with plasma acceleration gradients in the GeV/m range, EuPRAXIA aims to significantly reduce accelerator length, infrastructure footprint, and overall power consumption when compared to conventional facilities.

\subsection{Progress over the last two years}
The last two years have marked the transition of EuPRAXIA@SPARC\_LAB from conceptual design toward implementation readiness.

The Technical Design Report (TDR) was completed and published in February 2026 ~\cite{tdr2026}, providing a comprehensive validation of the facility concept, accelerator layout, FEL design, infrastructure, and implementation strategy. The international Review Committee concluded that the project objectives are realistic and that no critical showstoppers have been identified.

Major progress has been achieved in several areas as outlined hereafter.

A full-scale X-band accelerating structure prototype has been produced and prepared for high-power testing. Dedicated infrastructures such as the TEX facility are being used to validate RF performance, stability, and operational reliability.

Extensive R\&D activities at PLASMA\_LAB demonstrated operation of long discharge capillaries approaching EuPRAXIA parameters, including plasma densities and accelerating gradients required for FEL applications. Work is ongoing on plasma reproducibility, discharge stability, and beam-plasma coupling optimization.

The civil engineering and technical infrastructure designs have reached tender readiness, representing a major milestone toward construction. The implementation schedule foresees building completion around 2029 and first operational capability around 2031.

The EuPRAXIA Advanced Photon Sources (EuAPS) project, supported by the Italian PNRR programme, is expected to become operational by the end of 2026 and will represent the first operational user-oriented building block of the EuPRAXIA distributed infrastructure.

In parallel, strong progress has been made in project organization, industrial engagement, synchronization systems, diagnostics, and international coordination through the EuPRAXIA Preparatory Phase and EuPRAXIA Doctoral Network activities.

\subsection{Time scale for application}
EuPRAXIA@SPARC\_LAB is entering the implementation phase.

The current roadmap foresees:

\begin{itemize}
    \item start of building construction by the end of 2026,
    \item completion of major civil infrastructures around 2029,
    \item installation and commissioning of accelerator systems between 2029 and 2031,
    \item and initial user operation shortly thereafter.
\end{itemize}

The facility is conceived as a long-term European platform for advanced accelerator technologies and photon science. Future upgrades already under consideration include:

\begin{itemize}
    \item higher repetition rate operation (up to several hundred Hz),
    \item longer plasma stages for multi-GeV acceleration,
    \item enhanced average photon flux,
    \item advanced Compton and betatron radiation sources,
    \item experimental programs relevant to plasma-based linear collider development.
\end{itemize}

EuPRAXIA@SPARC\_LAB therefore represents both a near-term user facility and a strategic European investment toward the future generation of compact, sustainable accelerator infrastructures for science, industry, and high-energy physics.

\begin{figure}[h!tbp]
    \centering
    \includegraphics[width=1\linewidth]{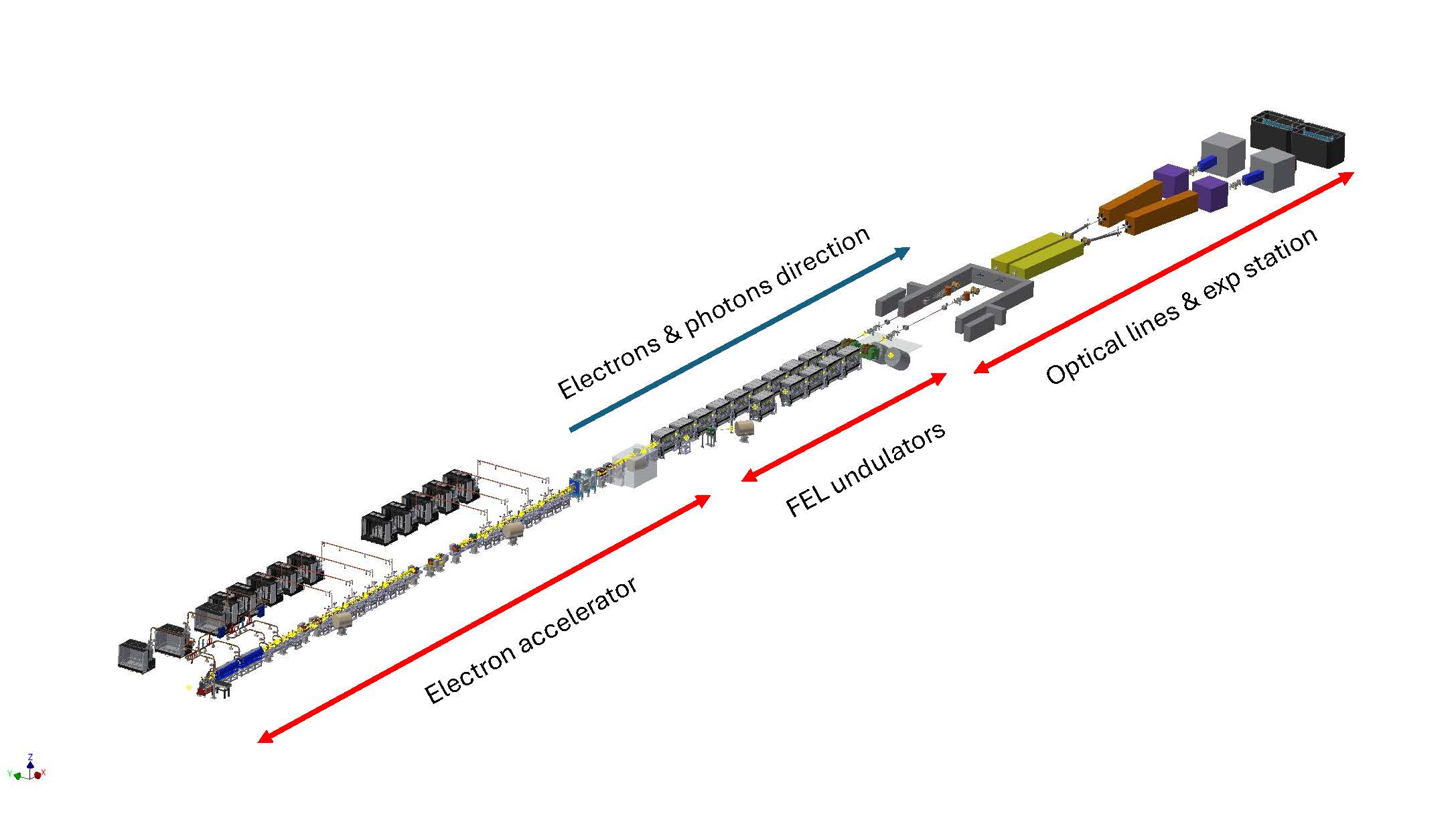}
    \caption{Layout of the accelerator and beamlines}
    \label{01fig:layout}
\end{figure}

\newpage




\chapterimage{figures/images_head_icep}
\addcontentsline{toc}{chapter}{Bibliography}
\printbibliography


\chapterimage{figures/images_head_feuillep}
\glsaddall

\printglossaries

\addcontentsline{toc}{chapter}{Glossary}
\vfill\eject


\chapterimage{figures/images_head_gravurep} 

\chapter{Committees and Participants}

 \textbf{Organizing committee} composed of Members of the ICFA ANA panel and of Livio Verra (INFN - Frascati National Laboratory).

\vspace{0.5cm}
\noindent \textbf{Local Organizing committee (INFN - LNF)}\\
Livio Verra, \textit{Chair}\\
Martina Carillo\\
Maria Rita Ferrazza\\
Francesco Fransesini\\
Mario Galletti\\
Manueala Giabbai\\
Giulia Latini\\
MariaChiara Mercuri\\
Gianmarco Parise\\
Giulia Vinicola\\

\vspace{0.5cm}
\noindent \textbf{List of workshop attendees}:
66 participants\\
Erik Adli, University of Oslo\\
Emanuele Angelo Bagnaschi, INFN LNF\\
Timothy Barklow, SLAC\\
Gholamreza Bazrafshan Delijani, DESY\\
Carlo Benedetti, LBNL\\
Allen Caldwell, Max Planck Institute for Physics\\
Pierluigi Campana, INFN-LNF\\
Martina Carillo, INFN-LNF\\
Oksana Chubenko, Northern Illinois University\\
Laura Corner, Cockcroft Institute, University of Liverpool\\
Brigitte Cros, LPGP CNRS Université Paris Saclay\\
Richard D'Arcy, University of Oxford\\
Romain Demitra, INFN-LNF\\
Claudio Di Giulio, INFN-LNF\\
Keegan Downham, University of California, Santa Barbara\\
Pierre Drobniak, University of Oslo\\
Antonio Falone, INFN-LNF\\
John Farmer, Max Planck Institute for Physics\\
Massimo Ferrario, INFN-LNF\\
Maria Rita Ferrazza, INFN-LNF\\
Gaetano Fiore, INFN-NA\\
Francesco Fransesini, INFN-LNF\\
Mario Galletti, INFN-LNF\\
Alessandro Gallo, INFN-LNF\\
Almantas Galvanauskas, University of Michigan\\
Spencer Gessner, SLAC\\
Manuela Giabbai, INFN-LNF\\
Luca Giannessi, INFN-LNF\\
Paola Gianotti, INFN-LNF\\
Leonida Antonio Gizzi, CNR - INO, INFN-PI\\
Thomas Grismayer, Insituto Superio Técnico, Lisbon University\\
Edda Gschwendtner, CERN\\
Marcel Hohmann, DESY\\
Verena Kain, CERN\\
Alexei Kanareykin, Euclid Techlabs LLC\\
Alexander Knetsch, SLAC\\
Giulia Latini, INFN-LNF\\
Wei Lu, Institute of High Energy Physics\\
Andreas Maier, DESY\\
Francesco Massimo, LPGP - CNRS\\
Mariachiara Mercuri, INFN-LNF\\
Sebastian Meuren, LULI, CNRS, Ecole Polytechnique\\
Patric Muggli, Max-Planck-Institut für Physik\\
Pietro Musumeci, UCLA\\
Pedro Oliveira, CLF\\
Toby Opferkuch, SISSA\\
Jens Osterhoff, LBNL\\
Gianmarco Parise, INFN-LNF\\
Parth Patil, University of Hamburg\\
Carlo Alberto Piccione, Ocem Power Electronics\\
Mikhail Polyanskiy, BNL\\
John Power, Argonne National Laboratory\\
Alexander Pukhov, Heinrich-Heine-University of Duesseldorf\\
Shriyansh Ranjan, DESY\\
Angira Rastogi, LBL\\
Lars Reichwein, Forschungszentrum Jülich\\
Dmitrii Samoilenko, DESY\\
Bruno Spataro, INFN-LNF\\
Federica Stocchi, INFN-LNF\\
Doug Storey, SLAC\\
Maxence Thévenet, DESY\\
Marlene Turner, CERN\\
Livio Verra, INFN-LNF\\
Giulia Vinicola, INFN-LNF\\
Marija Vranic, Instituto Superior Técnico, University of Lisbon\\
Matthew Wing, UCL\\
\cleardoublepage

\end{document}